\PassOptionsToPackage{table,dvipsnames}{xcolor}
\documentclass[reprint,superscriptaddress,amsmath,amssymb,aps,pra,twocolumn]{revtex4-2}

\usepackage[braket, qm]{qcircuit}

\usepackage[normalem]{ulem}
\usepackage{svg}
\usepackage{xr}
\usepackage{graphicx}%
\usepackage{makecell}
\usepackage{dcolumn}%
\usepackage{tabularx}
\usepackage{multirow}
\usepackage{bm}%
\usepackage{hyperref}
\hypersetup{breaklinks=true, colorlinks=true, linkcolor={blue!90!black}, citecolor={blue!90!black}, urlcolor={blue!90!black}}
\usepackage{braket}
\usepackage[nameinlink]{cleveref}
\usepackage{stmaryrd}
\usepackage{subfiles}
\usepackage{bibunits}

\let\oldaddcontentsline\addcontentsline

\newcommand{\starttocentries}{\let\addcontentsline\oldaddcontentsline}

\usepackage{amsthm, amsmath, amsfonts, amssymb}
\usepackage{braket}
\usepackage{graphicx}%
\usepackage{dcolumn}%
\usepackage[dvipsnames]{xcolor}

\usepackage{times}
\usepackage{titlesec}
\usepackage{cleveref}
\usepackage[utf8]{inputenc}
\usepackage{verbatim}
\usepackage{soul}
\usepackage{booktabs}
\usepackage{xr}
\usepackage{blindtext}
\usepackage{titlesec}
\usepackage{enumitem}
\usepackage{mathtools}
\usepackage{slashed}
\usepackage{comment}
\usepackage{newtxmath}
\usepackage{amsfonts}
\usepackage{makecell}
\usepackage{tabularx}
\usepackage{comment}
\usepackage{epstopdf}
\usepackage{float}
\usepackage{color}
\usepackage{multirow}
\usepackage{physics}
\usepackage{empheq}
\usepackage{bbm}
\usepackage{mdframed}
\usepackage{enumitem}
\usepackage{bbm}
\usepackage{array}
\usepackage{wasysym}
\usepackage{tikz}
\usetikzlibrary{
  arrows.meta,
  backgrounds,
  calc,
  decorations.pathreplacing,
  fit,
  matrix,
  positioning,
  shapes.geometric
}
\usepackage{microtype}

\definecolor{PlaqBlue}{HTML}{1F5A89}
\definecolor{PlaqBlueLight}{HTML}{E6F0F7}
\definecolor{PlaqRed}{HTML}{A63D40}
\definecolor{PlaqRedLight}{HTML}{F7E8E8}
\definecolor{PlaqGold}{HTML}{B07B19}
\definecolor{PlaqGoldLight}{HTML}{F7F0DD}
\definecolor{PlaqGreen}{HTML}{2D7054}
\definecolor{PlaqGreenLight}{HTML}{E5F2EB}
\definecolor{PlaqGray}{HTML}{5C6670}

\newcolumntype{Y}{>{\raggedright\arraybackslash}X}
\newcolumntype{C}{>{\centering\arraybackslash}X}

\providecommand{\ket}[1]{\lvert #1\rangle}
\providecommand{\bra}[1]{\langle #1\rvert}

\newcommand{\compactdots}{.\mkern-1mu.\mkern-1mu.}

\tikzset{
  qwire/.style={line width=0.62pt,PlaqGray},
  classical/.style={double,double distance=1.2pt,line width=0.35pt,PlaqGray},
  gate/.style={draw=PlaqBlue,fill=PlaqBlueLight,rounded corners=1.4pt,
    line width=0.7pt,minimum height=7mm,minimum width=12mm,
    align=center,font=\footnotesize},
  routegate/.style={draw=PlaqGreen,fill=PlaqGreenLight,rounded corners=1.4pt,
    line width=0.7pt,minimum height=9mm,minimum width=17mm,
    align=center,font=\footnotesize},
  selectgate/.style={draw=PlaqRed,fill=PlaqRedLight,rounded corners=1.4pt,
    line width=0.7pt,minimum height=9mm,minimum width=17mm,
    align=center,font=\footnotesize},
  phasegate/.style={draw=PlaqGold,fill=PlaqGoldLight,rounded corners=1.4pt,
    line width=0.7pt,minimum height=7mm,minimum width=12mm,
    align=center,font=\footnotesize},
  ctrl/.style={circle,fill=black,inner sep=1.55pt},
  octrl/.style={circle,draw=black,fill=white,line width=0.6pt,inner sep=1.65pt},
  flow/.style={-{Latex[length=2.2mm,width=1.5mm]},line width=0.75pt},
  annotation/.style={font=\scriptsize,align=center,text=PlaqGray},
  site/.style={circle,draw=PlaqGray,fill=white,line width=0.7pt,inner sep=2.35pt},
  bus/.style={circle,draw=PlaqRed,fill=PlaqRedLight,line width=0.8pt,inner sep=2.7pt}
}

\usepackage[figure,boxed,linesnumbered]{algorithm2e}
\DontPrintSemicolon
\SetKwInOut{Input}{Input}
\SetKwInOut{Output}{Output}
\SetKwComment{Comment}{/* }{ */}
\SetKw{Parameters}{Parameters : }
\SetKwFunction{SingleInvert}{SingleInvert}
\SetKwFunction{ForwardLearn}{ForwardLearn}
\IncMargin{1em}
\let\oldnl\nl
\newcommand{\nlnonumber}{\renewcommand{\nl}{\let\nl\oldnl}}

\makeatletter
\newcommand{\removelatexerror}{\let\@latex@error\@gobble}
\makeatother

\crefname{theorem}{thm.}{thm.}
\crefname{section}{sec.}{sec.}
\crefname{corollary}{cor.}{cor.}
\crefname{lemma}{lemma}{lemma}
\Crefname{theorem}{Thm.}{Thm.}
\Crefname{section}{Sec.}{Sec.}
\Crefname{figure}{Fig.}{Fig.}
\Crefname{equation}{Eqn.}{Eqn.}

\newtheorem{remark}{Remark}

\newmdtheoremenv{protocol_framed}{Protocol}
\newmdtheoremenv{algorithm_framed}{Algorithm}

\titleformat{\paragraph}
{\normalfont\normalsize\bfseries}{\theparagraph}{1em}{}
\titlespacing*{\paragraph}
{0pt}{3.25ex plus 1ex minus .2ex}{1.5ex plus .2ex}

\usepackage{oplotsymbl}

\makeatletter

\newcommand{\startappendixcontents}{%
  \let\saved@addtocontents\addtocontents
  \renewcommand{\addtocontents}[2]{%
    \def\toc@target{##1}\def\toc@main{toc}%
    \ifx\toc@target\toc@main
      \saved@addtocontents{apc}{##2}%
    \else
      \saved@addtocontents{##1}{##2}%
    \fi
  }%
}

\newcommand{\appendixtableofcontents}{%
  \begingroup
    \renewcommand{\tocname}{Appendix contents}%
    \let\saved@starttoc\@starttoc
    \def\@starttoc##1{\saved@starttoc{apc}}%
    \tableofcontents
  \endgroup
}

\let\saved@toc@writedimen\toc@writedimen
\def\toc@writedimen#1{%
  \ifdim\csname tocdim@#1\endcsname
      <\csname tocmax@#1\endcsname\relax
    \csname tocdim@#1\endcsname
      =\csname tocmax@#1\endcsname\relax
  \fi
  \expandafter\xdef\csname tocmax@#1\endcsname{%
    \the\csname tocdim@#1\endcsname}%
  \saved@toc@writedimen{#1}%
}

\makeatother

\begin{document}

 \title{Low-frustration block encodings by patch routing}

\author{Minzhao Liu}
\thanks{Equal contributions}
\affiliation{IonQ Inc, College Park, MD, 20740, USA}

\author{Abeynaya Gnanasekaran}
\thanks{Equal contributions}
\affiliation{IonQ Inc, College Park, MD, 20740, USA}

\author{Ananth Kaushik}
\affiliation{IonQ Inc, College Park, MD, 20740, USA}

\author{Siddharth Hariprakash}
\affiliation{IonQ Inc, College Park, MD, 20740, USA}

\author{Alireza Marefat Khah}
\affiliation{IonQ Inc, College Park, MD, 20740, USA}

\author{Martin Roetteler}
\affiliation{IonQ Inc, College Park, MD, 20740, USA}

\author{Grant Salton}
\affiliation{IonQ Inc, College Park, MD, 20740, USA}

\date{\today}

\begin{abstract}
Block-encoding has emerged as a core primitive in many quantum algorithms, including quantum simulation. In this work, we design efficient and extensible methods for generating low-frustration block encodings of frustrated lattice Hamiltonians built from repeated local interaction patches through a technique we call ``patch routing''. We construct block encodings for a full Hamiltonian by first block-encoding a small local patch and then using a SWAP-style routing network to coherently apply the local patch encoding across the lattice. The expensive, local block-encoding cost is paid only once per query and is independent of lattice size, whereas the comparatively inexpensive routing network grows linearly with the lattice size, dominating at large sizes. Our method allows for a hierarchical structure in which larger patches can be block-encoded with small normalization factors, albeit at a higher (local, non-extensive) block-encoding cost, thereby reducing overall query complexity in an algorithm. We exploit symmetries and algebraic structures in local interactions to construct efficient block encodings, and we apply these techniques to the Heisenberg, Fermi--Hubbard, and honeycomb spin-liquid models. For a $120\times 120$ square-lattice $J_1$--$J_2$ Heisenberg model near maximal frustration ($J_2/J_1=0.4-0.6$), our methods result in a $15-17\times$ reduction in the leading $T$ cost per qubitized-QPE run (ignoring initial state preparation and overlap) relative to prior state-of-the-art Pauli LCU constructions. Corresponding reductions are approximately $3\times$ for the $120\times120$ Fermi--Hubbard model at half-filling. For time evolution of the honeycomb spin-liquid model on $120\times120$ unit cells, the leading cost is reduced by $5\times$. Our framework provides a systematic way forward for larger and more accurate fault-tolerant simulations of real-world quantum systems.
\end{abstract}

\maketitle

\tableofcontents
\clearpage

\section{Introduction}

Block encoding methods have emerged in recent years as core primitives for many quantum algorithms~\cite{low2019qubitization,gilyen2019quantum}, such as qubitization-based quantum phase estimation (QPE), quantum linear system solvers, and algorithms based on quantum singular value transformation (QSVT). At its core, block encoding is a technique that allows for the use of a non-unitary operator within quantum circuits by encoding the operator in the top left block of a larger unitary.
Explicit quantum circuits for constructing block encodings have been developed in several cases of interest~\cite{clader2022quantum,quantumAlgorithmsWiki,camps2024explicit,Kane:2024odt,Hariprakash:2023tla,liu2026lowgatecountblockencodingssecondquantized,Antonioli:2026izv,Nibbi:2026rhu,Rhodes:2024zbr,Draper:2026bcj,Liu:2024hmm} with different tradeoffs between circuit cost and normalization. For quantum simulation, one is often interested in constructing efficient block encodings of Hamiltonians for use in fault-tolerant algorithms for time evolution or ground state preparation, and a particularly interesting class of Hamiltonians for quantum simulation is lattice-based models. Such lattice Hamiltonians often arise as effective descriptions of many-body physics, condensed matter systems, high energy physics, or materials sciences problems. Quantum simulation of these lattice models are among the primary targets for early fault-tolerant quantum computers, with applications including ground state energy estimation, low energy state preparation, and time evolution~\cite{babbush2018encoding,lin2020near,low2019qubitization,bartschi2026potential}. The cost of using these algorithms depends on both the cost to implement each query to the block encoding and the number of queries required in the overarching algorithm.

\subsection{Pauli LCU block encodings}

One standard approach to block-encoding uses a technique called linear combination of unitaries (LCU)~\cite{childs2012hamiltonian,Berry2015}, in which the Hamiltonian is expanded as a linear combination of Pauli strings $H=\sum_p \lambda_p P_p$, where $P_p$ is a (possibly multi-qubit) Pauli string and the $\lambda_p$ are scalar coefficients such that the coefficient $\ell_1$-norm is $\Lambda=\sum_p \abs{\lambda_p}$. A block encoding is then built from standard primitives~\cite{quantumAlgorithmsWiki}. The so-called preparation oracle, \textsc{Prep}, creates a quantum state $\textsc{Prep}\ket{0}=\frac{1}{\sqrt{\Lambda}}\sum_p \sqrt{\vert\lambda_p\vert}\ket{p}$ (ignoring the garbage) encoding the coefficients. The selection oracle, \textsc{Sel}, applies Pauli string $P_p$ conditioned on an input register being in state $p$: $\textsc{Sel} = \sum_p \operatorname{sign}(\lambda_p)P_p\otimes\ketbra{p}{p}$. Then, the operator $\textsc{Prep}^\dagger\,\textsc{Sel}\,\textsc{Prep}$ is a block encoding of the normalized Hamiltonian $H/\Lambda$. Then, a low-energy state preparation algorithm typically calls the block encoding $O(\Lambda)$ times.

While the Pauli LCU method provides a general and simple construction for constructing block encodings, it is often highly inefficient for lattice Hamiltonians. First, the normalization constant is the $\mathbb{L}_1$ norm of the coefficients and can exceed the optimal normalization given by the spectrum of the Hamiltonian. Second, the \textsc{Sel} oracle is a series of multi-controlled operators, resulting in a large $T$-gate count and depth, even for simple two-dimensional Hamiltonians with nearest-neighbor interactions and only two bonds per site. 

One standard approach to reduce the cost per query to the block encoding, known as unary iteration~\cite{babbush2018encoding}, works by reducing the cost of the \textsc{Sel} oracle. Instead of using an independent multi-controlled gate for each term in the LCU, unary iteration reuses the control logic when iterating over the terms, resulting in a cost of $K-1$ Toffoli gates for $K$ terms~\cite{babbush2018encoding}. This results in a Toffoli count given by the number of terms, not the number of qubits representing the physical system, which is inefficient. A further reduction in the cost can be obtained for lattice Hamiltonians by exploiting the fact that the selection oracle has a product structure \cite{yoshioka2024hunting}. For instance, for a 2-local Hamiltonian, the selection oracle can iterate over the system qubits twice to implement each family of the two-body Pauli product terms. The idea was first implicitly introduced for the electronic systems in \cite{babbush2018encoding} and then extended to spin systems \cite{yoshioka2024hunting}, where it was termed as a product-wise construction. For each family of two-body Pauli products, this implementation uses two unary-iteration passes over the system qubits, one for each endpoint. With $f$ such families, the selection cost is approximately $8f(N_s-1)$ $T$ gates, where $N_s$  is the number of qubits required to represent the system (system size).

\subsection{Block encoding normalization and shift}

Another way to reduce the simulation cost is to reduce the number of queries to the block encoding, which depends on the block-encoding normalization $\Lambda$ and energy shift. More concretely, suppose that $L\leq E_{\min}$ and $U\geq E_{\max}$ are bounds on the spectrum of $H$. Consider a block encoding that embeds the normalized Hamiltonian $\widebar{H}$ on a block:
\begin{align}
\widebar{H}=\frac{H-\Delta I}{\Lambda},\qquad
\Delta=\frac{U+L}{2},\qquad
\Lambda=\frac{U-L}{2},
\label{eq:intro_shift_normalization}
\end{align}
such that the spectrum of $\widebar{H}$ now lies in the range $[-1,1]$. If the bounds are tight, this transformation gives the smallest possible normalization $\Lambda_\mathrm{opt}=(E_{\max}-E_{\min})/2$. Now, consider the problem of energy estimation using QPE with energy error $\epsilon_E$, which can be used for ground state preparation with energy error $\epsilon_E$. Qubitization relates the normalized ground state energy $x_0$ to the eigenphase $\theta_0$ of the qubitized walk operator as $x_0 = \cos \theta_0$~\cite{low2019qubitization}. For sufficiently small phase errors and a ground state energy in the interior of the normalized interval, $|\delta E|\approx\Lambda_{\mathrm{eff}}|\delta\theta|$, where
\begin{align}
\Lambda_{\mathrm{eff}}
&=\left|\frac{dE}{d\theta}\right|_{E_{\min}} =\Lambda|\sin\theta_0|\nonumber\\
&=\sqrt{\Lambda^2-(E_{\min}-\Delta)^2}
=\sqrt{(E_{\min}-L)(U-E_{\min})}.
\label{eq:intro_effective_normalization}
\end{align}
Since the query count of the phase estimation algorithm scales as $\Lambda_{\mathrm{eff}}/\epsilon_E$ in this regime, tightening the spectral bounds reduces the overall cost.

A similar dependence on $\Lambda_{\mathrm{eff}}$ is observed in eigenstate filtering algorithms based on QSP or QSVT. These algorithms apply a polynomial that retains the desired eigenstate and suppresses the others~\cite{lin2020near}. The degree $d$ of the QSP polynomial controls how sharply the filter can change with the qubitization phase $\theta$, where the normalized energy $x = \cos \theta$. For ground state filtering, the polynomial must distinguish the ground energy from the first excited state that is separated by a gap $G$. In the interior regime discussed above, and for a sufficiently small gap, the corresponding phase separation is approximately $G/\Lambda_{\mathrm{eff}}$, and the required degree scales as $\widetilde{\mathcal{O}} (\Lambda_{\mathrm{eff}}/G)$~\cite{thibodeau2023nearly}. As the number of block-encoding queries is proportional to the polynomial degree, tighter spectral bounds can reduce the filtering costs as well.

Further savings are possible when the target eigenvalues lie sufficiently near an endpoint of the normalized interval. For instance, at $x_0=-1$, the phase to energy relation is $|\delta E|=\Lambda(1-\cos\delta\theta)\approx\Lambda(\delta\theta)^2/2$, so resolving the energy near this endpoint requires $\mathcal{O}(\sqrt{\Lambda/\epsilon_E})$ queries. Such improved endpoint resolution is used in nearly-frustration-free ground state preparation~\cite{thibodeau2023nearly}. These ideas also underlie spectral gap amplification~\cite{somma2013spectral} and sum-of-squares spectral amplification (SOSSA)~\cite{low2025fast,king2026sossa}. This provides additional motivation to obtain tighter spectral bounds.

In order to truly realize the savings from tight bounds, we also need a circuit construction that encodes $\widebar{H}$ at the chosen normalization. However, since constructing an arbitrary unitary as an optimal block encoding is impractical, we often cannot improve the normalization without incurring a significantly higher cost. A useful construction must balance achieving tighter bounds, and keeping the cost of the block encoding low.

\subsection{A patch-routing block encoding}

Lattice Hamiltonians can be written as a sum of local interaction terms, $H = \sum_k H_k$, where each $H_k$ is defined on some set of sites in a neighborhood, which we refer to as a patch (e.g., a bond, a triangle, a square plaquette, or larger patch). Each patch Hamiltonian can contain two-body interactions, many-body interactions or a combination of them. By decomposing the Hamiltonian into patches, one can obtain spectral bounds on the local operators and use them to construct block encodings of the full Hamiltonian. Using the extreme eigenvalues of the local operators, we define upper and lower bounds as
\begin{align*}
L=\sum_k e_{\min}(H_k)\leq E_{\min},\qquad
U=\sum_k e_{\max}(H_k)\geq E_{\max}.
\end{align*}
The lower bound $L$ is exact/tight for frustration-free Hamiltonians, where the ground state minimizes every local term in a valid decomposition of the Hamiltonian. Ground state energy estimation is trivial for such Hamiltonians as the energies of the local terms can be computed exactly and summed up to give the energy of the full Hamiltonian. Note, however, that ground state preparation itself can be a non-trivial task. While the techniques in this work apply to frustration-free Hamiltonians, we focus on \textit{frustrated} Hamiltonians for which $E_{\min}>L$.

A number of valid decompositions exist for any given lattice Hamiltonian. We can choose our patches to cover a plaquette, part of a plaquette, or several plaquettes. For each patch, we then choose how to group its interactions into the local Hamiltonian terms $H_k$. For example, the interactions within a square plaquette can be treated as separate bond terms, grouped into triangle terms, or treated jointly as one four-spin term. The spectral bounds depend on this grouping. When bonds are treated separately, the sum of their minimum energies does not account for the frustration between them. A triangle term includes this frustration in its minimum energy and can therefore give a tighter lower bound. Larger patches allow us to form larger groups of interactions and account for more of the frustration within each local term. Intuitively, larger patches can better represent the bulk physics of the full Hamiltonian by capturing additional frustration. 

While larger patches can improve the spectral bounds, we need a local encoding that achieves these bounds to obtain the promised query savings. Constructing such an encoding is a non-trivial task in general, and its circuit cost usually increases with the patch size. This cost becomes particularly important when the \textsc{Sel} operation applies a controlled local encoding at each addressed patch. Unary iteration can efficiently enumerate all patches, but a direct implementation still contains a copy of the local circuit at each location. Thus, the local encoding cost is multiplied by the number of patches, which is proportional to the system size $N_s$. Any additional controls can increase this cost further. As a result, the increase in circuit cost can far outweigh the savings in query count.

\begin{figure*}[t]
\centering
\includegraphics[width=0.8\textwidth]{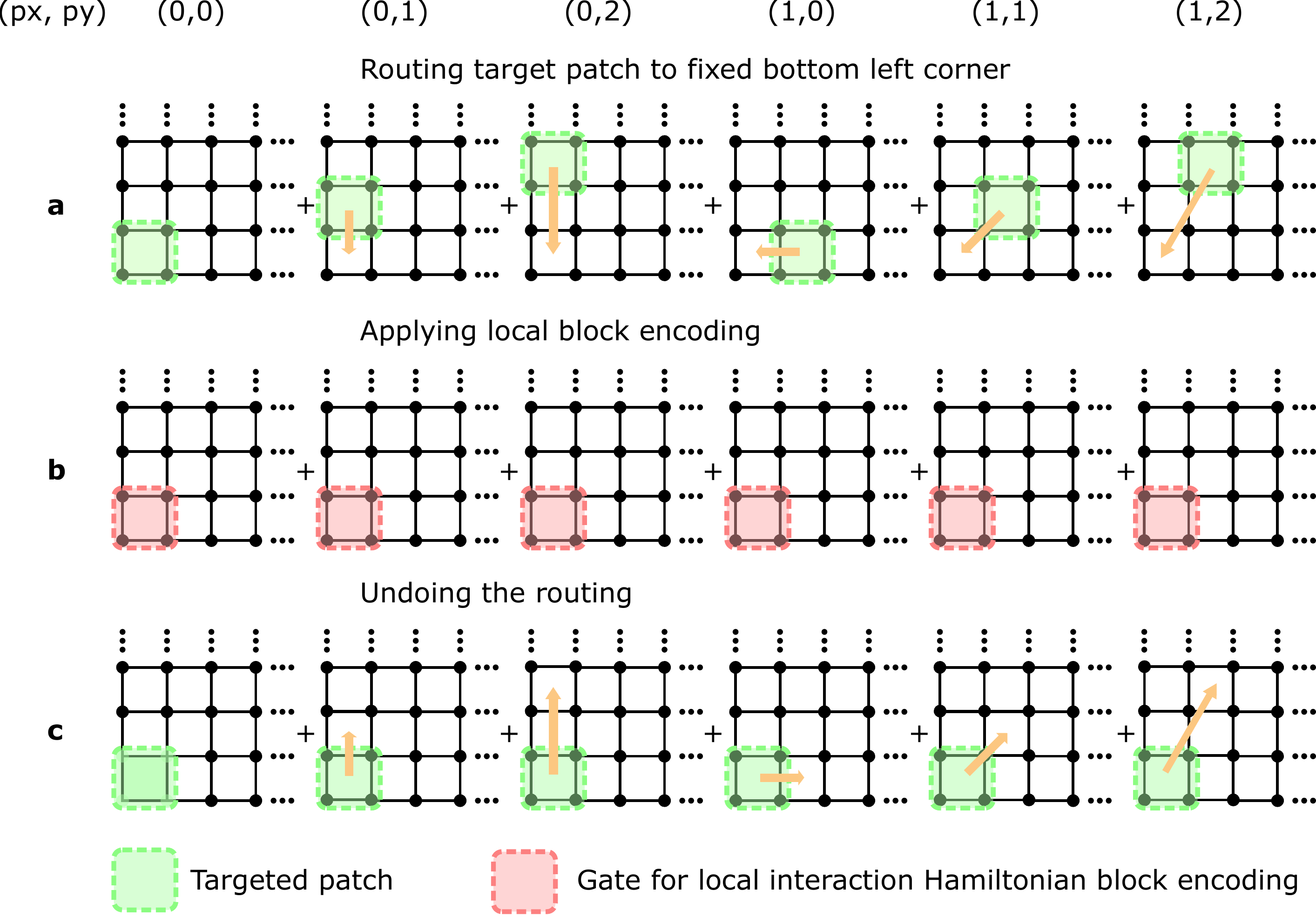}
\caption{\textbf{An example of our block encoding via routing local patches on the square lattice.} \textbf{a} The routing step. The register $(p_x,p_y)$ specifies coherently which target local patch (green) to route to the bottom left corner. \textbf{b} Apply the gates that implement the local interaction Hamiltonian block encoding on it. \textbf{c}  The unrouting step, which coherently swaps the patch back controlled on $(p_x, p_y)$.}
\label{fig:routing}
\end{figure*}

In this work, we introduce a routing-based approach to address this tradeoff, illustrated in Fig.~\ref{fig:routing}. We coherently route the selected patch to a fixed location, apply the local block encoding once, and reverse the routing. When every patch Hamiltonian is, up to a shift and rescaling, a translated copy of a single reference operator, as in translation-invariant lattice models, the same local circuit serves every patch. The routing circuit depends on the lattice geometry and the size of the patch, and its cost scales linearly with the system size $N_s$.  Even without normalization savings, the routed block encoding is often cheaper than state-of-the-art product-wise Pauli LCU constructions implemented using unary iteration.

For a fixed patch size, the local encoding cost is independent of $N_s$ and is added to the routing cost. This allows us to use more elaborate local encodings that achieve tighter spectral bounds without repeating the expensive local circuit at every patch location. We use this approach to construct a hierarchy of block encodings for standard Hamiltonians and evaluate the total cost from both the number of queries and the cost per query. For the lattice Hamiltonians considered here, our routing-based constructions outperform state-of-the-art approaches based on Pauli LCU constructions that exploit the product structure of interaction terms and use unary iteration for efficient implementation. Moreover, our methodology can be easily extended to other lattice Hamiltonians. We summarize our contributions next.

\subsection{Our contributions}

Below we briefly summarize our main contributions:
\begin{itemize}
    \item We construct coherent routing circuits for both qubit and fermionic systems on regular lattices with arbitrary unit cells. The resulting block encoding is often itself cheaper than existing methods, and has improved normalization and shift.
    \item We give a spectral method for constructing local block encodings, with exact constructions for small patch Hamiltonians. 
    The extreme eigenvalues of these patches determine the local shift and normalization, and unitary dilation gives an encoding of the resulting Hermitian contraction. We exploit commuting observables, degenerate eigenspaces, and spin and lattice symmetries to reduce the $T$-cost of the local circuits.
    \item For an initial block encoding with suboptimal normalization (e.g., combining smaller encodings to obtain an initial encoding of a larger patch), we use quantum eigenvalue transformation (QET) to improve its normalization. This provides an alternative and general approach when directly constructing the spectral encoding is too difficult or expensive.
    \item We provide pedagogical analysis and constructions for the Heisenberg and Fermi--Hubbard models on square lattices, and the spin-liquid Hamiltonian on a honeycomb lattice. This allows us to demonstrate square lattice, general lattice, long range, and fermionic routing. It also allows us to demonstrate the ability to reduce residual frustration and overall query complexity when exploiting the hierarchy of larger local patches. To the best of our knowledge, these results also allow us to obtain the best known $T$ counts for qubitization-based QPE of the Heisenberg and Fermi--Hubbard models.
\end{itemize}

The rest of the paper is structured as follows. We first introduce the methods used in this work in \Cref{sec:methods}, including patch-based block encodings, routing, and methods to build local patch block encodings. We then provide pedagogical analysis and constructions for the three systems mentioned above in \Cref{sec:heisenberg}, \Cref{sec:nn-hubbard}, and \Cref{sec:spin_liquids}. Gate accounting and concrete resource estimates for the Heisenberg and Fermi--Hubbard models are collected in~\Cref{app:cost_accounting,app:routing_cost_accounting}.

\section{Methods}
\label{sec:methods}

In this section, we develop the patch routing construction introduced above. We first show how the spectral bounds of grouped interactions determine the global shift and normalization. We then construct routing circuits that allow a single local encoding to serve every translated patch, before describing how to build local circuits that achieve the chosen normalization. For convenience, we have included a summary of definitions for some of the more important quantities used throughout the work in~\Cref{app:notation}.

\subsection{Patched block encodings}
Let $H$ be a lattice Hamiltonian written as a sum over patches,
\begin{align}
H=\sum_{p}H_p,
\end{align}
where $p$ specifies the patch and $H_p$ is a Hamiltonian that acts on that patch. In general, the patches can overlap and hence, the overlapping interactions on a patch are weighted so that summing over the patches produces the original Hamiltonian. Our goal is to construct an efficient block encoding of $H$ using the local encodings of the patch Hamiltonians.  Within each patch, we can encode all the interactions jointly or combine the encodings of smaller interaction terms. Treating all the interactions jointly can give tighter spectral bounds. On the other hand, constructing an encoding that achieves the resulting normalization can have a higher local circuit cost as we will discuss later.

For a chosen grouping of interactions, we first use the local spectral bounds to determine the shift and normalization that we want the encoding to achieve. Each patch Hamiltonian is shifted and normalized as follows,
\begin{align}
\widebar{H}_p=\frac{H_p-\delta_p I}{\lambda_p},\qquad \|\widebar{H}_p\|\leq 1.
\label{eq:methods_local_normalization}
\end{align}
Here $\delta_p, \lambda_p$ are computed  based on the spectral bounds $\ell_p\leq e_{\min}(H_p)$ and $u_p\geq e_{\max}(H_p)$ as $\delta_p=(u_p+\ell_p)/2$ and $\lambda_p=(u_p-\ell_p)/2$. If these bounds are exact, we get the optimal shift and normalization for the patch. Using the bounds on the extreme eigenvalues of a sum of Hermitian matrices, $
\sum_p e_{\min}(H_p)\leq E_{\min}(H)$ and $
\sum_p e_{\max}(H_p)\geq E_{\max}(H)
$, we obtain the global shift $\Delta$ and normalization $\Lambda$ simply by summing over the local ones,
\begin{align}
\Delta=\sum_p\delta_p,\qquad \Lambda=\sum_p\lambda_p,\qquad
\frac{H-\Delta I}{\Lambda}=\sum_p\frac{\lambda_p}{\Lambda}\widebar{H}_p.
\label{eq:methods_global_normalization}
\end{align}
The shifted and normalized Hamiltonian is therefore a weighted sum of the normalized patch operators, with weights $\lambda_p/\Lambda$. We use these weights to define the \textsc{Prep} oracle on the  address register as, 
\begin{align}
\textsc{Prep}\ket{0}=\sum_p\sqrt{\frac{\lambda_p}{\Lambda}}\ket{p}\equiv\ket{\mathcal L}.
\label{eq:methods_prep}
\end{align}
For translated copies of the same patch Hamiltonian, we use the same local shift and normalization at every location. In this case, $\ket{\mathcal L}$ is a uniform superposition over the patches. 

Achieving the chosen normalization can require an expensive local circuit, since a simple Pauli LCU does not always attain the normalization set by the local spectral bounds. In the standard implementation of the \textsc{Sel} oracle, a controlled copy of this circuit is applied at each patch location. This multiplies the local encoding cost by the number of patches and can make the overall cost prohibitive. To sidestep this issue, we coherently route the selected patch to a fixed location, apply the local block encoding once, and reverse the routing. Define $\Pi_p$ as the unitary that brings patch $p$ to a fixed location on the lattice. Controlling on the patch address register, the routing unitary is 
\begin{align}
R=\sum_p\ket{p}\bra{p}\otimes\Pi_p.
\label{eq:methods_routing}
\end{align}
Then, define the \textsc{Sel} oracle as,
\begin{align}
\textsc{Sel}=R^\dagger(I\otimes B_\mathrm{ref})R=\sum_p\ket{p}\bra{p}\otimes\Pi_p^\dagger B_\mathrm{ref}\Pi_p.
\label{eq:methods_select}
\end{align}
where $B_\mathrm{ref}$ is the block encoding of the normalized patch operator at the fixed location, $\widebar{H}_{\mathrm{ref}}$. Since patches are translated copies of the same local Hamiltonian, $\widebar{H}_p=\Pi_p^\dagger \widebar{H}_{\mathrm{ref}}\Pi_p$.  Note that the local block encoding $B_\mathrm{ref}$ is applied without controls on the address register. Thus, applying $\textsc{Prep}$, $\textsc{Sel}$, and $\textsc{Prep}^\dagger$ gives a block encoding of the shifted and normalized Hamiltonian,
\begin{align}
(\bra{\mathcal L}\otimes\bra{0_w})\textsc{Sel}(\ket{\mathcal L}\otimes\ket{0_w})
=\sum_p\frac{\lambda_p}{\Lambda}\widebar{H}_p
=\frac{H-\Delta I}{\Lambda},
\label{eq:methods_global_signal_block}
\end{align}
where $\ket{0_w}$ is the signal state of the local encoding. The patch-address register also forms part of the global signal register. 

To compare the different constructions, we track both the normalization and the $T$-cost of each block encoding. Let $T_{\textsc{Prep}}$ be the cost of preparing the state $\ket{\mathcal L}$ over the patch address registers, and $T_{\textsc{BE}}^p$ the cost of the local encoding circuit $B_\mathrm{ref}$. The superscript $p$ denotes the patch encoding, whose cost is the same at every translated location. We use $T_R$ for the combined cost of routing and reverse routing. Thus, the cost of the full block encoding is,
\begin{align}
T_{\textsc{BE}}&=2T_{\textsc{Prep}}+T_R+T_{\textsc{BE}}^p.
\label{eq:methods_t_cost}
\end{align}
The circuit construction and cost analysis for \textsc{prep} are detailed in~\Cref{app:cost_conventions}. The resulting bound from~\Cref{eq:methods_uniform_prep_t_cost} is
\begin{align*}
T_{\textsc{Prep}}\leq32\lceil\log_2M\rceil+8.
\end{align*}
In the following subsections, we discuss circuit constructions for routing (\Cref{sec:routing}) and local encodings (\Cref{sec:local_encoding}) and track their costs along with the achieved normalization.

\begin{remark}
\label{rem:reflection}
For QPE using qubitization, the typical $\textsc{Prep}^\dagger\,\textsc{Sel}\,\textsc{Prep}$ block encodings must be a reflection, which means we need $\textsc{Sel}^2=I$, which implies $[R^\dagger(I\otimes B_{\mathrm{ref}})R]^2=I$ and $B_{\mathrm{ref}}^2=I$. This is to say that the local patch block encoding must also be a reflection. We ensure that all the local block encodings we construct in this work satisfy this requirement. 
\end{remark}

\subsection{Routing}
\label{sec:routing}

The goal of the routing circuit is to bring all the spins of a selected patch to a fixed location, while keeping the patch address coherent. We construct this circuit first for a spin system on a square lattice. This can be adapted to arbitrary lattices and we present the honeycomb lattice as a pedagogical example. We then demonstrate the extension of the idea to route larger patches. We also provide a modification of the routing construction to fermionic systems in the Jordan–Wigner representation. All the constructions described here have a leading-order cost of $8N_s$, where $N_s$ is the number of qubits used to represent the system, which we refer to as the system size.

\begin{figure*}[t]
\centering
\input{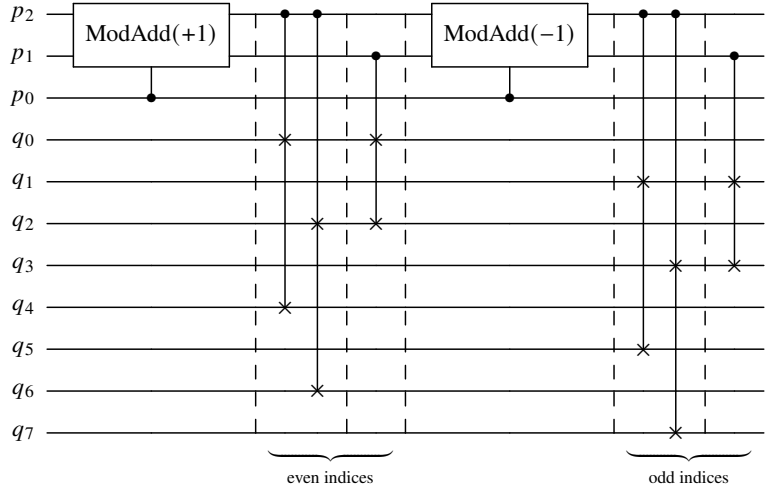}
\caption{\textbf{Routing an adjacent pair on a chain of eight qubits.} The first ModAdd block, controlled on $p_0$, increments the upper two address bits modulo four. These updated bits control the even-index network, which routes $e_{t+s}$ to position $0$. The inverse increment then restores the original address bits, which control the odd-index network and route $o_t$ to position $1$. The braces identify the two networks, each containing three CSWAPs, and dashed lines separate the routing layers.}
\label{fig:routing_1d}
\end{figure*}

\subsubsection{Routing for spin systems on a square lattice}
\label{subsec:square_lattice_routing}

First, consider a one-dimensional periodic chain of $M$ system qubits, with $M$ even. The positions are numbered from $0$ to $M-1$, and an address register of $\lceil\log_2 M\rceil$ qubits holds $p_x$. Say our goal is to bring the adjacent pair at positions $(p_x,p_x+1)$ to the front of the chain. All positions are understood to be modulo $M$. If we require the pair to arrive in the order $(0,1)$, a qubit can move to either destination depending on whether it is the first or second member of the selected pair. We instead let the pair arrive at $(s,1-s)$, where the parity $s=p_x\bmod2$. Thus, an even-indexed qubit always moves to position $0$ and an odd-indexed qubit always moves to position $1$. No qubit needs to cross between the two parity classes. This allows us to split the routing into two independent networks, each acting on $M/2$ qubits.

Let's construct the routing network. First, we identify which member of the pair $(p_x, p_x+1)$ the two networks act on. Write $p_x=2t+s$ and label the even and odd positions as $e_k=2k$ and $o_k=2k+1$. The selected pair is then
\begin{align}
(p_x,p_x+1)=
\begin{cases}
(e_t,o_t),&s=0,\\
(o_t,e_{t+1}),&s=1,
\end{cases}
\label{eq:square_routing_pair}
\end{align}
where all subscripts on $e$ and $o$ are taken modulo $M/2$. In both cases, the even member is $e_{t+s}$ and the odd member is $o_t$. Hence, one network routes $e_{t+s}$ to $e_0$, while the other routes $o_t$ to $o_0$. 

Each network is a sequence of CSWAPs controlled on the bits of its address. The even network uses $(t+s)\bmod(M/2)$ and the odd network uses $t$. The value of $t$ is already available in the address register by omitting its least significant bit, while computing $t+s$ requires a conditional modular increment. The modular reduction handles the pair $(M-1,0)$ across the periodic boundary. These addresses are computed coherently, used as controls for routing, and uncomputed when they are no longer needed. This is illustrated in~\Cref{fig:routing_1d} on a chain of eight qubits. The assumption that $M$ is even ensures that every adjacent pair contains one even and one odd position, including the pair across the periodic boundary. For odd $M$, both positions in $(M-1,0)$ are even, so the same two-network construction does not apply directly.

We now apply this construction to an $M\times M$ periodic square lattice with $N_s=M^2$ spins.  First, we treat each column as a unit and route columns $p_x$ and $p_x+1$ to columns $s$ and $1-s$, where $s=p_x\bmod2$. Then, within these two destination columns, we route rows $p_y$ and $p_y+1$ to rows $u$ and $1-u$, where $u=p_y\bmod2$. The four selected spins therefore move as follows:
\begin{align}
(p_x,p_y)&\longrightarrow(s,u),\nonumber\\
(p_x,p_y+1)&\longrightarrow(s,1-u),\nonumber\\
(p_x+1,p_y)&\longrightarrow(1-s,u),\nonumber\\
(p_x+1,p_y+1)&\longrightarrow(1-s,1-u).
\label{eq:square_routing_destination}
\end{align}
The destination order differs from the original patch order by a column reflection when $s=1$ and a row reflection when $u=1$. If the patch Hamiltonian is invariant under both reflections, the same local encoding can be applied in all four cases. This is in fact the case for the Heisenberg Hamiltonian in~\Cref{sec:heisenberg} as both reflections leave the patch Hamiltonian unchanged due to symmetry.

If the patch operator is not invariant, we restore its original order by exchanging the destination columns controlled on $s$ and the destination rows controlled on $u$, then reverse these exchanges after the local encoding. For Fermi--Hubbard Hamiltonians, these geometric exchanges must be fermionic mode permutations that preserve the Jordan--Wigner parity strings, as described below. The bounds derived in~\Cref{app:square_lattice_routing_cost}, with and without this order correction, are
\begin{align*}
T_R^{\square}&\leq8N_s+16\left\lceil\frac{\log_2N_s}{2}\right\rceil-64,\\
T_R^{\square,\mathrm{ord}}&\leq8N_s+16\left\lceil\frac{\log_2N_s}{2}\right\rceil-32.
\end{align*}
These reproduce~\Cref{eq:square_routing_t_cost,eq:square_routing_t_cost_ordered} for $N_s=M^2$ and even $M\geq4$.

\subsubsection{Routing of general lattice systems}
\label{subsec:honeycomb_lattice_routing}

A $D$-dimensional lattice with a finite repeating unit cell can be indexed by $D$ integer coordinates. Although the primitive lattice vectors need not be orthogonal and each cell may contain multiple sites, the cells form a regular $D$-dimensional array in these coordinates. Consequently, the routing construction described above for the square lattice can be readily extended to route any lattice systems. We provide a pedagogical example for the honeycomb lattice.

\begin{figure*}[t]
\centering
\includegraphics[width=0.7\textwidth]{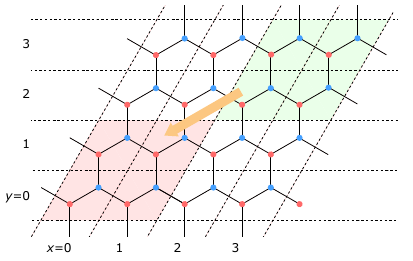}
\caption{\textbf{Routing on the honeycomb lattice.} Each rhombus unit cell contains two spins. The selected $2\times2$ block of cells is shown in green and the fixed destination in red, for address $(p_x,p_y)=(2,2)$. The block contains eight spins, including the six vertices of one hexagonal face. The even and odd routing networks act on whole cells.}
\label{fig:honeycomb}
\end{figure*}

The honeycomb lattice can be described by rhombus unit cells, each containing two spins, as shown in Fig.~\ref{fig:honeycomb}. Indexing these cells by $(x,y)$ gives the same two-dimensional address structure as the square lattice. We can therefore use the same routing construction, with a whole unit cell as the object moved by each swap. A selected $2\times2$ block of cells contains eight spins and is routed to a fixed block of four cells. 

As in the square lattice, the parity of the address determines whether the destination block is reflected in its rows or columns, and it is necessary to restore the order. The two destination columns of cells are exchanged controlled on the least significant bit of $p_x$, and the destination rows are exchanged controlled on the parity of $p_y$. Apply each exchange to both spins in a cell, and reverse the corrections after the local encoding. This restores the original spatial relationships. For $N_s=2M^2$ and even $M\geq4$, the bound in~\Cref{eq:honeycomb_fh_routing_t_cost}, derived in~\Cref{app:square_lattice_routing_cost}, is
\begin{align*}
T_R^{\,\hexago}\leq8N_s+16\left\lceil\frac{\log_2N_s-1}{2}\right\rceil-32.
\end{align*} 

\subsubsection{Long-range routing}
\label{subsec:long_range_routing}

Some systems have interactions that span more than two unit cells in either coordinate direction. In addition, as discussed earlier, larger patches can reduce the residual frustration not captured by the local encodings and thereby reduce the number of oracle calls. We therefore extend the routing construction to a $U\times U$ patch.

First, consider a one-dimensional periodic chain of $M$ qubits where $U$ divides $M$. We divide its positions into $U$ groups according to their residue modulo $U$. Group $g$ contains the qubits at positions $U\ell+g$, with $g=0,\ldots,U-1$ and $\ell=0,\ldots,L-1$, where $L=M/U$. Every section of $U$ consecutive qubits contains exactly one qubit from each group. We route that qubit to the first position of its group, which is position $g$ in the full chain. No qubit needs to cross between the groups.

To determine the address used by each group, write the section address as
\begin{align}
p&=Ut+s,\qquad 0\leq s<U.
\label{eq:uxu_address_decomposition}
\end{align}
The section starts at group $s$ in group position $t$. It passes through groups $s,\ldots,U-1$ at that position and groups $0,\ldots,s-1$ at the next position. Hence, the selected qubit of group $g$ has group address
\begin{align}
a_g&=
\begin{cases}
t,&g\geq s,\\
(t+1)\bmod L,&g<s.
\end{cases}
\label{eq:uxu_group_address}
\end{align}
A swap tree of the form used in~\Cref{subsec:square_lattice_routing} brings this qubit from group position $a_g$ to group position zero. Applying the $U$ independent trees therefore routes the qubits at positions $p+i$ to positions $(p+i)\bmod U$, for $i=0,\ldots,U-1$.

The quotient $t$, remainder $s$, comparison bits and group addresses are computed coherently from $p$. We retain these addresses and the arithmetic work until inverse routing is complete. In two dimensions, we first route the $U$ selected columns and then route the $U$ selected rows within those destination columns. The routed patch differs from its original order by cyclic shifts of its columns and rows, determined by $s_x=p_x\bmod U$ and $s_y=p_y\bmod U$. We restore the order before the local encoding and undo these corrections afterwards.

The complete gate accounting, including local order restoration and address arithmetic, is given in~\Cref{app:routing_longrange_fermion}. With $S_U=\sum_{j=0}^{\lceil\log_2U\rceil-1}[U-\gcd(U,2^j)]$, the resulting bound~\Cref{eq:uxu_routing_t_cost} is
\begin{align*}
T_R^{(U,q)}
&\leq 8N_s + 16qUS_U - 8qU^2 + \mathcal D_U(N_s) \\
& + 8(U-1)\Biggl[
    2\left\lceil
      \frac{\log_2(N_s/q)}{2} - \log_2 U
    \right\rceil + \lceil\log_2 U\rceil - 3
    \Biggr].
\end{align*}
Here $N_s=qM^2$, and $\mathcal D_U$ is the address-conversion bound in~\Cref{eq:uxu_address_conversion_bound}.

\subsubsection{Routing for fermionic systems on a square lattice}
\label{subsec:fermionic_spl}

While spins are natively encoded using qubits and swapping spins is straightforwardly accomplished by swapping qubits, swapping fermions is not as straightforward. This affects the construction of the routing network which is built on top of controlled swaps. This section constructs the routing network for fermionic systems that preserves the fermionic anticommutation relations. Here $N_s$ is the number of qubits representing the system, equivalently the number of spin-orbitals. The complete routing bound is derived in~\Cref{app:routing_longrange_fermion}.

\paragraph{Fermionic-qubit mapping}
Let us first specify the notation in the number basis,
\begin{align*}
\vert n\rangle_a=c_a^{\dagger n}\vert{\mathrm{vac}}\rangle,
\end{align*}
where $c_a^\dagger$ is the fermionic creation operator for mode $a$ and $\vert \mathrm{vac}\rangle$ is the vacuum state. Since $n_a\in\{0,1\}$, each mode can be represented by a qubit. Since there are $M^2$ sites and each site has a spin-up and a spin-down mode, there are a total of $N_s=2M^2$ qubits.

We additionally need to specify an ordering of the fermionic modes before we have a well-defined Jordan--Wigner representation. We use an ordering that is different from what is commonly employed, which we choose to make routing easier. Specifically, assume $M$ is even, define an integer index
\begin{align*}
a=a(\sigma,x,y)&=\sigma M^2+(x\bmod 2)\frac{M^2}{2}\\
&+(y\bmod 2)\frac{M^2}{4}+\left\lfloor\frac y2\right\rfloor\cdot\frac M2+\left\lfloor\frac x2\right\rfloor.
\end{align*}
In this ordering, all spin-up qubits are listed first. Within each spin collection, modes with even $x$ are listed first and odd $x$ are listed second. Within each spin and $x$-parity, modes with even $y$ are listed first and odd $y$ are listed second. The remaining order is in a row-major fashion. It turns out that separating the sites by the parity of $x$ is convenient for the construction of the Fermionic routing network, which we discuss later.

We now have a well defined Jordan--Wigner representation, with
\begin{align*}
c_a&=\left(\prod_{k<a}Z_k\right)\frac{X_a+iY_a}{2},
&n_a&=\frac{I-Z_a}{2},
&Z_{(a,b)}&=\prod_{a<k<b}Z_k
\end{align*}
and $a<b$. Then, we have
\begin{align*}
c_a^\dagger c_b+c_b^\dagger c_a
&=\frac12\left(X_aZ_{(a,b)}X_b+Y_aZ_{(a,b)}Y_b\right),\\
D_r&=\frac14Z_{r\uparrow}Z_{r\downarrow}\nonumber.
\end{align*}

We construct the fermionic routing network in the same fashion as the routing network for the Heisenberg model. However, in the Jordan--Wigner representation, routing via CSWAP is no longer correct. Nevertheless, the correct fermionic swap operation between modes $i$ and $j$, denoted as $f_{\mathrm{swap},ij}$, that maintains the proper anti-symmetrization is defined in Ref.~\cite{babbush2018low}:
\begin{align*}
f_{\mathrm{swap},ij}
=1+c_i^\dagger c_j+c_j^\dagger c_i-c_i^\dagger c_i-c_j^\dagger c    _j.
\end{align*}
One can check that in the Jordan--Wigner number basis representation,
\begin{align*}
f_{\mathrm{swap},ij}|\compactdots,n_i,\compactdots,n_j,\compactdots\rangle=(-1)^{n_in_j+(n_i\oplus n_j)\Pi_{ij}}
|\compactdots,n_j,\compactdots,n_i,\compactdots\rangle,
\end{align*}
where $\Pi_{ij}=\bigoplus_{i<r<j}n_r$.

The fermionic routing network uses the same forward and reverse swap structure as the spin construction, replacing each CSWAP by a controlled-$f_{\mathrm{swap}}$. The final routing cost~\Cref{eq:honeycomb_fh_routing_t_cost} is
\begin{align*}
T_R^{\mathrm{FH}}\leq8N_s+16\left\lceil\frac{\log_2N_s-1}{2}\right\rceil-32.
\end{align*}
For $2\times2$ routing, eight routing groups ( 2 spins$\times2x$-parity$\times2y$-parity) are routed separately. Different spin, $x$-parity, and $y$-parity collections are completely independent as specified by the mode ordering we chose above. The selected fermionic modes will be routed to modes $iM^2+j(M^2/2)+k(M^2/4)$ for $i,j,k\in\{0,1\}$. These modes are not consecutive and applying a local block encoding to them would be expensive. We therefore need to apply eight additional $f_{\mathrm{swap}}$ gates to route them to modes $0,1,\dots,7$. We neglect this part of the routing network since the cost is small.

We would still have the plaquette index-controlled column swaps separately for the even and odd $x$ columns. In the fermionic routing network here, we make this explicit by ordering the even columns first. After column swaps, we implement row swaps on the first two columns just like the Heisenberg case.

\paragraph{Implementation of controlled-$f_{\mathrm{swap}}$}

\begin{figure}[htbp]
\begin{gather}\nonumber
\scalebox{1.0}{
\Qcircuit @C=1.0em @R=0.2em @!R { \\
	 	\nghost{{c} :  } & \lstick{{c} :  } & \qw & \ctrl{2} & \ctrl{1} & \qw & \qw & \qw & \ctrl{2} & \qw  & \qw \\
	 	\nghost{{n_i} :  } & \lstick{{n_i} :  } & \ctrl{1} & \qw & \ctrl{0} & \ctrl{3} & \qw & \targ & \qw & \ctrl{1} & \qw \\
	 	\nghost{{n_j} :  } & \lstick{{n_j} :  } & \targ & \ctrl{2} & \qw & \qw & \qw & \qw & \ctrl{2} & \targ & \qw \\
	 	\nghost{{\Pi_{ij}} :  } & \lstick{{\Pi_{ij}} :  } & \qw & \qw & \qw & \qw & \ctrl{1} & \qw & \qw & \qw & \qw \\
	 	\nghost{\mathrm{anc}=0 :  } & \lstick{\mathrm{anc}=0 :  } & \qw & \targ & \qw & \ctrl{0} & \ctrl{0} & \ctrl{-3} & \targ & \qw & \qw \\
\\ }}
\end{gather}
\caption{\textbf{Circuit for controlled-$f_{\mathrm{swap},ij}$.}
}
    \label{fig:controlled_Gamma}
\end{figure}
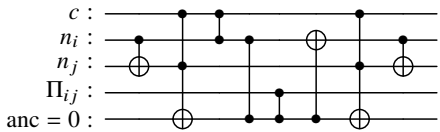

We construct controlled-$f_{\mathrm{swap},ij}$ assuming the interval parity $\Pi_{ij}$ is precomputed and available.  The action of controlled-$f_{\mathrm{swap},ij}$ needs to be
\begin{align*}
\vert 0,n_i,n_j,\Pi_{ij},0\rangle&\rightarrow\vert 0,n_i,n_j,\Pi_{ij},0\rangle,\\
\vert 1,n_i,n_j,\Pi_{ij},0\rangle&\rightarrow(-1)^{n_in_j+(n_i\oplus n_j)\Pi_{ij}}\vert 1,n_j,n_i,\Pi_{ij},0\rangle,
\end{align*}
where the first bit is the control bit and the last bit is a clean ancilla. We claim the circuit in Fig.~\ref{fig:controlled_Gamma} implements the gate. We note that since $\mathrm{anc}$ is a clean ancilla initialized and uncomputed to $\vert 0\rangle$, the two Toffoli gates in the gadget are logical AND and measurement-based inverse of logical AND, respectively. The first computes the required predicate and the second erases it.

Except the beginning and ending CNOTs, all gates are either controlled by $c$ or $\mathrm{anc}$. For $c=0$, we know $\mathrm{anc}$ stays as $0$, so the only non-trivial gates are the beginning and ending CNOTs which cancel each other. This means controlled on $c=0$, the gate is indeed identity as desired. For $c=1$, since $\mathrm{anc}$ is initialized to $0$, it suffices to enumerate all values of the $n_i,n_j,\Pi_{ij}$ registers and verify its behavior. Namely,
\begin{align*}
\vert 0,0,\Pi_{ij}\rangle&\rightarrow \vert 0,0,\Pi_{ij}\rangle,\\
\vert 0,1,\Pi_{ij}\rangle&\rightarrow (-1)^{\Pi_{ij}}\vert 1,0,\Pi_{ij}\rangle,\\
\vert 1,0,\Pi_{ij}\rangle&\rightarrow (-1)^{\Pi_{ij}}\vert 0,1,\Pi_{ij}\rangle,\\
\vert 1,1,\Pi_{ij}\rangle&\rightarrow -\vert 1,1,\Pi_{ij}\rangle.
\end{align*}
Detailed resource accounting for this gadget is given in~\Cref{app:fermionic_gadget_cost}.

\paragraph{Computation of $\Pi_{ij}$}

Let us focus on a fixed spin and $x,y$ parity group for the column swap stage. Denote the column index within this group as $t$ (i.e., $t=\lfloor x/2\rfloor$). During the column swap stage, there are $\lceil \log_2 \frac{M}{2}\rceil$ tiers of controlled fermionic swaps, each controlled by one qubit of the register $t$. This is exactly the same structure as the swap network in Ref.~\cite{low2024trading}. Each $f_{\mathrm{swap}}$ acts on all qubits between the target swapping qubits due to the need for the interval parity $\Pi$. Nevertheless, within each fixed tier, no $f_{\mathrm{swap}}$ acts on overlapping qubits. For each interval, compute $\Pi$ by CNOTs on its target qubits. Uncomputation of $\Pi$ must follow after $\Pi$ is used. One could also cache certain parity results from the previous tier to use in the next tier instead of uncomputing them eagerly to reduce the CNOT count at the expense of some ancillas.

\subsection{Local block encoding}
\label{sec:local_encoding}

After routing, the selected patch is available at a fixed location. We now construct its local block encoding. The main advantage of our approach is that this circuit is applied only once per query, regardless of the system size. We can therefore spend more resources on the local encoding to achieve a better shift and normalization and reduce the overall query complexity. This also motivates us to consider progressively larger local patches, which can capture more of the Hamiltonian's frustration locally. Less residual frustration implies more optimal normalization and shift of the full Hamiltonian. 

For a fixed patch size, the local encoding cost is independent of the system size, so routing dominates asymptotically. However, a naive implementation can still make the local circuit expensive enough that the benefits of our approach appear only for very large lattices. Efficient local constructions are therefore important for realizing these savings at reasonable system sizes. In this section, we provide general ideas that exploit the structure of the patch Hamiltonian to improve the shift and normalization while keeping the local circuit cost low.

We first determine the shift and normalization for the patch Hamiltonian from its spectral bounds. If the extreme eigenvalues are known exactly, this gives the optimal shift and normalization for the patch. We then need a local circuit that achieves this normalization. A simple construction is to decompose the shifted patch Hamiltonian into Pauli terms and use LCU. However, its normalization is set by the $\mathcal{L}_1$ norm of the Pauli coefficients, which can be bigger than the desired normalization. In order to realize the promised query savings from the tight spectral bounds, the constructed circuit must also achieve the corresponding normalization. We first describe how to do this using unitary dilation, and then discuss alternatives when constructing the full spectral basis becomes expensive.

\subsubsection{Spectral block encoding}
\label{subsec:spectral_encoding}

The first step is to compute the extreme eigenvalues of the patch Hamiltonian $H_p$. For small interaction groups, these can often be found analytically by using symmetries and structure of the Hamiltonian. This can be used to define the optimal $\delta_p$ and $\lambda_p$ as in~\Cref{eq:methods_local_normalization}. If the exact extrema are not available, rigorous enclosing bounds $\ell_p$ and $u_p$ still give a valid normalization. Such bounds can be obtained by bounding the smaller symmetry blocks separately and by combining bounds for smaller interaction groups. Numerical estimates can also be used if their errors are bounded so that the interval contains the full spectrum. We use a combination of these techniques for the applications discussed in~\Cref{sec:heisenberg,sec:nn-hubbard,sec:spin_liquids}. For the construction below, write
\begin{align}
A\equiv\widebar{H}_p=\frac{H_p-\delta_p I}{\lambda_p},\qquad A=A^\dagger,\qquad \|A\|\leq1.
\label{eq:local_contraction}
\end{align}
Here we assume $\lambda_p>0$.

If $A$ is a unitary, we can directly construct its circuit. If not, we can always leverage the fact that the shifted and normalized operator $A$ is a Hermitian contraction. Hence, $I-A^2$ is positive semidefinite, allowing us to use Halmos unitary dilation~\cite{halmos1950normal, robinson2018julia} to embed $A$ into a unitary using one signal qubit $w$,
\begin{equation}
\begin{aligned}
B(A)=
\begin{pmatrix}
A&\sqrt{I-A^2}\\
\sqrt{I-A^2}&-A
\end{pmatrix},\\
(\bra{0_w}\otimes I)B(A)(\ket{0_w}\otimes I)=A.
\label{eq:local_dilation}
\end{aligned}
\end{equation}
Now, $B(A)$ is a Hermitian unitary and gives the required local block encoding. 

While unitary dilation gives a general expression for the encoding, we still need to construct its circuit. This can be highly non-trivial in general. To understand what this entails, let us write the spectral decomposition in terms of the distinct eigenvalues $a_j$ of $A$ and their projectors $P_j$:
\begin{align}
A=\sum_{j=1}^{r}a_jP_j,\qquad
\sum_jP_j=I,\qquad P_jP_k=\delta_{jk}P_j.
\label{eq:local_spectral_decomposition}
\end{align}
Then,
\begin{align}
B(A)
&=Z_w\otimes A+X_w\otimes\sqrt{I-A^2}\nonumber\\
&=\left[I_w\otimes A-iY_w\otimes\sqrt{I-A^2}\right]
  (Z_w\otimes I)\nonumber\\
&=\left[\sum_j
  \left(a_jI_w-i\sqrt{1-a_j^2}Y_w\right)\otimes P_j\right]
  (Z_w\otimes I)\nonumber\\
&=\left[\sum_j R_y(2\arccos a_j)\otimes P_j\right]
  (Z_w\otimes I).
\end{align}
To finish the construction, we need to map the eigenbasis to the computational basis in order to apply the $Y$ rotation controlled on the correct eigenspace by controlling on that computational basis state. Let $V_p$ be such a transformation. The transformed projectors $Q_j=V_pP_jV_p^\dagger$ are then diagonal in the computational basis. This gives the general block-encoding form as,
\begin{equation}
\begin{aligned}
B(A) ={}& (I_w \otimes V_p^\dagger) \\
&\times \Bigl[
  \sum_j R_y(2\arccos a_j)\otimes Q_j
  \Bigr] \\
&\times (I_w\otimes V_p)(Z_w\otimes I).
\end{aligned}
\label{eq:local_multiplexed_dilation}
\end{equation}
Thus, in order to construct the circuit, we need to find a simple basis change, identify the distinct eigenspaces and apply the corresponding rotation on the signal qubit. The number of distinct rotation angles depends only on the number of distinct eigenvalues as the eigenstates in a degenerate eigenspace use the same one. One has the choice of choosing any orthonormal basis within that eigenspace to simplify the basis transformation. 

The main task is therefore to find a simple basis change and a simple way to distinguish the eigenspaces. A generic diagonalization need not give a low $T$-cost circuit. Instead, we can leverage  the symmetries and structure of the local Hamiltonian to simplify the construction. We use the following ideas to simplify the local constructions in~\Cref{sec:heisenberg,sec:nn-hubbard,sec:spin_liquids}.
\begin{enumerate}
    \item \textit{Spin identities.} Some shifted and normalized interactions are directly implementable unitaries. For example, $\operatorname{SWAP}_{ij}=(I+X_iX_j+Y_iY_j+Z_iZ_j)/2$, so an isotropic two-spin interaction can be shifted and normalized to a SWAP gate~\cite{MukhopadhyayWiebeZhang2023}. 
    \item \textit{Commuting Hermitian operators.} When the Hamiltonian is a linear combination of commuting Hermitian operators, their common orthonormal eigenbasis diagonalizes the Hamiltonian. Its eigenvalues are given by the same linear combination of the eigenvalues of these operators. For spin interactions, these operators can include the squared total spin and the squared spins of nested or disjoint subsets of the patch. Clebsch--Gordan transforms provide a structured way to transform between the computational basis and this coupled-spin basis. See~\Cref{app:clebsch_gordan} for details. 
    \item \textit{Structure of the coupling matrix.} For a two-spin interaction with a real symmetric exchange matrix, a common rotation of the two spins reduces the interaction to XYZ form, which is diagonal in the Bell basis~\cite{berry2019qubitization}. We can therefore construct the eigenbasis by diagonalizing the smaller coupling matrix. Additional structure in this matrix can reduce the number of variable rotations needed for the spin basis change. See~\Cref{app:real_symmetric_exchange} for details.
    \item \textit{Anticommuting Pauli groups.} For pairwise anticommuting Hermitian Pauli strings $P_j$, a real linear combination $G=\sum_jc_jP_j$ obeys $G^2=(\sum_jc_j^2)I$. Hence, $G/\sqrt{\sum_jc_j^2}$ is a Hermitian unitary. When this unitary has an efficient circuit, it gives an encoding with normalization $\sqrt{\sum_jc_j^2}$ instead of the Pauli LCU value $\sum_j|c_j|$~\cite{ZhaoYuan2021}. See~\Cref{app:anticommuting_pauli_groups} for details.
    \item \textit{Degenerate eigenvalues. } After the basis change, all computational basis states corresponding to the same eigenvalue require the same rotation on the signal qubit. We can therefore apply one rotation controlled on membership in the entire eigenspace. Eigenvalues at the normalized endpoints $\pm1$ require no arbitrary-angle ancilla rotations, simplifying the construction. 
    
    \item \textit{Symmetry reduction.} Conserved quantities divide the Hamiltonian into blocks that can be diagonalized separately, reducing the size of the required basis transformations~\cite{Loaiza_2023}. Blocks that are already diagonal need no further transformation. Symmetry-related blocks can sometimes use the same basis change, allowing parts of the circuit to be shared~\cite{LoaizaIzmaylov2023BLISS}.  
\end{enumerate}

These ideas can be combined within the same patch. A symmetry basis can reduce the diagonalization to smaller blocks, while degeneracy simplifies the controls for rotations on the signal qubit. We can also split $H_p=\sum_gH_{p,g}$ into smaller interaction groups, encode each group using a suitable construction, and combine the encodings through LCU~\cite{gilyen2019quantum}. This gives a local encoding $B_p$ of $(H_p-\delta_p I)/\lambda_p$, where $\delta_p=\sum_g\delta_{p,g}$ and $\lambda_p=\sum_g\lambda_{p,g}$. The corresponding local LCU cost is $T_{\textsc{BE}}^p=2T_{\textsc{Prep}}^p+T_{\textsc{Sel}}^p$. While this avoids constructing a common eigenbasis for the full patch Hamiltonian, the resulting normalization is the sum of the group normalizations and can be larger than that obtained by encoding all the interactions jointly. The grouping and local constructions should therefore balance the circuit cost against the normalization they achieve.

\subsubsection{Extensions using QET}
\label{subsec:qet}

Quantum eigenvalue transformation (QET) provides a way to improve the normalization without constructing the full eigenbasis. Let $U_0^p$ be an initial Hermitian local block-encoding unitary whose signal block is
\begin{align}
A_p^{(0)}=\frac{H_p-\delta_p I}{\beta_p}.
\label{eq:local_qet_initial_encoding}
\end{align}
Suppose tighter spectral bounds give a normalization $\lambda_p$ satisfying $\|H_p-\delta_p I\|\leq\lambda_p<\beta_p$, for the same shift $\delta_p$. We need to rescale the encoded eigenvalues by $\rho_p=\beta_p/\lambda_p$. QET provides a phase sequence that maps the encoded operator $A_p^{(0)}$ to a polynomial $P_d(A_p^{(0)})$ of degree $d$, provided the polynomial has the required parity and is bounded in magnitude by one on $[-1,1]$~\cite{low2019qubitization,gilyen2019quantum}.

We choose a real odd polynomial such that
\begin{align}
\left|P_d(x)-(1-s)\rho_p x\right|\leq\epsilon_{\mathrm{QET}},
\qquad |x|\leq\rho_p^{-1},
\label{eq:local_qet_target}
\end{align}
while $|P_d(x)|\leq1$ for every $x\in[-1,1]$. Here $s>0$ is a small headroom parameter. Exact amplification over a continuous interval is not possible at finite degree: a polynomial that agrees exactly with $\rho_p x$ on an interval would agree with it everywhere and would exceed one outside the approximation interval when $\rho_p>1$. The headroom keeps the target values below one in magnitude, providing a margin for a bounded finite-degree approximation.

Let $U_{\mathrm{QET}}^{(d)}$ denote the real-polynomial QET circuit constructed in~\Cref{app:qet_cost_accounting} with the above polynomial. Then, the QET circuit block encodes $A_p^{\mathrm{QET}}$ for
\begin{align}
A_p^{\mathrm{QET}}
&\equiv(\bra{0_{\mathrm{sig}}}\otimes I)U_{\mathrm{QET}}^{(d)}(\ket{0_{\mathrm{sig}}}\otimes I)
=P_d(A_p^{(0)}),\nonumber\\
&\left\|A_p^{\mathrm{QET}}-\frac{1-s}{\lambda_p}(H_p-\delta_p I)\right\|
\leq\epsilon_{\mathrm{QET}}.
\label{eq:local_qet_error}
\end{align}
The effective local normalization is therefore $\tilde\lambda_p=\lambda_p/(1-s)$. For QET to improve the normalization, we require $(1-s)\rho_p>1$, or equivalently $0<s<1-\rho_p^{-1}$. The global shift and normalization are
\begin{align}
\Delta=\sum_p\delta_p,\qquad
\Lambda=\sum_p\tilde\lambda_p=\sum_p\frac{\lambda_p}{1-s}.
\label{eq:local_qet_global_normalization}
\end{align} 
For $N_l$ translated copies of the same patch, $\Delta=N_l\delta_p$ and $\Lambda=N_l\lambda_p/(1-s)$, where $N_l$ is the number of patches or unit cells in the lattice. 

For fixed $s$ and $\epsilon_{\mathrm{QET}}$, the required degree can be determined by
\begin{equation}
\begin{aligned}
&d_\star(\rho_p,s,\epsilon_{\mathrm{QET}}) =
\min_{\substack{
    d\in\{1,3,5,\ldots\}\\
    P_d\in\mathbb R[x]}} d \\[2pt]
&\text{subject to}\\
&\quad\deg P_d \leq d, \\
&\quad P_d(-x) = -P_d(x), \\
&\quad \max_{|x|\leq 1}|P_d(x)| \leq 1, \\
&\quad \max_{|x|\leq \rho_p^{-1}}
    \bigl|P_d(x)-(1-s)\rho_p x\bigr|
    \leq \epsilon_{\mathrm{QET}}.
\end{aligned}
\label{eq:local_qet_degree_optimization}
\end{equation}
We can search over odd degrees using constrained minimax or Chebyshev-basis convex optimization, with the polynomial bounds verified over the full intervals. The corresponding phase vector is then synthesized for the QET circuit. In practice, the headroom $s$ is optimized: decreasing $s$ improves the normalization, while  increasing $s$ reduces $d_\star$.

The circuit construction and cost accounting are given in~\Cref{app:qet_cost_accounting}. If the source $U_0^p$ uses $a$ signal qubits and $T_0^p$ $T$ gates, QET uses $a+1$ output signal qubits and has the uncontrolled cost bound of~\Cref{eq:local_qet_t_cost},
\begin{align*}
T_{\textsc{BE}}^p
\leq d_\star T_0^p+(d_\star+1)\bigl[4(a-1)+T_{\mathrm{rot}}\bigr],
\qquad d_\star\geq1.
\end{align*}
Here $T_{\mathrm{rot}}$ is the cost of one synthesized axial rotation (i.e. a single-qubit rotation $e^{-i\theta P/2}$ about a Pauli axis $P\in\{X,Y,Z\}$). The controlled-walk estimates use this QET cost with the additive control overhead omitted as specified in~\Cref{app:qet_cost_accounting}.

\begin{remark}

Both local constructions in this section yield reflections, as required by~\Cref{rem:reflection}.
\begin{enumerate}
    \item \textit{Spectral encodings} (\Cref{subsec:spectral_encoding}). If $A$ is unitary, as for the SWAP of an isotropic bond or a normalized anticommuting Pauli group, we construct its circuit directly. Since $A$ is also Hermitian, $A^2=I$. Otherwise, we use the dilation $B(A)$ of~\Cref{eq:local_dilation}, which is Hermitian and unitary, so $B(A)^2=I$. When we combine such encodings $B_k$ through LCU, or move the selected interaction group to a fixed position with the routing of~\Cref{eq:methods_select} applied inside the patch, the resulting local encoding has the form $B_p=V^\dagger\big(\sum_k\ket{k}\!\bra{k}\otimes B_k\big)V$ where $V$ is unitary. Hence $B_p^2=V^\dagger\big(\sum_k\ket{k}\!\bra{k}\otimes B_k^2\big)V=I$.
    \item \textit{QET} (\Cref{subsec:qet}). QET transforms the initial encoding $U_0^p$ of~\Cref{eq:local_qet_initial_encoding}, which in this work is an LCU of spectral encodings and therefore a reflection. With the use of symmetric phases and one extra signal qubit, the real-part completion of~\Cref{app:qet_cost_accounting} makes the QET circuit a reflection as well.
\end{enumerate}
Hence $B_{\mathrm{ref}}^2=I$ and $\textsc{Sel}^2=I$, so each walk step needs one call to the block encoding, or one controlled call in QPE. A general block encoding would instead need controlled calls to both the encoding and its inverse~\cite{low2019qubitization}, which doubles the local cost.
\end{remark}

Throughout this work, we quote the block-encoding cost $T_{\textsc{BE}}$ as the cost per query, since the walk operator and its controlled version cost only slightly more. This is due to the following argument. Since $\textsc{Prep}^\dagger\,\textsc{Sel}\,\textsc{Prep}$ is a reflection, the qubitized walk operator~\cite[Corollary~9]{low2019qubitization} is
\begin{align}
W=\left(2\ketbra{0}{0}-I\right)\textsc{Prep}^\dagger R^\dagger(I\otimes B_{\mathrm{ref}})R\,\textsc{Prep},
\label{eq:walk_operator}
\end{align}
where the signal reflection $2\ketbra{0}{0}-I$ acts on the global signal register. To control $W$, it suffices to control the signal reflection and a few gates of $B_{\mathrm{ref}}$. All other subcircuits, namely $\textsc{Prep}$, the routing $R$ and the local preparations, routings and basis changes, cancel with their inverses when the control is off and hence do not require controls. The same holds in a QET encoding for all gates except the middle factor $X_bU_0^p$ of~\Cref{eq:local_qet_conjugation}. The few remaining gates are Paulis, whose controlled versions are Clifford, and gates that are already controlled by a few local qubits (\Cref{app:heisenberg_cost_accounting,app:fh_cost_accounting,app:honeycomb_cost_accounting}). In particular, for the encodings presented in~\Cref{sec:heisenberg,sec:nn-hubbard,sec:spin_liquids}, the logical ANDs that form these controls all share one operand qubit and thus one additional temporary AND (between this operand qubit and $W$'s control qubit) adds the desired control. The controlled block encoding therefore costs at most $T_{\textsc{BE}}+4$ $T$ gates. The signal reflection is a multi-controlled $Z$ gate on the global signal register. For $M=120$, the largest lattice in~\Cref{sec:overall-cost}, this register has at most $23$ qubits, and the signal reflection costs at most $84$ $T$ gates, or $88$ with the control~\cite{gidney2018halving}. A controlled walk thus costs at most $T_{\textsc{BE}}+92$ $T$ gates, less than $0.1\%$ more than $T_{\textsc{BE}}$.

In the following sections, we apply this methodology to construct a hierarchy of local encodings for the square-lattice $J_1$--$J_2$ Heisenberg Hamiltonian, the square-lattice Fermi--Hubbard Hamiltonians, and the honeycomb spin-liquid Hamiltonian, along with some extensions of these models. For all three applications, we consider an $M\times M$ periodic array with $N_l=M^2$ unit cells. For the square-lattice Heisenberg model, the number of system qubits $N_s=N_l$, whereas for the Fermi--Hubbard model and honeycomb spin model, $N_s=2N_l$.

\section{Block encodings for the Heisenberg model}
\label{sec:heisenberg}

We first consider the $J_1$--$J_2$ Heisenberg model on a periodic $M\times M$ square lattice, with $M\geq4$ even and $N=M^2$ spins. The Hamiltonian is
\begin{align}
H=J_1\sum_{\langle i,j\rangle}\boldsymbol{\sigma}_i\cdot\boldsymbol{\sigma}_j
+J_2\sum_{\langle\!\langle i,j\rangle\!\rangle}\boldsymbol{\sigma}_i\cdot\boldsymbol{\sigma}_j,\label{eq:heisenberg_pauli_convention}
\end{align}
where $\boldsymbol{\sigma}_i=(X_i,Y_i,Z_i)$ is the Pauli vector. Here, $\langle i,j\rangle$ and $\langle\!\langle i,j\rangle\!\rangle$ denote nearest-neighbor (NN) and next-nearest-neighbor (NNN) bonds, respectively. We assume $J_1,J_2>0$.

We write the Hamiltonian as a sum over the $N_l=N$ translated plaquettes, $H=\sum_p H^\square_p$, where $p=(p_x,p_y)$ and $p_x,p_y\in\{0,\ldots,M-1\}$. Each NN bond belongs to two plaquettes, while each NNN bond belongs to one. Hence, we assign half of each NN coupling and all of each NNN coupling to the patch Hamiltonian,
\begin{align*}
H^\square_{p_x,p_y}&=\frac{J_1}{2}\Bigl(\boldsymbol{\sigma}_{p_x,p_y}\cdot\boldsymbol{\sigma}_{p_x,p_y+1}+\boldsymbol{\sigma}_{p_x,p_y}\cdot\boldsymbol{\sigma}_{p_x+1,p_y}\\
&+\boldsymbol{\sigma}_{p_x+1,p_y+1}\cdot\boldsymbol{\sigma}_{p_x,p_y+1}+\boldsymbol{\sigma}_{p_x+1,p_y+1}\cdot\boldsymbol{\sigma}_{p_x+1,p_y}\Bigr)\\
&+J_2\Bigl(\boldsymbol{\sigma}_{p_x,p_y}\cdot\boldsymbol{\sigma}_{p_x+1,p_y+1}+\boldsymbol{\sigma}_{p_x+1,p_y}\cdot\boldsymbol{\sigma}_{p_x,p_y+1}\Bigr),
\end{align*}
where all coordinates are taken modulo $M$. We use the square-lattice routing construction in~\Cref{subsec:square_lattice_routing} to bring the selected plaquette to the sites $(0,0),(0,1),(1,0),(1,1)$, with $(0,0)$ the bottom-left site, up to row and column reflections. We write $H^\square_{\mathrm{ref}}$ for the patch operator on these reference sites. The patch Hamiltonian is invariant under row and column reflections, since each reflection permutes the four equally weighted NN bonds among themselves and exchanges the two equally weighted NNN bonds. Thus, the routing needs no order correction.

Next, we construct a hierarchy of local block encodings by choosing different groupings of the interactions. We start with individual bonds and then consider triangle terms, which already attain the optimal shift and normalization of the full plaquette. Finally, we enlarge the patch to $3\times3$ sites to capture part of the frustration between neighboring plaquettes. We compare these constructions through their achieved normalization and local $T$-cost.

\subsection{Bond-wise local block encoding}
\label{sec:heisenberg_bond}

The smallest local interaction is the two-body bond-wise term. Define a bond term with a generic coupling $J>0$,
\[
H^b_{ij}(J)=J\boldsymbol{\sigma}_i\cdot\boldsymbol{\sigma}_j=J(2P_{ij}-I),
\]
where $P_{ij}$ swaps qubits $i,j$. The SWAP operator has eigenvalue $+1$ on the triplet sector and $-1$ on the singlet sector. Hence, the bond Hamiltonian has two distinct eigenvalues, $e_{\min}=-3J$ and $e_{\max}=+J$. 
The optimal shift and normalization are therefore $\delta_b=-J$ and $\lambda_b=2J$.
The shifted and normalized operator is
\[
A^b_{ij}
=\frac{H^b_{ij}(J)-\delta_{b}I}
{\lambda_b}=
\frac{J(2P_{ij}-I)+J I}{2J}
=P_{ij}.
\]
Thus, the optimal block encoding of a bond is simply a SWAP gate, with no dilation needed. Using the weighted couplings in $H_p^\square$, the normalized NN and NNN bond operators and their shifts and normalizations are
\begin{gather*}
A^b_{ij,\mathrm{NN}}=\frac{H^b_{ij}(J_1)+J_1 I}{2J_1}=P_{ij},\,
\delta_{b,\mathrm{NN}}=-\frac{J_1}{2},\,
\lambda_{b,\mathrm{NN}}=J_1\\
A^b_{ij,\mathrm{NNN}}=\frac{H^b_{ij}(J_2)+J_2 I}{2J_2}
=P_{ij},\, \delta_{b,\mathrm{NNN}}=-J_2,\,\lambda_{b,\mathrm{NNN}}=2J_2
\end{gather*}

We use LCU to combine the six bond encodings within the patch. We then have the local circuit $B_{\mathrm{ref}}$ encoding $(H^\square_{\mathrm{ref}}-\delta_{p,b}I)/\lambda_{p,b}$, where $\delta_{p,b}=-2(J_1+J_2)$ and $\lambda_{p,b}=4(J_1+J_2)$. Summing over the $N$ patches gives the global shift and normalization,
\begin{align*}
\Delta_b=-2N(J_1+J_2),\qquad
\Lambda_b=4N(J_1+J_2).
\end{align*}

\Cref{app:heisenberg_bondwise_cost_accounting} gives the gate-level circuit and the block-encoding cost of
\begin{align*}
T_{\textsc{BE},b}^\square\leq8N+80\lceil\log_2M\rceil+2T_{\mathrm{rot}}(\epsilon)+16,
\end{align*}
where $N=M^2$ and $T_{\mathrm{rot}}(\epsilon)$ is the $T$ cost of one axial rotation synthesized to accuracy $\epsilon$. \Cref{app:heisenberg_anisotropic} discusses the natural extension of this framework to an anisotropic XXZ exchange $J(X_iX_j+Y_iY_j+\kappa Z_iZ_j)$.

\subsection{Triangular local block encoding}
\label{sec:heisenberg_triangle}

Since the lower spectral bound of the bond-wise encoding does not account for the frustration between overlapping bonds, the next level of the hierarchy groups the interactions into three-body terms on triangles. This grouping already attains the optimal shift and normalization of the full four-spin plaquette operator (\Cref{appendix:optimal_plaquette_normalization}), so we do not construct a separate plaquette encoding.

We define a right-angle triangle as an ordered triplet $(i,j,k)$ in which $j$ is the right-angle vertex, so the pairs $(i,j)$ and $(j,k)$ are NN bonds and the pair $(i,k)$ is an NNN bond. We decompose $H^\square_{\mathrm{ref}}$ into the four right-angle triangles with right angles at the lower-left, lower-right, upper-right, and upper-left corners, indexed by $\tau\in\{0,1,2,3\}$,
\begin{align}
H^\square_{\mathrm{ref}}&=H^\triangle_{(0,1),(0,0),(1,0)}+H^\triangle_{(0,0),(1,0),(1,1)}\nonumber \\&+H^\triangle_{(1,0),(1,1),(0,1)}+H^\triangle_{(1,1),(0,1),(0,0)}\equiv\sum_{\tau=0}^3H^\triangle_\tau.
\label{eq:triangle_decomposition}
\end{align}
Each NN bond lies in two of the four triangles of a plaquette and in two plaquettes, while each NNN bond lies in two of the four triangles and in only one plaquette:
\begin{align}
H^\triangle_{i,j,k}=\frac{J_1}{4}(\boldsymbol{\sigma}_{i}\cdot\boldsymbol{\sigma}_{j}+\boldsymbol{\sigma}_{j}\cdot\boldsymbol{\sigma}_{k})+\frac{J_2}{2}\boldsymbol{\sigma}_{i}\cdot\boldsymbol{\sigma}_{k}.
\label{eq:triangle_term}
\end{align}

Next, we compute the extreme eigenvalues of $H^\triangle_{i,j,k}$ using commuting Hermitian operators, as in~\Cref{subsec:spectral_encoding}, with the coupled-spin operators of~\Cref{app:cg_operators}. Let $\mathbf S_a=\boldsymbol\sigma_a/2$ be the spin at vertex $a$, so that $\boldsymbol\sigma_a\cdot\boldsymbol\sigma_b=4\mathbf S_a\cdot\mathbf S_b$ and $H^\triangle_{i,j,k}=J_1\mathbf S_j\cdot(\mathbf S_i+\mathbf S_k)+2J_2\,\mathbf S_i\cdot\mathbf S_k$. Define $\mathbf L=\mathbf S_i+\mathbf S_k$ as the total spin of the diagonal and $\mathbf F=\mathbf L+\mathbf S_j$ as the total spin of the triangle. With $(1,2,3)=(i,j,k)$, the coupled-spin identities in~\Cref{eq:cg_three_spin_identities} give
\begin{align}
H^\triangle_{i,j,k}=\frac{J_1}{2}\mathbf F^2+\left(J_2-\frac{J_1}{2}\right)\mathbf L^2-\left(\frac{3J_1}{8}+\frac{3J_2}{2}\right)I.
\label{eq:triangle_coupled_form}
\end{align}
The operators $\mathbf L^2$ and $\mathbf F^2$ commute, and their joint eigenspaces, which we call sectors, are $(\ell,f)=(0,\tfrac12)$, $(1,\tfrac12)$ and $(1,\tfrac32)$, with dimensions two, two and four (\Cref{tab:cg_three_spin_sectors}). On these sectors, $H^\triangle_{i,j,k}$ takes the values
\begin{align}
e_1=-\frac{3J_2}{2},\qquad e_2=-J_1+\frac{J_2}{2},\qquad e_3=\frac{J_1+J_2}{2},
\label{eqn:triangular_patch_eigenvalues}
\end{align}
respectively. The maximum is always $e_3$, while the minimum is $e_2$ for $2J_2\leq J_1$ and $e_1$ otherwise. The optimal shift and normalization of $A^\triangle_{i,j,k}=(H^\triangle_{i,j,k}-\delta_\triangle I)/\lambda_\triangle$ are therefore
\begin{align}
2J_2\leq J_1:&\quad
\delta_\triangle=\frac{2J_2-J_1}{4},
&\lambda_\triangle&=\frac{3J_1}{4},\nonumber\\
2J_2>J_1:&\quad
\delta_\triangle=\frac{J_1-2J_2}{4},
&\lambda_\triangle&=\frac{J_1+4J_2}{4}.
\label{eq:A_triangle_normalization}
\end{align}

We combine the four triangles within the patch as follows. The local preparation is a uniform superposition over a two-qubit triangle selection register, prepared with two Hadamards. Write the right-angle vertex of triangle $\tau$ as 	$(x_\tau,y_\tau)$ and store these two bits in the register. Controlled on $x_\tau$ and $y_\tau$, a column reflection and a row reflection of the routed patch bring triangle $\tau$ to the reference triangle $\tau=0$. We then apply the encoding of $A^\triangle_{(0,1),(0,0),(1,0)}$ once and undo the reflections. Since $H^\triangle_{i,j,k}$ is symmetric under the exchange of $i$ and $k$, the order in which the two legs arrive does not matter. The resulting local circuit $B_{\mathrm{ref}}$ encodes $(H^\square_{\mathrm{ref}}-4\delta_\triangle I)/(4\lambda_\triangle)$. Since this is a shifted and normalized plaquette operator, the encoding keeps the reflection invariance that the routing requires. Summing over the $N$ patches gives the global shift and normalization,
\begin{equation}
\begin{aligned}
\Delta_\triangle&=\begin{cases}
N(2J_2-J_1),&2J_2\leq J_1,\\
N(J_1-2J_2),&2J_2>J_1,
\end{cases}\\
\Lambda_\triangle&=\begin{cases}
3NJ_1,&2J_2\leq J_1,\\
N(J_1+4J_2),&2J_2>J_1.
\end{cases}
\end{aligned}
\label{eq:triangle_global_normalization}
\end{equation}

\subsubsection{Circuit construction}
\label{sec:triangle_interaction}

For $J_2\neq J_1/2$, the triangle term has three distinct eigenvalues. The optimal shift and normalization send the extreme ones to $\pm1$, while the third lies strictly inside the interval, so $A^\triangle_{i,j,k}$ is not a unitary. Let $\Pi_{(\ell,f)}$ denote the projectors onto the sectors. We call $(1,\tfrac32)$, which carries the highest eigenvalue, the top sector, and we call the sectors of the lowest and the intermediate eigenvalue the low and the middle sector, with projectors $\Pi_{\mathrm{low}}$ and $\Pi_{\mathrm{mid}}$. For $r=J_2/J_1\leq\frac12$, $\Pi_{\mathrm{low}}=\Pi_{(1,\frac12)}$ and $\Pi_{\mathrm{mid}}=\Pi_{(0,\frac12)}$, and the two are exchanged for $r>\frac12$. Then
\begin{equation}
\begin{aligned}
A^\triangle_{i,j,k}&=\Pi_{(1,\frac32)}-\Pi_{\mathrm{low}}+\xi \Pi_{\mathrm{mid}},\\
\xi&=\begin{cases}
(1-8r)/3,&r\leq\frac12,\\
(4r-5)/(1+4r),&r>\frac12.
\end{cases}
\end{aligned}
\label{eq:triangle_A_projectors}
\end{equation}
We encode this operator with the spectral encoding of~\Cref{eq:local_multiplexed_dilation}, with $Q_{\mathrm{low}}=V_p\Pi_{\mathrm{low}}V_p^\dagger$ and $Q_{\mathrm{mid}}=V_p\Pi_{\mathrm{mid}}V_p^\dagger$. The top sector needs no gate and the low sector needs only the sign flip $I-2Q_{\mathrm{low}}$ on the system. Only the middle sector requires the synthesized rotation $R_y(2\arccos\xi)$ on $w$.

The basis change is $V_p=V_{\mathrm{CG}}^\dagger$, with $V_{\mathrm{CG}}$ the three-spin Clebsch--Gordan circuit of~\Cref{app:cg_transforms} in~\Cref{eq:cg_three_spin_circuit} and the qubits $(1,2,3)$ taken as $(i,j,k)$. This circuit maps each sector to a fixed set of computational basis states. We show in~\Cref{app:cg_transforms} that each set is uniquely identified by two parities of the computational bits. As a result, one logical AND of these parities flags the middle sector and controls its rotation, and the sign flip on the low sector is a Clifford gate (\Cref{app:heisenberg_triangular_cost_accounting}). Apart from the basis changes, the only non-Clifford operations in the encoding of $A^\triangle_{i,j,k}$ are therefore this logical AND and the controlled rotation of $w$.

At $J_2=J_1/2$, the two $f=\tfrac12$ eigenvalues coincide and $\xi=-1$, so $A^\triangle_{i,j,k}=2\Pi_{(1,\frac32)}-I$ is a reflection. After the basis change, this reflection is the Pauli operator $Z_jZ_k$ (\Cref{app:heisenberg_triangular_cost_accounting}), so $A^\triangle_{i,j,k}=V_{\mathrm{CG}}Z_jZ_kV_{\mathrm{CG}}^\dagger$, and its circuit omits the signal qubit $w$ and the logical AND.

\Cref{app:heisenberg_triangular_cost_accounting} gives the gate-level circuit description and derives the following bounds,
\begin{align*}
T_{\textsc{BE}}^p&=40+6T_{\mathrm{rot}}(\epsilon),\\
T_{\textsc{BE},\triangle}^\square&\leq8N+80\lceil\log_2M\rceil+6T_{\mathrm{rot}}(\epsilon)-8.
\end{align*}
Compared with the bond-wise encoding, a query costs $4T_{\mathrm{rot}}(\epsilon)-24$ more $T$ gates for $J_2\neq J_1/2$, while the global normalization is smaller by the factor $\Lambda_b/\Lambda_\triangle$. This factor equals $4(1+r)/3$ for $r\leq\frac12$ and $4(1+r)/(1+4r)$ for $r>\frac12$, with its maximum of two at $r=\frac12$.

\Cref{app:heisenberg_onsite} discusses the natural extension of the triangular local block-encoding framework to include an on-site field $h\sum_iZ_i$.

\subsection{Hierarchical $3\times3$ local block encoding}
\label{sec:heisenberg_three_by_three}

Since the triangular encoding already attains the optimal shift and normalization of the plaquette operator (\Cref{appendix:optimal_plaquette_normalization}), a smaller normalization requires a patch that is larger than a plaquette. We therefore enlarge the patch to $3\times3$ sites, which contain four overlapping plaquettes. For a patch with coordinate label $p=(p_x,p_y)$, define
\begin{align*}
\mathcal P_p
=\{p+(n_x,n_y):n_x,n_y\in\{0,1,2\}\},
\end{align*}
where coordinates are evaluated modulo $M$. It is useful to distinguish four orbits of interactions under the reflection and rotation symmetries of the patch. There are eight NN bonds on the perimeter, four NN bonds joining the center to an edge midpoint, four NNN bonds joining the center to a corner, and four NNN bonds joining adjacent edge midpoints. Denote these sets by $\mathcal E_{p,\mathrm{per}}$, $\mathcal E_{p,\mathrm{in}}$, $\mathcal E_{p,\mathrm{rad}}$, and $\mathcal E_{p,\mathrm{cyc}}$, respectively. A decomposition that respects these symmetries assigns one weight to each orbit. With $\mathbf c=(c_{\mathrm{per}},c_{\mathrm{in}},c_{\mathrm{rad}},c_{\mathrm{cyc}})$, it reads
\begin{align}
H^{3\times3}_p(\mathbf c)
&=J_1\left(
c_{\mathrm{per}}\!\!\sum_{(i,j)\in\mathcal E_{p,\mathrm{per}}}\!\!
\boldsymbol\sigma_i\cdot\boldsymbol\sigma_j
+c_{\mathrm{in}}\!\!\sum_{(i,j)\in\mathcal E_{p,\mathrm{in}}}\!\!
\boldsymbol\sigma_i\cdot\boldsymbol\sigma_j
\right)\nonumber\\
&+J_2\left(
c_{\mathrm{rad}}\!\!\sum_{(i,j)\in\mathcal E_{p,\mathrm{rad}}}\!\!
\boldsymbol\sigma_i\cdot\boldsymbol\sigma_j
+c_{\mathrm{cyc}}\!\!\sum_{(i,j)\in\mathcal E_{p,\mathrm{cyc}}}\!\!
\boldsymbol\sigma_i\cdot\boldsymbol\sigma_j
\right).
\label{eq:three_by_three_general_decomposition}
\end{align}
Every NN bond occurs in four translated patches as a perimeter bond and in two translated patches as an inner bond. Similarly, every NNN bond occurs in two translated patches as a radial bond and in two as a cyclic bond. Consequently, the constraints
\begin{align}
4c_{\mathrm{per}}+2c_{\mathrm{in}}=1,
\qquad
2c_{\mathrm{rad}}+2c_{\mathrm{cyc}}=1
\label{eq:three_by_three_reconstruction_constraints}
\end{align}
give the exact reconstruction $H=\sum_{p}H^{3\times3}_{p}(\mathbf c)$. They leave the freedom to redistribute the interaction weight among the four orbits without changing the global Hamiltonian. We restrict attention to nonnegative weights, so that each patch Hamiltonian remains an antiferromagnetic Heisenberg model. The fully polarized state is then an eigenstate of every bond term with that term's largest eigenvalue. Every such patch Hamiltonian therefore has the same upper edge $J_1(8c_{\mathrm{per}}+4c_{\mathrm{in}})+4J_2(c_{\mathrm{rad}}+c_{\mathrm{cyc}})=2(J_1+J_2)$, so a redistribution can lower the patch normalization only by raising the patch ground energy.

A direct spectral encoding of the nine-spin patch needs its full eigenbasis, and the cost of such a basis change grows quickly with the patch size (\Cref{app:cg_gate_cost}). We therefore choose weights for which the patch Hamiltonian is a uniform average of operators that already have an optimal encoding. Let $q=(q_x,q_y)\in\{0,1\}^2$ label the four $2\times2$ subplaquettes contained in $\mathcal P_p$. We define the hierarchical patch Hamiltonian
\begin{align}
H^{3\times3}_p
=\frac{1}{4}\sum_{q_x,q_y\in\{0,1\}}
H^\square_{p_x+q_x,p_y+q_y}.
\label{eq:hierarchical_patch_definition}
\end{align}
Translation invariance immediately gives $\sum_pH^{3\times3}_p=\frac14\sum_q\sum_pH^\square_{p+q}=H$. A perimeter NN bond lies in one subplaquette, an inner NN bond lies in two, and every NNN bond lies in one. Hence~\Cref{eq:hierarchical_patch_definition} is the member of~\Cref{eq:three_by_three_general_decomposition} with $c_{\mathrm{per}}=\frac18$ and $c_{\mathrm{in}}=c_{\mathrm{rad}}=c_{\mathrm{cyc}}=\frac14$. These weights need not minimize the patch normalization, and we leave the optimization over $\mathbf c$ open.

Each term $H^\square_{p+q}$ in~\Cref{eq:hierarchical_patch_definition} is a translated plaquette operator, and hence all four terms have the optimal shift $\delta_2=4\delta_\triangle$ and normalization $\lambda_2=4\lambda_\triangle$ (\Cref{appendix:optimal_plaquette_normalization}). We denote the optimal shift and normalization of $H^{3\times3}_p$ by $\delta_3$ and $\lambda_3$. We obtain $\delta_3$ and $\lambda_3$ by exact diagonalization of the nine-spin operator and list their values in~\Cref{tab:hierarchical_patch_spectrum}. At $r=0.4$, $0.5$ and $0.6$, $(\delta_3,\lambda_3)/J_1=(0.177,2.623)$, $(0.334,2.666)$ and $(0.470,2.730)$, so $\lambda_3$ lies $13\%$, $11\%$ and $20\%$ below the plaquette value $\lambda_2$. The operators $H^{3\times3}_p$ and $H^\square_p$ share the upper edge of the fully polarized state, $\delta_2+\lambda_2=\delta_3+\lambda_3=2(J_1+J_2)$. The smaller normalization of the $3\times3$ patch therefore comes from its higher ground energy, which reflects the frustration between neighboring plaquettes.

We route the nine spins of $\mathcal P_p$ with the construction of~\Cref{subsec:long_range_routing} at $U=3$. This routing requires $M$ to be a multiple of three with $M\geq6$, instead of the even $M$ of the plaquette routing. Every translated patch contains one site from each residue class $(x\bmod3,y\bmod3)$, and each class has its own routing tree. The routed patch arrives with its columns and rows cyclically shifted by $p_x\bmod3$ and $p_y\bmod3$. Since the patch Hamiltonian is invariant under row and column reflections, it is enough to undo these shifts up to a reflection. We achieve this through controlled exchanges of two columns or two rows, and we reverse them after the local encoding. \Cref{app:three_by_three_routing} gives this circuit and derives the routing cost
\begin{align*}
T_R^{3\times3}\leq8N+16\lceil\log_2M\rceil+32\lceil\log_2(M/3)\rceil-24.
\end{align*}

\subsubsection{Local block encoding of the $3\times3$ patch}

We build the local encoding from the triangular plaquette encoding of~\Cref{sec:heisenberg_triangle} in three steps. We first average this encoding over the four subplaquettes, which reproduces the plaquette shift $\delta_2$ and normalization $\lambda_2$. We then recenter the average at $\delta_3$ and amplify it toward $\lambda_3$ with the QET construction of~\Cref{subsec:qet}.

\paragraph{Plaquette averaging}

Let $A^\square_p=(H^\square_p-\delta_2I)/\lambda_2$, and let $U_2$ denote the plaquette encoding $B_{\mathrm{ref}}$ of~\Cref{sec:heisenberg_triangle}. We prepare a two-qubit register in the uniform superposition $\frac12\sum_q\ket{q}$. Controlled on $q$, a subplaquette routing moves subplaquette $q$ onto the four input qubits of $U_2$ (\Cref{fig:3x3_plaquette_swap}). We then apply $U_2$ once, undo the routing and undo the preparation. The resulting circuit $U_{\mathrm{av}}$ has the signal block
\begin{align}
A_{\mathrm{av}}
=\frac14\sum_q A^\square_{p+q}
=\frac{H^{3\times3}_p-\delta_2 I}{\lambda_2}.
\label{eq:hierarchical_average_signal_operator}
\end{align}

\paragraph{Recentering}

The operator in~\Cref{eq:hierarchical_average_signal_operator} is centered at $\delta_2$ rather than at $\delta_3$. We recenter it with a two-term LCU of $A_{\mathrm{av}}$ and the identity,
\begin{equation}
\begin{aligned}
A^{(0)}_3&=\frac{\lambda_2}{\beta_3}A_{\mathrm{av}}-\frac{\delta_3-\delta_2}{\beta_3}I=\frac{H^{3\times3}_p-\delta_3I}{\beta_3},\\
\beta_3&=\lambda_2+\delta_3-\delta_2=2\lambda_2-\lambda_3,
\end{aligned}
\label{eq:hierarchical_recentered_operator}
\end{equation}
where the last equality uses the shared upper edge, $\delta_3-\delta_2=\lambda_2-\lambda_3$, which is positive because the $3\times3$ patch has the higher ground energy. We denote the resulting Hermitian circuit by $B^{(0)}_3$ (\Cref{app:heisenberg_hierarchical}). Since $\beta_3>\lambda_2$, the recentering alone increases the normalization, and it is useful only together with the amplification below.

\paragraph{Local QET}

The recentered encoding has normalization $\beta_3$, while the optimal one is $\lambda_3$. We therefore apply the QET construction of~\Cref{subsec:qet} to the source $U_0^p=B^{(0)}_3$, so that $\beta_p=\beta_3$, $\delta_p=\delta_3$ and $\lambda_p=\lambda_3$ in~\Cref{eq:local_qet_initial_encoding}. The amplification ratio is $\rho_3=\beta_3/\lambda_3=2\lambda_2/\lambda_3-1$. For a headroom $0<s<1-\rho_3^{-1}$ and the degree $d_\star$ of~\Cref{eq:local_qet_degree_optimization},~\Cref{eq:local_qet_global_normalization} gives
\begin{align}
\Delta_3=N\delta_3,\qquad\Lambda_3=\frac{N\lambda_3}{1-s}.
\label{eq:hierarchical_global_normalization}
\end{align}
The global encoding represents $(H-\Delta_3I)/\Lambda_3$ up to the QET error $\epsilon_{\mathrm{QET}}$ in operator norm (\Cref{eq:local_qet_error,eq:local_qet_global_error}) with normalization smaller than the triangular value $\Lambda_\triangle=N\lambda_2$ when $s<1-\lambda_3/\lambda_2$. Without QET, we use $B^{(0)}_3$ directly, with $\Delta_3=N\delta_3$ and $\Lambda_3=N\beta_3$, which corresponds to $s=1-\lambda_3/\beta_3$.

\Cref{app:heisenberg_hierarchical} gives the gate-level construction and derives the global cost for odd $d_\star\geq1$,
\begin{align*}
T_{\textsc{BE},3\times3}^\square
&\leq 8N + 124d_\star + 12 + 80\lceil\log_2 M\rceil \\
&\quad+ 32\lceil\log_2(M/3)\rceil + (9d_\star+1)T_{\mathrm{rot}}(\epsilon).
\end{align*}

\section{Block encodings for the Fermi--Hubbard model}
\label{sec:nn-hubbard}

We consider the nearest-neighbor, particle--hole-symmetric
Fermi--Hubbard model on an $M\times M$ square lattice with periodic boundary
conditions (even $M\geq4$),
\begin{align}
H
={}&-t\sum_{\langle r,r'\rangle}\sum_{\sigma\in\{\uparrow,\downarrow\}}
\left(c_{r\sigma}^{\dagger}c_{r'\sigma}
+c_{r'\sigma}^{\dagger}c_{r\sigma}\right)
+U\sum_rD_r,
\label{eq:nn-fh}\\
D_r={}&
\left(n_{r\uparrow}-\frac12\right)
\left(n_{r\downarrow}-\frac12\right).
\nonumber
\end{align}
Equation~\eqref{eq:nn-fh} is the conventional half-filled Hubbard Hamiltonian, with chemical potential $\mu=U/2$, up to a known scalar.

\subsection{Bond-wise block encoding}
\label{sec:nn-fh-bondwise}

For a NN edge $b=\langle r,r'\rangle$, define the two-site bond operator
\begin{align}
H_b={}&-t\sum_{\sigma}T_{rr',\sigma}
+\frac U4(D_r+D_{r'}),\nonumber\\
T_{rr',\sigma}={}&c_{r\sigma}^{\dagger}c_{r'\sigma}
+c_{r'\sigma}^{\dagger}c_{r\sigma}.
\label{eq:nn-fh-bond-term}
\end{align}
Every site is incident on four NN edges, so the on-site interaction is exactly
reconstructed and $H=\sum_bH_b$.

It is well known that the spectrum of the Fermi--Hubbard model is symmetric about zero, which also holds for a single bond. Exact diagonalization of its four fermionic modes (which we discuss in more detail in Section~\ref{sec:FH_bond_wise_diagonalization}) gives
\begin{align}
\lambda_b=\|H_b\|
=\sqrt{4t^2+\frac{U^2}{64}}.
\label{eq:nn-fh-local-bond-range}
\end{align}
There are $2M^2$ undirected bonds. The global bond-wise normalization is therefore
\begin{align}
\Lambda_b=2M^2\sqrt{4t^2+\frac{U^2}{64}}.
\label{eq:nn-fh-bond-normalization}
\end{align}
This is tighter than treating every spin-resolved hopping string and every on-site Pauli as a separate LCU term, which has block-encoding normalization
\begin{align}
\Lambda_{\mathrm{sep}}=M^2\left(4|t|+\frac{|U|}{4}\right)=2M^2\sqrt{4t^2+\frac{U^2}{64}+\frac{|t|\cdot|U|}{2}},
\label{eq:nn-fh-separated-normalization}
\end{align}
which can be found in Eq.~15 of Ref.~\cite{kan2025resource}.

\subsubsection{Implementation of the interaction}\label{sec:FH_bond_wise_diagonalization}

For the bond on the first two sites of the lattice $A$ and $B$, we use the mode index $0,1,2,3$ to represent the spin-up mode on $A$, the spin-down mode on $A$, the spin-up mode on $B$, and the spin-down mode on $B$, respectively. Under Jordan--Wigner transformation, the local Hamiltonian is
\begin{equation}
\begin{aligned}
H_b&=-\frac{t}{2}\left(X_0Z_1X_2+Y_0Z_1Y_2+X_1Z_2X_3+Y_1Z_2Y_3\right)\\
&+\frac{U}{16}(Z_0Z_1+Z_2Z_3).
\end{aligned}
\end{equation}

Define $a=U/8$. Since $H_b$ conserves the total spin-up and spin-down electrons $N_\uparrow=n_0+n_2$ and $N_\downarrow=n_1+n_3$, we can split the 16-dimensional space into nine sectors. This makes diagonalization much easier. With the diagonalization circuit $V$, the Hamiltonian becomes
\begin{align}
&V^\dagger H_b V\\
&\quad=\operatorname{diag}(a,t,t,a,-t,-a,-a,-t,-t,-\lambda_b,-a,-t,\lambda_b,t,t,a),\nonumber
\end{align}
which can be block encoded using standard methods. Discussion on the construction and the cost analysis is in \Cref{app:fh_cost_accounting}. The cost of the global block encoding is
\begin{align}
T_{\textsc{Be},b}
&\leq16M^2+80\lceil\log_2M\rceil+8T_{\mathrm{rot}}+344,
\end{align}
as is shown in \Cref{eq:fh-bond-be-cost}

\subsection{A $2\times2$ plaquette block encoding}
\label{sec:nn-fh-plaquette}

Let $\mathcal P_p$ be the four sites of the plaquette anchored at $p$, and let
$\mathcal E_p$ be its four perimeter edges.  Define
\begin{align}
H_p^\square
={}&-\frac t2\sum_{\{r,r'\}\in\mathcal E_p}\sum_\sigma
T_{rr',\sigma}
+\frac U4\sum_{r\in\mathcal P_p}D_r.
\label{eq:nn-fh-plaquette-term}
\end{align}
Every NN edge belongs to two translated plaquettes and every site belongs to
four, so $H=\sum_{p\in\mathbb Z_M^2}H_p^\square$.

As with the bond-wise case, the local Hamiltonian spectrum is symmetric about zero. Hence
\begin{align}
\lambda_\square=\left\|H_p^\square\right\|=\max_k|E_{k}|,\quad\delta_\square=0.
\end{align}
The half-width is available in closed form.  Set
\begin{align}
\tau=|t|,\qquad v=\frac{|U|}{4}.
\end{align}
For $(t,U)\neq(0,0)$, it turns out that
\begin{align}
\lambda_\square
=2\sqrt{\frac{4\tau^2+v^2}{3}}
\cos\left[
\frac13\arccos\left(
\frac{3\sqrt3\,\tau^2v}{(4\tau^2+v^2)^{3/2}}
\right)\right],
\label{eq:nn-fh-plaquette-closed-form}
\end{align}
which we can verify numerically and analytically. We discuss its verification in \Cref{sec:fh-2x2-verification}

\subsubsection{Implementation of the local interaction}

The $2\times 2$ plaquette already has eight qubits (two times multiplicity from different spins), and directly implementing the exact optimal block encoding by diagonalizing the Hamiltonian is challenging. We similarly consider using QET as in the $3\times 3$ block encoding for the Heisenberg model.

The block encoding to which we apply QET is the uniform average of the four normalized bond block encodings on the plaquette, which has normalization $2\lambda_b$ and shift $0$. This is because
\begin{align}
A_\square
=\frac14\sum_{b=\{r,r'\}\in\mathcal E_p}\frac{H_b}{\lambda_b}
=\frac{H_p^\square}{2\lambda_b}.
\end{align}
The factor of two is because $H_p^\square$ has half the contribution from each bond because every bond appears in two plaquettes. Further, since the block encoding is already centered, there is no need for additional centering; the normalization factor to be amplified is $2\lambda_b$. Since the optimal normalization is $\lambda_\square$ instead of $2\lambda_b$, we need to rescale the eigenvalues by $\rho=2\lambda_b/\lambda_\square$. For QET with headroom $s$ and degree $d_\star$, by \Cref{eq:local_qet_global_normalization}, the global shift and normalization are
\begin{align*}
\Delta_\square=0,\quad\Lambda_\square=\frac{M^2\lambda_\square}{1-s}.
\end{align*}
Discussion on the construction and the cost analysis is in \Cref{app:fh_cost_accounting}. Finally, the block encoding has cost
\begin{align}
T_{\textsc{BE},\square}
&\leq16M^2+80\lceil\log_2 M\rceil+368d_\star-8+(9d_\star+1)T_{\mathrm{rot}},
\end{align}
as is shown in \Cref{eq:fh_plaquette_global_qet_cost}.

\section{Block encodings for the spin-liquid Hamiltonian on a honeycomb lattice}
\label{sec:spin_liquids}

We next consider honeycomb spin Hamiltonians motivated by the search for Kitaev quantum spin liquids~\cite{kitaev2006anyons}. Models for candidate materials such as $\alpha$-$\mathrm{RuCl}_3$ include Heisenberg, Kitaev, and symmetric off-diagonal exchange interactions~\cite{kim2015kitaev,rau2014generic,rau2014trigonal}. We include interactions through third neighbors and a uniform magnetic field~\cite{yadav2016kitaev}. Comparing ground-state predictions and magnetization dynamics with experimental data motivates quantum simulation of these models~\cite{bartschi2026potential}.

We first define the Hamiltonian on a periodic lattice of $M\times M$ unit cells, with $M$ even, giving $N_s=2M^2$ spins and $K=M^2$ hexagonal faces.  We use spin-$\tfrac12$ operators $\mathbf S_i=\boldsymbol{\sigma}_i/2$ and assume translation-invariant couplings. Let $\mathcal E_n^\gamma$ denote the set of $n$th-neighbor bonds with exchange label $\gamma\in\{x,y,z\}$. This label specifies the spin component entering the Kitaev interaction. For a second-neighbor bond, the two NN bonds along the path connecting its endpoints have different colors. We assign the remaining color as its exchange label $\gamma$. To write the off-diagonal interactions, we define the other two spin axes for each bond type,
\begin{align}
(\alpha_\gamma,\beta_\gamma)
=
\begin{cases}
(y,z), & \gamma=x,\\
(z,x), & \gamma=y,\\
(x,y), & \gamma=z.
\end{cases}
\label{eq:complementary_spin_axes}
\end{align}
Thus, $\{\alpha_\gamma,\beta_\gamma,\gamma\}=\{x,y,z\}$ for each
fixed value of $\gamma$.  Then, the nearest-neighbor bond Hamiltonian is defined as,
\begin{align}
H^b_{ij}(\gamma,1)
={}&
J_1\,\mathbf S_i\cdot\mathbf S_j
+K_{1,\gamma}S_i^\gamma S_j^\gamma\nonumber \\
&
+\Gamma_{1,\gamma}
\left(
S_i^{\alpha_\gamma}S_j^{\beta_\gamma}
+S_i^{\beta_\gamma}S_j^{\alpha_\gamma}
\right)\nonumber \\
&+\Gamma'_{1,\gamma}
\Big[
S_i^\gamma
 \left(S_j^{\alpha_\gamma}+S_j^{\beta_\gamma}\right)
+
\left(S_i^{\alpha_\gamma}+S_i^{\beta_\gamma}\right)
S_j^\gamma
\Big].
\label{eq:honeycomb_nn_bond}
\end{align}
Here $J_1$ is the isotropic Heisenberg coupling, $K_{1,\gamma}$ is the Kitaev coupling, and $\Gamma_{1,\gamma},\Gamma'_{1,\gamma}$ couple different spin components  depending on the bond label $\gamma$. For second- and third-neighbor bonds, we include Heisenberg and Kitaev interactions,
\begin{align}
H^b_{ij}(\gamma,2)
&=
J_2\,\mathbf S_i\cdot\mathbf S_j
+K_{2,\gamma}S_i^\gamma S_j^\gamma,
\label{eq:honeycomb_nnn_bond}\\
H^b_{ij}(\gamma,3)
&=
J_3\,\mathbf S_i\cdot\mathbf S_j
+K_{3,\gamma}S_i^\gamma S_j^\gamma.
\label{eq:honeycomb_third_neighbor_bond}
\end{align}
The full Hamiltonian is therefore
\begin{align}
H
={}&
\sum_{\gamma=x,y,z}
\sum_{(i,j)\in\mathcal E_1^\gamma}
H^b_{ij}(\gamma,1)
+
\sum_{\gamma=x,y,z}
\sum_{(i,j)\in\mathcal E_2^\gamma}
H^b_{ij}(\gamma,2)\nonumber \\
&+
\sum_{\gamma=x,y,z}
\sum_{(i,j)\in\mathcal E_3^\gamma}
H^b_{ij}(\gamma,3)
+
A\sum_i(S_i^c)^2
+\sum_i\mathbf h\cdot\mathbf S_i.
\label{eq:extended_honeycomb_hamiltonian}
\end{align}

Here $A$ is the single-ion anisotropy coefficient and $\mathbf h$ is the effective magnetic field in the spin frame. To relate it to the applied field, we use the crystallographic axes $(\hat a,\hat b,\hat c)$, where $\hat a$ and $\hat b$ lie in the lattice plane and $\hat c$ is normal to it. It is related to the spin frame through, 
\begin{align}
\begin{bmatrix}\hat a\\\hat b\\\hat c\end{bmatrix}
=D\begin{bmatrix}\hat x\\\hat y\\\hat z\end{bmatrix},
\qquad
D=\begin{bmatrix}
1/\sqrt6 & 1/\sqrt6 & -2/\sqrt6\\
1/\sqrt2 & -1/\sqrt2 & 0\\
1/\sqrt3 & 1/\sqrt3 & 1/\sqrt3
\end{bmatrix},
\label{eq:honeycomb_frame}
\end{align}
Writing the applied field as $\mathbf H=(H_a,H_b,H_c)$ and the anisotropic $g$ tensor as $G=\operatorname{diag}(g_a,g_b,g_c)$ in this frame, the Zeeman term becomes
\begin{align}
\sum_i\mathbf h\cdot\mathbf S_i
&=\mu_B\sum_i\left(g_aS^a_iH_a+g_bS^b_iH_b+g_cS^c_iH_c\right),
\nonumber \\
\mathbf h&=\mu_B\,D^{T}G\,\mathbf H,
\label{eq:honeycomb_zeeman}
\end{align}
where $\mu_B$ is the Bohr magneton. Also, $S^c_i=(S^x_i+S^y_i+S^z_i)/\sqrt3$, so $(S^c_i)^2=I/4$. Hence, the single-ion term is the constant $AN_s I/4$. It shifts $\Delta$ and leaves $\Lambda$ untouched, so we can remove it from the circuit and restore it as a known energy shift later. 

We use the honeycomb routing construction in~\Cref{subsec:honeycomb_lattice_routing} to bring a selected face to a fixed location. The router moves four unit cells containing eight spins, six of which form the hexagonal face as shown in~\Cref{fig:honeycomb}.  We construct a hierarchy of local block encodings, starting with separate bond terms and then treating larger groups of interactions jointly, including on larger patches. 

\subsection{Bond-wise local block encoding}
\label{subsec:honeycomb_bondwise}

In this section, we construct a local block encoding of the hexagonal plaquette with six spins that form  a complete interaction graph of 15 bonds. We first set the Zeeman field to zero and encode each bond separately. A hexagonal face contains six NN bonds, six second-neighbor bonds, and three third-neighbor bonds. Each NN bond belongs to two faces, while each second- and third-neighbor bond belongs to one. Hence, the field-free patch Hamiltonian is
\begin{align}
H_p^{\hexago}
=\sum_{g=0}^{14}w_gH^b_{i_gj_g}(\gamma_g,n_g),\qquad
w_g=\begin{cases}
\tfrac12,&n_g=1,\\
1,&n_g=2,3,
\end{cases}
\label{eq:honeycomb_patch_decomposition}
\end{align}
where $g$ labels the bond of type $\gamma_g$ with interaction on the $n_g$-th neighbor. The two other routed spins are not acted on by this encoding. To find the optimal shift and normalization of a bond, we write its exchange couplings as a matrix,
\begin{align*}
H^b_{ij}(\gamma, n)=\sum_{a,b\in\{x,y,z\}}M_{ab} (\gamma, n)\,S^a_iS^b_j ,
\end{align*}
where $\mathbf{M}(\gamma, n)$ is a $3\times3$ matrix of couplings, 
\begin{align}
&\mathbf M(x,1)=
\begin{pmatrix}
J_1+K_{1,x} & \Gamma'_{1,x} & \Gamma'_{1,x}\\
\Gamma'_{1,x} & J_1 & \Gamma_{1,x}\\
\Gamma'_{1,x} & \Gamma_{1,x} & J_1
\end{pmatrix},
\nonumber \\
&
\mathbf M(y,1)=
\begin{pmatrix}
J_1 & \Gamma'_{1,y} & \Gamma_{1,y}\\
\Gamma'_{1,y} & J_1+K_{1,y} & \Gamma'_{1,y}\\
\Gamma_{1,y} & \Gamma'_{1,y} & J_1
\end{pmatrix},
\nonumber \\
&
\mathbf M(z,1)=
\begin{pmatrix}
J_1 & \Gamma_{1,z} & \Gamma'_{1,z}\\
\Gamma_{1,z} & J_1 & \Gamma'_{1,z}\\
\Gamma'_{1,z} & \Gamma'_{1,z} & J_1+K_{1,z}
\end{pmatrix},
\label{eq:honeycomb_coupling_matrix}
\end{align}
Note that the three matrices are simply cyclic permutations of the each other, with the corresponding couplings. For $n=2,3$, $\mathbf M(\gamma,n)=J_nI+K_{n,\gamma}e_\gamma e_\gamma^T$.

Each coupling matrix is real and symmetric. A common rotation of the two spins therefore reduces the bond Hamiltonian to an XYZ interaction, whose three terms commute and share the Bell eigenbasis. This follows from the general construction for real symmetric exchange matrices in~\Cref{app:real_symmetric_exchange}. Let $\tilde J_x,\tilde J_y,\tilde J_z$ be the eigenvalues of $\mathbf M$, suppressing the bond labels. For NN bonds, the three coupling eigenvalues $\{\tilde J_x,\tilde J_y,\tilde J_z\}$ form the set:
\begin{align}
\left\{
J_1-\Gamma_{1,\gamma},
J_1+\frac{K_{1,\gamma}+\Gamma_{1,\gamma}}{2}\pm
\sqrt{\left(\frac{\Gamma_{1,\gamma}-K_{1,\gamma}}{2}\right)^2\!+2(\Gamma'_{1,\gamma})^2}
\right\}
\label{eq:honeycomb_exchange_eigenvalues}
\end{align}
For second- and third-neighbor bonds, the original coupling matrix is already diagonal. We can take $O=I$ and $U=I$, with $\tilde J_\gamma=J_n+K_{n,\gamma}$ and $\tilde J_a=J_n$ for $a\ne\gamma$. Finally, the XYZ interaction is diagonal in the Bell basis, with energies
\begin{align}
(e_{\Psi^-},e_{\Phi^-},e_{\Phi^+},e_{\Psi^+})
=\frac14\left(-\Sigma,\ \Sigma-2\tilde J_x,\ \Sigma-2\tilde J_y,\ \Sigma-2\tilde J_z\right).
\label{eq:honeycomb_bond_spectrum}
\end{align}
where $\Sigma=\tilde J_x+\tilde J_y+\tilde J_z$. Taking the minimum and maximum of these four energies gives the exact spectral bounds. In particular, define $\bar\Sigma=|\tilde J_x|+|\tilde J_y|+|\tilde J_z|$, $m_b=\min_c|\tilde J_c|$, and $s_b=\operatorname{sign}(\tilde J_x\tilde J_y\tilde J_z)$. Then
\begin{align}
e_{\min}&=\frac{-\bar\Sigma+(1-s_b)m_b}{4},\qquad
e_{\max}=\frac{\bar\Sigma-(1+s_b)m_b}{4}
\end{align}
The optimal shift and normalization of the bond Hamiltonian $H_{ij}^b(\gamma, n)$ are 
\begin{align}
\delta_b(\gamma, n)=-\frac{s_bm_b}{4},\qquad
\lambda_b(\gamma, n)=\frac{\bar\Sigma-m_b}{4}.
\label{eq:honeycomb_bond_normalization}
\end{align}

We now encode the shifted and normalized bond operator $A_{ij}=(H^b_{ij}(\gamma,n)-\delta_b(\gamma,n)I)/\lambda_b(\gamma,n)$. $A_{ij}$ can have at most four distinct eigenvalues, with at least one at $+1$ and $-1$. Some choices of couplings give a unitary directly, without requiring dilation.  For a real nonzero coupling $J$, let $s_J=\operatorname{sign}(J)$. The following cases have simple circuits in the original spin basis:
\begin{align}
\mathbf M=JI
&\quad\Longrightarrow\quad A_{ij}=s_JP_{ij},
\nonumber \\
\mathbf M=Je_\gamma e_\gamma^T
&\quad\Longrightarrow\quad A_{ij}=s_J\sigma_i^\gamma\sigma_j^\gamma,\nonumber \\
\mathbf M=J(I-2e_\gamma e_\gamma^T)
&\quad\Longrightarrow\quad A_{ij}=-s_JP_{ij}\sigma_i^\gamma\sigma_j^\gamma.
\label{eq:honeycomb_reflected_encoding}
\end{align}
The first recovers the Heisenberg SWAP encoding in~\Cref{sec:heisenberg}. The second is a pure Kitaev bond. The third has $K_{n,\gamma}=-2J_n$ and no off-diagonal exchange. 

For a general bond, we use the spectral encoding from~\Cref{sec:local_encoding}. Let $O$ be a proper rotation that diagonalizes $\mathbf M$ and $u$ its corresponding single-spin unitary, with $U=u_i\otimes u_j$. The eigenbasis transformation is
\begin{align}
V_{ij}=(UC_{\mathrm{Bell}})^\dagger,
\label{eq:honeycomb_bond_eigenbasis}
\end{align}
where $C_{\mathrm{Bell}}$ is a Clifford circuit that maps the computational basis to the Bell states in the order used in~\Cref{eq:honeycomb_bond_spectrum}. We combine the 15 bond encodings through local LCU to obtain $B_{\mathrm{ref}}$, encoding $(H_{\mathrm{ref}}^{\hexago}-\delta_{p,b}I)/\lambda_{p,b}$, where,
\begin{align}
\delta_{p,b}
&=\sum_{\gamma=x,y,z}
\left(\delta_b(\gamma,1)+2\delta_b(\gamma,2)+\delta_b(\gamma,3)\right),\\
\lambda_{p,b}
&=\sum_{\gamma=x,y,z}
\left(\lambda_b(\gamma,1)+2\lambda_b(\gamma,2)+\lambda_b(\gamma,3)\right).
\label{eq:honeycomb_bond_patch_normalization}
\end{align}
Summing over the $N_l=N_s/2$ patches gives the global shift and normalization,
\begin{align}
\Delta_b^{\hexago}=\frac{N_s}{2}\delta_{p,b}+\frac{AN_s}{4},\qquad
\Lambda_b^{\hexago}=\frac{N_s}{2}\lambda_{p,b}.
\label{eq:honeycomb_bond_global_normalization}
\end{align}
where we have included the known global shift from the single-ion term.

The additional symmetry of the NN coupling matrix under exchanging the two axes perpendicular to $\gamma$ reduces its spin basis change to one variable angle and fixed axis changes. The other bonds need no spin basis change. We give the preparation, selection circuits, and detailed cost accounting in~\Cref{app:subsec:honeycomb_bondwise}. The local and global bounds derived in~\Cref{eq:honeycomb_bond_local_t_cost,eq:honeycomb_bond_total_t_cost} are
\begin{align*}
T_{\textsc{BE}}^p&\leq200+114T_{\mathrm{rot}}(\epsilon),\\
T_{\textsc{BE},b}^{\hexago}
&\leq8N_s+80\left\lceil\frac{\log_2N_s-1}{2}\right\rceil
+114T_{\mathrm{rot}}(\epsilon)+184,
\end{align*}
where $N_s=2M^2$.

\subsubsection{Extension to the Zeeman field}

We now include the uniform Zeeman field by assigning its contribution to the NN bonds. Each site belongs to three patches and meets two NN bonds within each patch. Hence, each bond endpoint carries $1/6$ of the on-site field. For bond $g$ with sites $i,j$ and labels $(\gamma_g,n_g)$, define
\begin{align}
K^b_{ij}=w_gH^b_{ij}&(\gamma_g,n_g)+\eta_g\mathbf h\cdot(\mathbf S_i+\mathbf S_j),
\nonumber \\
(w_g,\eta_g)&=
\begin{cases}
\left(\tfrac12,\tfrac16\right), & n_g=1,\\[2pt]
(1,0), & n_g=2,3.
\end{cases}
\label{eq:honeycomb_bond_with_field}
\end{align}
For a fixed bond, we suppress the labels $(\gamma,n)$. We reuse the exchange-matrix analysis for $w_g\mathbf M$, whose eigenvalues we now denote by $\tilde J_c$. The rotation $O$ is unchanged by this positive rescaling and also rotates the field, giving $\mathbf f=\eta_g O^T\mathbf h$.

Since $\mathbf M$ is symmetric and the field couples equally to both endpoints, $K^b_{ij}$ commutes with the SWAP operator $P_{ij}$. Hence, it is block-diagonal in the singlet and triplet sectors. The total spin vanishes on the singlet, $(\mathbf S_i+\mathbf S_j)\ket{\Psi^-}=0$, so its energy is unchanged by the field, $E_-=-\Sigma/4$, where $\Sigma=\tilde J_x+\tilde J_y+\tilde J_z$. Using the Bell transform $C_{\mathrm{Bell}}$ defined above, in the order $(\ket{\Psi^-},\ket{\Phi^-},\ket{\Phi^+},\ket{\Psi^+})$, and writing $d_c=(\Sigma-2\tilde J_c)/4$, we obtain
\begin{align}
C_{\mathrm{Bell}}^\dagger U^\dagger K^b_{ij}UC_{\mathrm{Bell}}
=\begin{pmatrix}E_-&0\\0&\mathbf T\end{pmatrix},
\qquad
\mathbf T=
\begin{pmatrix}
d_x&f_z&-if_y\\
f_z&d_y&f_x\\
if_y&f_x&d_z
\end{pmatrix},
\label{eq:honeycomb_field_block}
\end{align}
The field mixes the triplet states, while the singlet remains unchanged. Let $E_0,E_1,E_2$ be the eigenvalues of $\mathbf T$, with closed-form expressions given in~\Cref{eq:honeycomb_zeeman_eigenvalues}. Then define $E_{\min}^{Z}$ and $E_{\max}^{Z}$ as the extreme eigenvalues of the bond operator. When the field vanishes, $\mathbf T$ is diagonal and we recover the field-free bond spectrum. The optimal shift and normalization of $K_{ij}^b(\gamma,n)$ can be computed using the extreme eigenvalues as $\delta_Z(\gamma,n)$ and $\lambda_Z(\gamma,n)$.

To complete the local encoding, we compute an orthonormal eigenvector matrix $Q$ of $\mathbf T$ classically, with $Q^\dagger\mathbf TQ=\operatorname{diag}(E_0,E_1,E_2)$. The full eigenbasis transformation is
\begin{align}
V_Z=\left[UC_{\mathrm{Bell}}\begin{pmatrix}1&0\\0&Q\end{pmatrix}\right]^\dagger, \,
V_ZK^b_{ij}V_Z^\dagger=\operatorname{diag} (E_-,E_0,E_1,E_2).
\label{eq:honeycomb_zeeman_eigenbasis}
\end{align}
We use the spectral encoding from~\Cref{subsec:spectral_encoding} for $A^Z_{ij}=(K^b_{ij}-\delta_Z(\gamma,n)I)/\lambda_Z(\gamma,n)$, with basis change $V_Z$. We reuse the local preparation oracle with the normalizations of $K^b_{ij}$ as the LCU weights, and sum their shifts and normalizations over the patch. The gate-level circuit and cost accounting in~\Cref{app:subsec:honeycomb_bondwise_zeeman,eq:honeycomb_zeeman_local_t_cost,eq:honeycomb_zeeman_global_t_cost} gives
\begin{align*}
T_{\textsc{BE},Z}^{p}&\leq200+258T_{\mathrm{rot}}(\epsilon),\\
T_{\textsc{BE},Z}^{\hexago}&\leq8N_s+80\left\lceil\frac{\log_2N_s-1}{2}\right\rceil+258T_{\mathrm{rot}}(\epsilon)+184.
\end{align*}

\subsection{Star-based local block encoding}
\label{subsec:honeycomb_star}

The bond-wise construction treats each interaction separately. We can instead group the interactions into larger terms, such as triangles, quadrilaterals or stars, and encode each group jointly. We have numerous such design choices and it is impossible to enumerate all of them. There are two conflicting effects at play here. Grouping terms can improve the normalization since $\lambda(A+B)\leq\lambda(A)+\lambda(B)$, where $\lambda$ denotes the optimal normalization obtained from the spectral bounds. Equality is achieved when the interactions are disjoint, while overlapping interactions can give a smaller normalization when encoded jointly. On the other hand, efficiently encoding a Hamiltonian representing a larger interaction group is challenging. A $k$-spin Hamiltonian is a $2^k\times2^k$ matrix, and constructing its eigenbasis can become expensive unless there is structure to exploit.

For the different groupings within the routed eight-spin register, the routing circuit and its cost remain the same. The local encoding cost is independent of the lattice size and is paid once per query. So, the effect of a smaller normalization matters more at large $M$, while a cheaper local encoding can be more important at small $M$. Note that it is not purely the number of spins that matters but also the geometry and structure of the interactions within each group. For example, a four-spin interaction graph can be a quadrilateral or a star face on a hexagonal lattice. 

\begin{figure*}[!ht]
\centering
\includegraphics[width=0.7\textwidth]{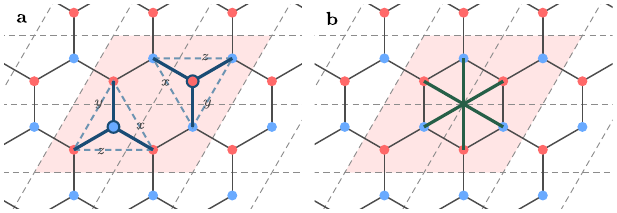}
\caption{\textbf{The honeycomb lattice decomposition.} The red rhombus is the routed $2\times2$ block of four cells and contains eight spins. \textbf{a}, The solid dark-blue spokes are the two nearest-neighbor stars, one bond of each color $x,y,z$ per star.  The leaves are mutually second neighbors and close into the dashed triangles.  The solid and dashed bonds together make each star a complete graph on four spins.  The sites drawn with a ring are the star centers; the two additional leaves lie outside the six-site face but complete the stars. \textbf{b}, The three green bonds are the mutually disjoint third-neighbor diagonals of the face.}
\label{fig:honeycomb-encoding}
\end{figure*}

Next, we provide a construction for the honeycomb lattice that improves the normalization over the bond-wise encoding while still being simple to implement. On the routed eight-spin patch, we form two disjoint four-spin stars, one centered on each sublattice, as shown in~\Cref{fig:honeycomb-encoding}. Each star contains three NN bonds connecting the center to its leaves and three second-neighbor bonds connecting the leaves. So encoding the star as one interaction can improve the normalization compared with encoding its bonds separately. The two stars are disjoint, so their optimal normalizations add exactly. The three third-nearest-neighbor bonds are also mutually disjoint.  They do, however, share spins with the stars. We encode these bonds separately and combine them with the two star encodings through LCU. Hence, this need not achieve the optimal normalization of the complete eight-spin patch. However, this allows us to construct a simpler local encoding circuit. 

Let us denote the two sublattices as $A$ and $B$, shown as red and blue spins in the figure. A block anchored at cell $p=(p_x,p_y)$ contains the spins $A(p_x+r,p_y+s)$ and $B(p_x+r,p_y+s)$, with $r,s\in\{0,1\}$. Once the block has been routed, we can drop the patch address and simply denote the eight qubits by $A_{rs}$ and $B_{rs}$. We define a star as a site and its three nearest neighbors. The routed register contains two stars defined as,
\begin{align}
\mathcal S_A
&=\{A_{11};B_{11},B_{01},B_{10}\},
&
\mathcal S_B
&=\{B_{00};A_{00},A_{10},A_{01}\},
\label{eq:honeycomb_two_stars}
\end{align}
where the centers are $A_{11}$ and $B_{00}$. In either star the three leaves may be labeled $\ell_x,\ell_y,\ell_z$ so that $(c,\ell_\gamma)$ is the nearest-neighbor bond of type $\gamma$. A star has three nearest-neighbor bonds and three second-neighbor bonds forming a complete graph. For a leaf pair $\{i,j\}$, let $\gamma_{ij}$ be its exchange label, so that $(i,j)\in\mathcal E_2^{\gamma_{ij}}$. For a star $\mathcal S=\{c;\ell_x,\ell_y,\ell_z\}$ define,
\begin{align}
H^{\star}_{\mathcal S}
={}\underbrace{\frac12\sum_{\gamma=x,y,z}
H^b_{c\ell_\gamma}(\gamma,1)}_{\text{nearest-neighbor bonds}}
&+\underbrace{\sum_{\{i,j\}\subset\mathcal S \setminus\{c\}}
H^b_{ij}(\gamma_{ij},2)}_{\text{second-neighbor bonds}} \nonumber \\&
+\underbrace{\frac14\sum_{i\in\mathcal S}\mathbf h\cdot\mathbf S_i}_{\text{field}} .
\label{eq:honeycomb_star_hamiltonian}
\end{align}
A nearest-neighbor bond lies in two translated stars, so it gets a weight of $1/2$, whereas a second-neighbor bond lies in exactly one translated star and gets a weight of $1$. A spin occurs in four translated stars, once as a center and three times as a leaf, so the field term gets a weight of $1/4$. Including the three third-neighbor bonds, the local Hamiltonian for the block at $p$ is,
\begin{align}
H^\star_p
={}&H^\star_{\mathcal S_A}+H^\star_{\mathcal S_B}
+\sum_{(i,j)\in\mathcal D}H^b_{ij}(\gamma_{ij},3).
\label{eq:honeycomb_two_star_patch}
\end{align}
Here $\mathcal D=\{(A_{11},B_{00}),(A_{01},B_{10}),(A_{10},B_{01})\}$, and $\gamma_{ij}$ is the exchange label of each third-neighbor bond. This gives a different decomposition from the six-spin face construction, since the stars include interactions on the two additional spins. Summing over all translated patches still gives $H=\sum_p H^\star_p+AN_s I/4$. We know the block encoding for the bonds, so the star is the only new term in this construction.

\subsubsection{Local block encoding of the star}
\label{sec:honeycomb_star_encoding}

In terms of the coupling matrices in~\Cref{eq:honeycomb_coupling_matrix},
\begin{align}
H^\star_{\mathcal S}&={}\frac18\sum_{\gamma, a, b\in\{x,y,z\}}\ 
M_{ab}(\gamma,1)\,\sigma^a_c\sigma^b_{\ell_\gamma}
\nonumber\\ &+\frac14 \sum_{\{i,j\}\subset\mathcal S\setminus\{c\}}\ \sum_{a,b\in\{x,y,z\}}
M_{ab}(\gamma_{ij},2)\,\sigma^a_i\sigma^b_j
\nonumber\\ &+\frac18\sum_{i\in\mathcal S} \sum_{a\in\{x,y,z\}} h_a \sigma_i^a.
\end{align}

We provide two constructions for the star. The first divides the Pauli terms into anticommuting groups. This gives a simple and cheap construction, but need not achieve the optimal normalization of the star. The second uses spectral decomposition and exploits symmetry under assumptions discussed below. This can give a tighter normalization while costing more rotations to implement. The two approaches can therefore favor different lattice sizes. Comparing their normalizations and circuit costs for a chosen application determines the exact crossover size.

\paragraph*{Approach 1: Anticommuting Pauli strings}

This approach is based on the fact that if $P_1, P_2, \dots, P_m$ are Pauli strings that pairwise anticommute, i.e., $P_jP_k = -P_k P_j$ for $j \neq k$, then, 
\begin{align}
G^2 = \left(\sum_j c_j P_j\right)^2
&=\sum_jc_j^2P_j^2+\sum_{j<k}c_jc_k\left(P_jP_k+P_kP_j\right)\nonumber\\ &
=\Vert c\Vert_2^2\,I .
\end{align}
Since the spectrum of $G$ is $\pm\|c\|_2$, the normalized operator $G/\|c\|_2$ is a Hermitian unitary and achieves the optimal normalization for this group.

We group the Pauli strings of the star Hamiltonian based on the geometry of the interactions. Two strings anticommute when they disagree on an odd number of qubits where both act nontrivially. On the same NN bond, the strings must differ in exactly one of the spin indices $a,b$. On different NN bonds, they share only the center spin, so their indices $a$ must differ. For second-neighbor bonds, $\mathbf M(\gamma,2)$ is diagonal, so every string has $a=b$. Strings on the same bond commute, so each second-neighbor group takes at most one string per bond. Strings on different leaf bonds share one spin, on which their Pauli operators must differ. Finally, the three field components on each site anticommute. Putting these together, we have the following ten groups, with three for the NN interactions, three for the second-nearest neighbor interactions and four for the field. For the NN bonds, we have,
\begin{align}
\mathcal G_1&=\left\{\sigma^x_c\sigma^b_{\ell_x}\right\}_b\cup\left\{\sigma^y_c\sigma^b_{\ell_y}\right\}_b\cup\left\{\sigma^z_c\sigma^b_{\ell_z}\right\}_b,
\nonumber\\
\mathcal G_2&=\left\{\sigma^y_c\sigma^b_{\ell_x}\right\}_b\cup\left\{\sigma^z_c\sigma^b_{\ell_y}\right\}_b\cup\left\{\sigma^x_c\sigma^b_{\ell_z}\right\}_b,
\nonumber\\
\mathcal G_3&=\left\{\sigma^z_c\sigma^b_{\ell_x}\right\}_b\cup\left\{\sigma^x_c\sigma^b_{\ell_y}\right\}_b\cup\left\{\sigma^y_c\sigma^b_{\ell_z}\right\}_b,
\label{eq:honeycomb_star_three_groups}
\end{align}
with each $b$ running over $x,y,z$. Each group holds one row of each coupling matrix $\mathbf M(\gamma,1)$. So its normalization is,
\begin{align}
\beta_k=\frac18\sqrt{\sum_{\gamma\in\{x,y,z\}}\ \sum_{b\in\{x,y,z\}}M_{\pi_k(\gamma)b}(\gamma,1)^2},
\qquad k=1,2,3,
\label{eq:honeycomb_star_nn_group_norms}
\end{align}
where $\pi_k(\gamma)$ is the row that group $k$ takes from color $\gamma$. $\pi_1$ is the identity and $\pi_2,\pi_3$ are the two cyclic shifts. The structure of the NN coupling matrices gives $\beta_2=\beta_3$, including for color-dependent couplings. For the second nearest neighbor bonds, we have,
\begin{align}
&\mathcal G_4=\left\{\sigma^z_{\ell_x}\sigma^z_{\ell_y},\ \sigma^x_{\ell_y}\sigma^x_{\ell_z},\ \sigma^y_{\ell_z}\sigma^y_{\ell_x}\right\},\nonumber\\ &
\mathcal G_5=\left\{\sigma^x_{\ell_x}\sigma^x_{\ell_y},\ \sigma^y_{\ell_y}\sigma^y_{\ell_z},\ \sigma^z_{\ell_z}\sigma^z_{\ell_x}\right\},\nonumber\\ &
\mathcal G_6=\left\{\sigma^y_{\ell_x}\sigma^y_{\ell_y},\ \sigma^z_{\ell_y}\sigma^z_{\ell_z},\ \sigma^x_{\ell_z}\sigma^x_{\ell_x}\right\}.
\label{eq:honeycomb_star_triangle_groups}
\end{align}
With $\mathbf M(\gamma,2)=J_2I+K_{2,\gamma}\,e_\gamma e_\gamma^T$,
\begin{align}
\beta_4=\frac14\sqrt{\sum_{\gamma}\left(J_2+K_{2,\gamma}\right)^2},
\qquad
\beta_5=\beta_6=\frac{\sqrt3}{4}\left\vert J_2\right\vert .
\label{eq:honeycomb_star_triangle_group_norms_closed}
\end{align}
Finally, the four field groups are,
\begin{align}
\mathcal G_7=\left\{\sigma^a_c\right\}_a,
\,
\mathcal G_8=\left\{\sigma^a_{\ell_x}\right\}_a,
\,
\mathcal G_9=\left\{\sigma^a_{\ell_y}\right\}_a,
\,
\mathcal G_{10}=\left\{\sigma^a_{\ell_z}\right\}_a,
\label{eq:honeycomb_star_field_groups}
\end{align}
with each $a$ running over $x,y,z$. Each field group holds the three components of $\mathbf h$ on one site, so $\beta_7=\beta_8=\beta_9=\beta_{10}=\|\mathbf h\|_2/8$. 

We can construct one normalized operator from each group. Let $G_g=\sum_{P_\nu\in\mathcal G_g}c_\nu P_\nu$, with $m_g$ strings, and set $\beta_g=\|c_g\|_2$ and $S_g=G_g/\beta_g$. Zero-weight groups are omitted. Since $S_g^2=I$, it is a Hermitian unitary and the star Hamiltonian is simply a linear combination of unitaries,
\begin{align}
H^\star_{\mathcal S}=\sum_g\beta_gS_g,
\qquad
\beta^{\star}=\sum_g\beta_g.
\label{eq:honeycomb_star_anti_lcu}
\end{align}
Each group has a spectrum symmetric about zero, so we use zero shift in this construction. We implement each $S_g$ using the anticommuting-group construction in~\Cref{app:anticommuting_pauli_groups} and combine these unitaries through local LCU. The repeated group normalizations simplify the preparation, whose circuit and cost are given in~\Cref{app:subsec:honeycomb_anti}. The cost accounting in~\Cref{app:subsec:honeycomb_anti,eq:honeycomb_star_anti_t_cost} gives
\begin{align*}
T_{\textsc{BE}}^\star\leq36+88T_{\mathrm{rot}}(\epsilon).
\end{align*}

\paragraph*{Approach 2: Symmetry-reduced spectral encoding} 

We now use the spectrum of the exchange Hamiltonian to obtain a tighter normalization. So far, we have allowed the couplings to depend on the bond color. Here, we assume $K_{1,\gamma}=K_1$, $\Gamma_{1,\gamma}=\Gamma_1$, $\Gamma'_{1,\gamma}=\Gamma'_1$, and $K_{2,\gamma}=K_2$. These assumptions give a cyclic symmetry that we use to simplify the eigenbasis construction. Split the star Hamiltonian into exchange and field terms as,
\begin{align}
H^\star_{\mathcal S}
&=H^\star_{\mathcal S,0}+H^\star_{\mathcal S,h},\qquad
H^\star_{\mathcal S,h}=\frac14\sum_{i\in\mathcal S}\mathbf h\cdot\mathbf S_i.
\label{eq:honeycomb_star_exchange_field_split}
\end{align}
The exchange term contains all six bonds of the star. We construct its spectral encoding first and then add the field through LCU.

With these assumptions, the exchange Hamiltonian is invariant under a combined cyclic permutation of the spin axes and the three leaves. Let $u_3$ rotate a spin by $2\pi/3$ about $(1,1,1)/\sqrt3$, cycling $X\to Y\to Z\to X$, and let $P_\ell$ cycle the leaves as $\ell_x\to\ell_y\to\ell_z\to\ell_x$. The combined operation is,
\begin{align}
u_3=\frac12\left[I-i(X+Y+Z)\right],\qquad
C_3=P_\ell u_3^{\otimes4}.
\label{eq:honeycomb_star_c3_definition}
\end{align}
Under $C_3$ (cyclic symmetry of order 3), each exchange bond becomes the corresponding bond of the next color. For example, $S_c^xS_{\ell_x}^x$ becomes $S_c^yS_{\ell_y}^y$. The same transformation holds for the off-diagonal exchange terms and the leaf triangle. Since the couplings are independent of color, the exchange Hamiltonian is unchanged,
\begin{align}
[C_3,H^\star_{\mathcal S,0}]=0.
\label{eq:honeycomb_star_c3_invariance}
\end{align}
Since commuting matrices preserve each other's eigenspaces, choosing a basis organized by the eigenspaces of $C_3$ makes $H^\star_{\mathcal S,0}$ block diagonal. Its three symmetry sectors have dimensions $4,6,6$, so we can diagonalize these smaller matrices separately. Let $F_{C_3}$ be the basis change that exposes the three symmetry sectors.   Additional reflection and time-reversal symmetries give a further reduction of eigenspaces. Reflection symmetry splits the four-dimensional sector into even and odd subspaces, each of dimension two. It also relates the two six-dimensional sectors, allowing us to choose bases in which their Hamiltonian matrices are identical. Time-reversal symmetry allows all these matrices to be chosen real. We collect these further basis changes into a unitary $G$ and define the complete transformation as $F_{D_3}=G F_{C_3}$. Then,
\begin{align}
F_{D_3}H^\star_{\mathcal S,0}F_{D_3}^\dagger
=A_+\oplus A_-\oplus B\oplus B,
\label{eq:honeycomb_star_d3_blocks}
\end{align}
where $A_\pm$ are real $2\times2$ matrices acting on the even and odd subspaces, and $B$ is a real $6\times6$ matrix. Since the two copies of $B$ are identical, the same eigenbasis transformation diagonalizes both. 

We compute the eigenvalues of $A_\pm$ analytically and diagonalize the real $6\times6$ matrix $B$ classically. Let $O_\pm$ and $O_B$ contain orthonormal eigenvectors of $A_\pm$ and $B$, respectively. Taking the minimum and maximum over these eigenvalues gives $\epsilon_{\min}$ and $\epsilon_{\max}$. We use these to define the optimal exchange shift $\delta^\star_0$, normalization $\lambda^\star_0$, and the shifted/ normalized operator $A^\star_0$. To share the circuit for $O_B$, we use a permutation $\mathcal P$ of the symmetry-basis labels so that one qubit distinguishes the two copies of $B$ and the other three use the same six labels within each copy. Define $\widehat O_B=O_B\oplus I_2$ on this three-qubit register. Let $U_A$ apply $O_+$ or $O_-$ to the remaining two labels for sector bit zero or one, respectively, leaving the $B$ subspaces unchanged. The complete eigenbasis transformation is then,
\begin{align}
V^\star_0=U_A^T(I_2\otimes\widehat O_B^T)\mathcal P F_{D_3}.
\label{eq:honeycomb_star_exchange_eigenbasis_main}
\end{align}
With this basis change and the normalized eigenvalues, we use the spectral encoding from~\Cref{subsec:spectral_encoding} to construct the exchange encoding $B^\star_0$. We give a detailed description of the basis construction and its circuit cost in~\Cref{appendix:star_additional_symmetries}.

The uniform field is unchanged by the leaf permutation, but the spin rotation cycles its components. The field term commutes with $C_3$ only when $h_x=h_y=h_z$. For a generic field, we avoid constructing a new eigenbasis for the complete star and instead reuse the four field-group unitaries $S_7,\dots,S_{10}$ from the anticommuting construction. Each has weight $\|\mathbf h\|_2/8$, giving the total field normalization $\lambda^\star_h=\|\mathbf h\|_2/2$. Their combination with the exchange encoding gives,
\begin{align}
A^\star_{\mathcal S}
&=\frac{H^\star_{\mathcal S}-\delta^\star_0I}{\lambda^\star}=\frac{\lambda^\star_0}{\lambda^\star}A^\star_0
+\frac{\lambda^\star_h}{4\lambda^\star}\sum_{g=7}^{10}S_g, \text{where } \lambda^\star=\lambda^\star_0+\lambda^\star_h.
\label{eq:honeycomb_symmetry_star_lcu}
\end{align}
Thus, the shift remains $\delta^\star_0$, while the normalization includes the field contribution. Detailed circuit construction and costing analysis are provided in~\Cref{app:subsec:honeycomb_symm_spec} and we get the following cost bounds from~\Cref{eq:honeycomb_exchange_star_encoding_cost,eq:honeycomb_star_controlled_exchange_cost,eq:honeycomb_symmetry_star_encoding_cost},
\begin{align*}
T^\star_{\textsc{BE},0}
&\leq126T_{\mathrm{rot}}(\epsilon)+252, \\
T_{\mathrm{c}B^\star_0}
&\leq142T_{\mathrm{rot}}(\epsilon)+316,\\
T_{\textsc{BE}}^\star
&\leq160T_{\mathrm{rot}}(\epsilon)+328.
\end{align*}

\subsubsection{Local block encoding of the third-neighbor bonds}

The third-neighbor bonds have the same Heisenberg--Kitaev form used in the bond-wise construction. Here, we use a two-term LCU that achieves their optimal shift and normalization. For a bond of type $\gamma$, write $Q_{ij}^\gamma=\sigma_i^\gamma\sigma_j^\gamma$. Using $\mathbf S_i=\boldsymbol\sigma_i/2$ and $\boldsymbol\sigma_i\cdot\boldsymbol\sigma_j=2P_{ij}-I$, we have,
\begin{align}
H^b_{ij}(\gamma,3)
=\frac{J_3}{2}P_{ij}+\frac{K_{3,\gamma}}4Q_{ij}^\gamma-\frac{J_3}{4}I.
\label{eq:honeycomb_third_neighbor_swap_form}
\end{align}
The SWAP $P_{ij}$ and the Pauli product $Q_{ij}^\gamma$ commute. Their joint eigenstates give the three possible energies,
\begin{align}
\left\{\frac{J_3+K_{3,\gamma}}4,\;
\frac{J_3-K_{3,\gamma}}4,\;
-\frac{3J_3+K_{3,\gamma}}4\right\}.
\label{eq:honeycomb_third_neighbor_spectrum}
\end{align}
Taking their minimum and maximum gives the optimal shift $\delta_{3,\gamma}=(e_{\max}+e_{\min})/2$ and normalization $\lambda_{3,\gamma}=(e_{\max}-e_{\min})/2$. Using $I=P_{ij}-Q_{ij}^\gamma+P_{ij}Q_{ij}^\gamma$, we can absorb the shift into the three unitaries,
\begin{align}
&H^b_{ij}(\gamma,3)-\delta_{3,\gamma}I
=c_PP_{ij}+c_QQ_{ij}^\gamma+c_{PQ}P_{ij}Q_{ij}^\gamma,\nonumber\\
&c_P=\frac{J_3}{4}-\delta_{3,\gamma},\,
c_Q=\frac{J_3+K_{3,\gamma}}4+\delta_{3,\gamma}, \,c_{PQ}=-\frac{J_3}{4}-\delta_{3,\gamma}.
\label{eq:honeycomb_third_neighbor_lcu}
\end{align}
The shift belongs to $\{-J_3/4,J_3/4,-(J_3+K_{3,\gamma})/4\}$, so at least one coefficient vanishes. The remaining coefficients satisfy $|c_P|+|c_Q|+|c_{PQ}|=\lambda_{3,\gamma}$. Thus, this LCU encodes $(H^b_{ij}(\gamma,3)-\delta_{3,\gamma}I)/\lambda_{3,\gamma}$ at the optimal normalization.

We can implement this simply using a two term LCU using the standard \textsc{Prep}, \textsc{Sel} approach. One qubit prepares the two coefficient magnitudes using one rotation. The coefficient signs are applied as phases in the selection circuit. For the pairs $\{P_{ij},P_{ij}Q_{ij}^\gamma\}$ and $\{Q_{ij}^\gamma,P_{ij}Q_{ij}^\gamma\}$, we factor out the common unitary. For the pair $\{P_{ij},Q_{ij}^\gamma\}$, the two values of the selection bit choose the two operations directly. Given the outer bond-selection flag, each case needs at most one temporary AND to combine it with the inner selection bit and one CSWAP. The remaining controlled Pauli products and sign gates are Clifford. Including the preparation and its inverse, each selected bond costs at most $2T_{\mathrm{rot}}(\epsilon)+8$. The three bond branches therefore contribute,
\begin{align}
T_{\textsc{BE}}^{3}
&\leq 6T_{\mathrm{rot}}(\epsilon)+24.
\label{eq:honeycomb_third_neighbor_t_cost}
\end{align}

\subsubsection{Combining stars and third-neighbor bonds}

We now combine the two stars and three third-neighbor bonds into a local encoding of the routed eight-spin patch. Let $j\in\{\mathrm{anti},\mathrm{spec}\}$ denote the star construction. Its shift and normalization are $(\delta^\star_{\mathrm{anti}},\lambda^\star_{\mathrm{anti}})=(0,\beta^\star)$ for the anticommuting construction and $(\delta^\star_{\mathrm{spec}},\lambda^\star_{\mathrm{spec}})=(\delta^\star_0,\lambda^\star)$ for the symmetry-reduced construction. These already include the field assigned to each star. The patch shift and normalization are,
\begin{align}
\delta_{p,j}=2\delta^\star_j+\sum_{\gamma=x,y,z}\delta_{3,\gamma},\nonumber\qquad \lambda_{p,j}=2\lambda^\star_j+\sum_{\gamma=x,y,z}\lambda_{3,\gamma}.
\end{align}

The local LCU therefore encodes $(H_p^\star-\delta_{p,j}I)/\lambda_{p,j}$. All $N_l=N_s/2$ translated patches use the same local encoding. Thus, the global shift and normalization are,
\begin{align}
\Delta_j=\frac{N_s}{2}\delta_{p,j}+\frac{AN_s}{4},\qquad
\Lambda_j=\frac{N_s}{2}\lambda_{p,j}.
\end{align}
The resulting global encoding represents $(H-\Delta_jI)/\Lambda_j$. The local preparation and circuit construction are given in~\Cref{app:subsec:honeycomb_combined}. Combining these costs gives,
\begin{align}
T_{\textsc{BE},j}^p
&=2T_{\textsc{Prep},j}^p+T_{\textsc{Sel},j}^p\leq\begin{cases}
100T_{\mathrm{rot}}(\epsilon)+108,&j=\mathrm{anti},\\
172T_{\mathrm{rot}}(\epsilon)+404,&j=\mathrm{spec}.
\end{cases}
\label{eq:honeycomb_two_star_local_t_cost}
\end{align}
The overall block-encoding costs are,
\begin{align}
T_{\textsc{BE},\mathrm{anti}}
&\leq8N_s+80\left\lceil\frac{\log_2N_s-1}{2}\right\rceil
+100T_{\mathrm{rot}}(\epsilon)+92,
\label{eq:honeycomb_two_star_anti_total_t_cost}\\
T_{\textsc{BE},\mathrm{spec}}
&\leq8N_s+80\left\lceil\frac{\log_2N_s-1}{2}\right\rceil
+172T_{\mathrm{rot}}(\epsilon)+388.
\label{eq:honeycomb_two_star_spec_total_t_cost}
\end{align}

\section{Numerical Comparison and Discussion}
\label{sec:overall-cost}

We compare the encoding approaches by accounting for both their query complexity and their cost per query. A tighter normalization can reduce the number of queries while requiring a more expensive local circuit. We evaluate this tradeoff across coupling parameters and lattice sizes for two applications, ground state energy estimation and time evolution.

\subsection{Low-energy state preparation}

Consider block encoding $U_A$ such that it encodes the normalized Hamiltonian $A=(H-\Delta I)/\Lambda$ with a normalized ground state energy $x$. For both QPE and QSP-based approaches, the problem of resolving $x$ becomes resolving $\theta$ with some precision for $x=\cos\theta$. Therefore, to understand the overall cost of low-energy-state preparation, we can evaluate a universal constant that is indicative of the cost for all methods. Namely, the universal cost constant is the inverse of the required resolution on $\theta$, which we call $\Lambda_{\mathrm{eff}}$
\begin{align*}
\Lambda_{\mathrm{eff}}\equiv\left|\frac{dE}{d\theta}\right|=\Lambda|\sin\theta|=\sqrt{\Lambda^2-(E-\Delta)^2}.
\end{align*}
The number of block-encoding queries should be proportional to $\Lambda_{\mathrm{eff}}$. The exact query count of the algorithm further depends on the initial state and the energy resolution target, and the constant factors depend on the algorithm. Therefore, to calculate the total $T$ cost, we need the ground state energy $E$ to plug into $\Lambda_{\mathrm{eff}}$.

The total number of queries to the block encoding is approximately
\begin{align*}
N_{\mathrm{QPE}}\approx\frac{\pi\Lambda_{\mathrm{eff}}}{2\Delta E}.
\end{align*}
The leading QPE cost in the number of $T$ gates is therefore approximately $N_{\mathrm{QPE}}$ multiplied by the cost per block encoding:
\begin{align*}
T_{\mathrm{QPE}}\approx\frac{\pi\Lambda_{\mathrm{eff}}}{2\Delta E}\cdot T_{\textsc{BE}}.
\end{align*}

However, for QET-based block encodings such as the $3\times3$ Heisenberg block encoding and the $2\times 2$ Fermi--Hubbard block encoding, the number of QPE iterations also depends on the optimization result. For a fixed $s$, the QET error can be chosen to be $0\leq r_{\mathrm{QET}}< 1$ of the total error budget, resulting in the choice
\begin{align*}
\epsilon_{\mathrm{QET}}=\frac{r_{\mathrm{QET}}\Delta E}{K}\cdot\frac{1-s}{\lambda_p},
\end{align*}
where $K=M^2$, with $\lambda_p=\lambda_3$ for Heisenberg and $\lambda_p=\lambda_\square$ for Fermi--Hubbard. The remaining $1-r_{\mathrm{QET}}$ of the error budget is still allocated to QPE, so
\begin{align}
N_{\mathrm{QPE}}&\approx\frac{\pi\Lambda_{\mathrm{eff}}}{2\times (1-r_{\mathrm{QET}})\Delta E}\nonumber \\
T_{\mathrm{QPE}}&\approx\frac{\pi\Lambda_{\mathrm{eff}}}{2(1-r_{\mathrm{QET}})\Delta E}\cdot T_{\textsc{BE}}.
\label{eq:overall_qpe_t_cost}
\end{align}

We assume a rotation synthesis error of $\epsilon=10^{-10}$, which results in a negligible fidelity impact for the applications below. The cost per rotation is estimated using
\begin{align*}
T_{\mathrm{rot}}=0.53\log_2(1/\epsilon)+4.86.
\end{align*}
This corresponds to the expected number of $T$ gates using the repeat-until-success approach of Ref.~\cite{kliuchnikov2023shorter}, which results in $\epsilon$ error in the diamond norm.

\subsubsection{The Heisenberg model}

We estimate the cost of resolving the ground-state energy of the periodic
$M\times M$ square lattice, with $N=M^2$, to the extensive accuracy
\begin{align*}
 \Delta E=0.01J_1N.
\end{align*}

The resource formula depends on the location of the target eigenvalue in the
encoded interval.  We therefore require representative ground-state energy
densities.  Most of the numerical literature writes the Hamiltonian using
spin operators $\mathbf S_i=\boldsymbol\sigma_i/2$.  At the same numerical
values of $J_1$ and $J_2$, the Hamiltonian in this manuscript is four times
that conventional spin Hamiltonian. We use the thermodynamic-limit
high-order coupled-cluster value at $r\equiv J_2/J_1=0.4$
\cite{Darradi2008ground,Aktersky2016singlet}, the two-dimensional-limit
$SU(2)$ DMRG estimate at $r=0.5$ \cite{Gong2014plaquette}, and the
thermodynamic-limit tensor-product-state estimate at $r=0.6$
\cite{Wang2016tensor}.  Table~\ref{tab:heis_energy_inputs} gives the extensive energies in our convention used below.

\begin{table}[ht]
\centering
\begin{tabular}{c|c|rrrr}
\hline\hline
$r$ & $e_0$ & $M=30$ & $M=60$ & $M=90$ & $M=120$ \\
\hline
$0.4$ & $-2.0942$ & $-1885$ & $-7539$ & $-16963$ & $-30156$ \\
$0.5$ & $-1.9872$ & $-1788$ & $-7154$ & $-16096$ & $-28616$ \\
$0.6$ & $-1.9092$ & $-1718$ & $-6873$ & $-15465$ & $-27492$ \\
\hline\hline
\end{tabular}
\caption{Ground-state energy proxies for the Heisenberg model. Here $e_0=E_0/N$. The entries for finite $M$ are $E_0=N e_0$.}
\label{tab:heis_energy_inputs}
\end{table}

These thermodynamic energy densities are used only as operating points for
estimating $\Lambda_{\mathrm{eff}}$ and are not prior-energy inputs to QPE. For the present lattice sizes, using a common thermodynamic density is adequate at the quoted $0.01J_1$ energy-density resolution.  It also avoids mixing boundary conditions and finite-size conventions from different
numerical studies.

The bond-wise and triangular block-encoding cost bounds are given by~\Cref{tab:heis_direct_be_costs}. For the $3\times 3$ plaquette local block encoding, however, the need for optimization leaves the cost uncertain depending on the problem input.
\begin{table}[t]
\centering
\small
\setlength{\tabcolsep}{3pt}
\begin{tabular}{c|rrrr}
\hline\hline
$M$ & $30$ & $60$ & $90$ & $120$\\
\hline
$\mathbb E[T_{\textsc{BE},b}]$
& $7.66\times10^{3}$
& $2.93\times10^{4}$
& $6.54\times10^{4}$
& $1.16\times10^{5}$\\
$\mathbb E[T_{\textsc{BE},\triangle}]$
& $7.73\times10^{3}$
& $2.94\times10^{4}$
& $6.55\times10^{4}$
& $1.16\times10^{5}$\\
\hline\hline
\end{tabular}
\caption{Expected $T$-gate counts per block-encoding query for the bond-wise ($b$) and triangular ($\triangle$) Heisenberg encodings on an $M\times M$ square lattice.}
\label{tab:heis_direct_be_costs}
\end{table}

For an encoding with extensive shift $N\delta$ and normalization $N\lambda$,
the energy proxies in Table~\ref{tab:heis_energy_inputs} give
\begin{align}
 \frac{\Lambda_{\mathrm{eff}}}{N}
 &=\sqrt{\lambda^2-(e_0-\delta)^2},\nonumber\\
 N_{\mathrm{QPE}}
 &\simeq \frac{\pi}{2(0.01)}\frac{\Lambda_{\mathrm{eff}}}{J_1N}
 =\frac{157.079633}{J_1}\times\sqrt{\lambda^2-(e_0-\delta)^2},
 \label{eq:heis_extensive_queries}
\end{align}
where $e_0=E_0/N$. The cancellation of $N$ in Eq.~\ref{eq:heis_extensive_queries} is a direct consequence of the extensive accuracy. The resulting direct query counts for the bond-wise and triangular block encodings are given by~\Cref{tab:heis_direct_queries}.
\begin{table}[t]
\centering
\small
\setlength{\tabcolsep}{3pt}
\begin{tabular}{c|rr|rr}
\hline\hline
& \multicolumn{2}{c|}{Bond-wise}
& \multicolumn{2}{c}{Triangular} \\
$r$
& $\Lambda_{\mathrm{eff}}/N$ & $N_{\mathrm{QPE}}$
& $\Lambda_{\mathrm{eff}}/N$ & $N_{\mathrm{QPE}}$ \\
\hline
$0.4$ & $5.5553$ & $873$ & $2.3264$ & $365$ \\
$0.5$ & $5.9139$ & $929$ & $2.2475$ & $353$ \\
$0.6$ & $6.2685$ & $985$ & $2.9392$ & $462$ \\
\hline\hline
\end{tabular}
\caption{Effective normalizations and estimated QPE query counts for the bond-wise and triangular Heisenberg encodings. Here $r=J_2/J_1$ and $N=M^2$.}
\label{tab:heis_direct_queries}
\end{table}

For the $3\times3$ block encoding, we shall denote the spectral half-width and midpoint of $H^{3\times 3}_p$ as $\lambda_3$ and $\delta_3$. For $J_1=1$ and $J_2/J_1=0.4,0.5,0.6$, exact diagonalization of the nine-spin
operator in Eq.~\ref{eq:hierarchical_patch_definition} gives the values in~\Cref{tab:hierarchical_patch_spectrum}.
\begin{table}[t]
\centering
\begin{tabular}{c|cc}
\hline\hline
$J_2/J_1$ & $\delta_3/J_1$ & $\lambda_3/J_1$ \\
\hline
$0.4$ & $0.176685$ & $2.623315$ \\
$0.5$ & $0.333576$ & $2.666424$ \\
$0.6$ & $0.470423$ & $2.729577$ \\
\hline\hline
\end{tabular}
\caption{Optimal local shifts $\delta_3$ and normalizations
$\lambda_3$ for the hierarchical $3\times3$ Heisenberg patch
defined in Eq.~\eqref{eq:hierarchical_patch_definition},
obtained by exact diagonalization. Both quantities are
reported in units of $J_1$.}
\label{tab:hierarchical_patch_spectrum}
\end{table}
Optimization of the QET headroom $s$ is needed to saturate $\lambda_3$ as closely as possible without incurring too high of a cost as discussed before. In particular, QET is not always advantageous, and when no QET is applied, we have $d=0$, $r_{\mathrm{QET}}=0$, and $s=1-\lambda_3/\beta_3$.

Table~\ref{tab:heis_qpe_t_counts} reports the expected total $T$ for all three hierarchy levels, as well as a baseline developed in Ref.~\cite{yoshioka2024hunting}. The advantage of the $3\times3$ block encoding is clear for large system sizes. At $M=120$, its expected $T$ count at the listed candidates is lower than that of the triangle encoding by approximately $31.9\%$, $29.8\%$, and $43.1\%$ for $r=0.4,0.5,0.6$, respectively. The $3\times 3$ encoding improves over the baseline by factors of $15.32$, $16.58$, and $16.78$ for $r=0.4,0.5,0.6$, respectively.
\begin{table*}[t]
\centering
\begingroup
\small
\setlength{\tabcolsep}{11pt}
\resizebox{\textwidth}{!}{%
\begin{tabular}{cc|r|r|r|rrrrrr}
\hline\hline
$r$ & $M$
& $T_{{\mathrm{QPE}},\mathrm P}$
& $T_{{\mathrm{QPE}},b}$
& $T_{{\mathrm{QPE}},\triangle}$
& $T_{{\mathrm{QPE}},3\times3}$
& $d$ & $r_{\mathrm{QET}}$ & $s$
& $N_{{\mathrm{QPE}},3\times3}$
& $T_{\textsc{BE},3\times3}$\\
\hline
0.4 & 30 & $2.76\!\times\!10^{7}$ & $6.69\!\times\!10^{6}$ & $\mathbf{2.82\!\times\!10^{6}}$ & $3.14\!\times\!10^{6}$ & 0 & 0.000000 & 0.223109 & 393 & $8.00\!\times\!10^{3}$\\
0.4 & 60 & $1.10\!\times\!10^{8}$ & $2.56\!\times\!10^{7}$ & $1.07\!\times\!10^{7}$ & $\mathbf{8.86\!\times\!10^{6}}$ & 19 & 0.062322 & 0.032624 & 248 & $3.57\!\times\!10^{4}$\\
0.4 & 90 & $2.48\!\times\!10^{8}$ & $5.71\!\times\!10^{7}$ & $2.39\!\times\!10^{7}$ & $\mathbf{1.74\!\times\!10^{7}}$ & 31 & 0.037423 & 0.018126 & 230 & $7.57\!\times\!10^{4}$\\
0.4 & 120 & $4.42\!\times\!10^{8}$ & $1.01\!\times\!10^{8}$ & $4.23\!\times\!10^{7}$ & $\mathbf{2.88\!\times\!10^{7}}$ & 37 & 0.030026 & 0.014711 & 225 & $1.28\!\times\!10^{5}$\\
\hline
0.5 & 30 & $2.98\!\times\!10^{7}$ & $7.12\!\times\!10^{6}$ & $\mathbf{2.73\!\times\!10^{6}}$ & $3.01\!\times\!10^{6}$ & 0 & 0.000000 & 0.200131 & 376 & $8.00\!\times\!10^{3}$\\
0.5 & 60 & $1.19\!\times\!10^{8}$ & $2.73\!\times\!10^{7}$ & $1.04\!\times\!10^{7}$ & $\mathbf{8.74\!\times\!10^{6}}$ & 21 & 0.053548 & 0.025670 & 241 & $3.63\!\times\!10^{4}$\\
0.5 & 90 & $2.68\!\times\!10^{8}$ & $6.08\!\times\!10^{7}$ & $2.31\!\times\!10^{7}$ & $\mathbf{1.73\!\times\!10^{7}}$ & 27 & 0.041862 & 0.019515 & 232 & $7.44\!\times\!10^{4}$\\
0.5 & 120 & $4.77\!\times\!10^{8}$ & $1.08\!\times\!10^{8}$ & $4.09\!\times\!10^{7}$ & $\mathbf{2.87\!\times\!10^{7}}$ & 39 & 0.028178 & 0.012718 & 223 & $1.29\!\times\!10^{5}$\\
\hline
0.6 & 30 & $3.19\!\times\!10^{7}$ & $7.54\!\times\!10^{6}$ & $\mathbf{3.57\!\times\!10^{6}}$ & $3.64\!\times\!10^{6}$ & 15 & 0.094363 & 0.058510 & 287 & $1.27\!\times\!10^{4}$\\
0.6 & 60 & $1.28\!\times\!10^{8}$ & $2.89\!\times\!10^{7}$ & $1.36\!\times\!10^{7}$ & $\mathbf{9.53\!\times\!10^{6}}$ & 25 & 0.039639 & 0.038582 & 253 & $3.76\!\times\!10^{4}$\\
0.6 & 90 & $2.87\!\times\!10^{8}$ & $6.44\!\times\!10^{7}$ & $3.02\!\times\!10^{7}$ & $\mathbf{1.85\!\times\!10^{7}}$ & 33 & 0.047080 & 0.023838 & 242 & $7.63\!\times\!10^{4}$\\
0.6 & 120 & $5.11\!\times\!10^{8}$ & $1.14\!\times\!10^{8}$ & $5.35\!\times\!10^{7}$ & $\mathbf{3.04\!\times\!10^{7}}$ & 47 & 0.024743 & 0.018415 & 232 & $1.31\!\times\!10^{5}$\\
\hline\hline
\end{tabular}%
}
\endgroup
\caption{QPE resource estimates at $\Delta E=0.01J_1N$ for the $M\times M$ square lattice Heisenberg model. $r=J_2/J_1$, and $T_{\mathrm{QPE}}$ is the total $T$ count of the ground state preparation algorithm using qubitized phase estimation. The Pauli LCU baseline $\mathrm P$ uses the product-wise construction of Ref.~\cite{yoshioka2024hunting}, retaining only the leading cost $T_{\textsc{BE},\mathrm P}=24N_s$. Here $N_s=N=M^2$, $\Lambda_{\mathrm P}=6N_s(|J_1|+|J_2|)$ and $\Delta_{\mathrm P}=0$. Total costs are estimated as $T_{\mathrm{QPE}}\approx N_{\mathrm{QPE}}T_{\textsc{BE}}$. The $d=0$ rows bypass QET. Bold entries indicate the lowest total cost in each row.}
\label{tab:heis_qpe_t_counts}
\end{table*}

\subsubsection{The Fermi--Hubbard model}
\label{sec:fh-overall-cost}

We next estimate the cost of resolving the half-filled nearest-neighbor
Fermi--Hubbard ground energy on the periodic
$M\times M$ square lattice. Numerical Hubbard results are usually quoted for
$H_{\mathrm{std}}=-tK+U\sum_r n_{r\uparrow}n_{r\downarrow}$.  At half filling,
the Hamiltonian in Eq.~\eqref{eq:nn-fh} differs by the known constant
$-UM^2/4$.  We therefore convert a standard ground-energy density according
to
\begin{align}
e_0\equiv\frac{E_0}{M^2}
=e_{0,\mathrm{std}}-\frac U4.
\label{eq:fh-energy-convention-conversion}
\end{align}
We use the thermodynamic-limit AFQMC values in Table II of Ref.~\cite{leblanc2015solutions}, and the converted values are provided in Table~\ref{tab:fh_energy_inputs}.
\begin{table}[h]
\centering
\begin{tabular}{c|c|rrrr}
\hline\hline
$U/t$ & $e_0/t$ & $M=30$ & $M=60$ & $M=90$ & $M=120$\\
\hline
$2$ & $-1.6763$ & $-1509$ & $-6035$ & $-13578$ & $-24139$\\
$4$ & $-1.8603$ & $-1674$ & $-6697$ & $-15068$ & $-26788$\\
$6$ & $-2.1568$ & $-1941$ & $-7764$ & $-17470$ & $-31058$\\
$8$ & $-2.5247$ & $-2272$ & $-9089$ & $-20450$ & $-36356$\\
\hline\hline
\end{tabular}
\caption{Ground-state energy proxies for the half-filled square-lattice
Fermi--Hubbard model in the particle--hole-symmetric convention.
Here $e_0=E_0/M^2$. The finite-$M$ entries are
$E_0/t=M^2 e_0/t$.}
\label{tab:fh_energy_inputs}
\end{table}

The rest of the analysis follows similar considerations in the Heisenberg case. The Fermi--Hubbard bond-wise block-encoding cost bounds are given by~\Cref{tab:fh_direct_be_costs}.
\begin{table}[t]
\centering
\small
\setlength{\tabcolsep}{3pt}
\begin{tabular}{c|rrrr}
\hline\hline
$M$ & $30$ & $60$ & $90$ & $120$ \\
\hline
$\mathbb E[T_{\textsc{BE},b}]$
& $1.53\times10^{4}$
& $5.86\times10^{4}$
& $1.31\times10^{5}$
& $2.31\times10^{5}$ \\
\hline\hline
\end{tabular}
\caption{Estimates of the expected $T$-gate count per block-encoding query for the bond-wise Fermi--Hubbard encoding on an $M\times M$ square lattice, with $N_s=2M^2$ system qubits.}
\label{tab:fh_direct_be_costs}
\end{table}
The resulting Fermi--Hubbard bond-wise effective normalizations and
query counts are given by~\Cref{tab:fh_direct_queries}.
\begin{table}[t]
\centering
\begin{tabular}{c|rr}
\hline\hline
$U/t$
& $\Lambda_{\mathrm{eff},b}/\left(M^2t\right)$
& $N_{\mathrm{QPE},b}$ \\
\hline
$2$ & $3.6661$ & $576$ \\
$4$ & $3.6796$ & $578$ \\
$6$ & $3.6876$ & $579$ \\
$8$ & $3.6913$ & $580$ \\
\hline\hline
\end{tabular}
\caption{Effective normalizations per lattice site (in units of $t$) and estimated QPE query counts for the bond-wise Fermi--Hubbard encoding.}
\label{tab:fh_direct_queries}
\end{table}

Table~\ref{tab:fh-qpe-t-counts} gives the costs for three representative interaction strengths for our methods and a baseline from Refs.~\cite{babbush2018encoding,yoshioka2024hunting}. QET becomes worthwhile when its additional constant-size local work costs less than the query saving produced by the tighter normalization. Additionally, at $M=120$, the plaquette encoding improves over the baseline by factors of $3.09$, $3.10$, and $3.28$ for $U/t=2,4,8$, respectively.
\begin{table*}[t]
\centering
\begingroup
\small
\setlength{\tabcolsep}{11pt}
\resizebox{\textwidth}{!}{%
\begin{tabular}{cc|r|r|rrrrrr}
\hline\hline
$U/t$ & $M$
& $T_{{\mathrm{QPE}},\mathrm P}$
& $T_{{\mathrm{QPE}},b}$
& $T_{{\mathrm{QPE}},\square}$
& $d$ & $r_{\mathrm{QET}}$ & $s$
& $N_{{\mathrm{QPE}},\square}$
& $T_{\textsc{BE},\square}$\\
\hline
2 & 30 & $1.18\!\times\!10^{7}$ & $8.82\!\times\!10^{6}$ & $\mathbf{6.83\!\times\!10^{6}}$ & 13 & 0.029825 & 0.143213 & 308 & $2.22\!\times\!10^{4}$\\
2 & 60 & $4.72\!\times\!10^{7}$ & $3.37\!\times\!10^{7}$ & $\mathbf{1.87\!\times\!10^{7}}$ & 27 & 0.040302 & 0.050639 & 254 & $7.35\!\times\!10^{4}$\\
2 & 90 & $1.06\!\times\!10^{8}$ & $7.53\!\times\!10^{7}$ & $\mathbf{3.69\!\times\!10^{7}}$ & 37 & 0.037766 & 0.033512 & 244 & $1.51\!\times\!10^{5}$\\
2 & 120 & $1.89\!\times\!10^{8}$ & $1.33\!\times\!10^{8}$ & $\mathbf{6.12\!\times\!10^{7}}$ & 41 & 0.027711 & 0.031879 & 241 & $2.54\!\times\!10^{5}$\\
\hline
4 & 30 & $1.31\!\times\!10^{7}$ & $8.86\!\times\!10^{6}$ & $\mathbf{7.27\!\times\!10^{6}}$ & 13 & 0.078486 & 0.096400 & 327 & $2.22\!\times\!10^{4}$\\
4 & 60 & $5.25\!\times\!10^{7}$ & $3.39\!\times\!10^{7}$ & $\mathbf{2.05\!\times\!10^{7}}$ & 25 & 0.049509 & 0.046376 & 284 & $7.23\!\times\!10^{4}$\\
4 & 90 & $1.18\!\times\!10^{8}$ & $7.55\!\times\!10^{7}$ & $\mathbf{4.06\!\times\!10^{7}}$ & 33 & 0.029952 & 0.037235 & 272 & $1.49\!\times\!10^{5}$\\
4 & 120 & $2.10\!\times\!10^{8}$ & $1.34\!\times\!10^{8}$ & $\mathbf{6.77\!\times\!10^{7}}$ & 41 & 0.020764 & 0.031803 & 266 & $2.54\!\times\!10^{5}$\\
\hline
8 & 30 & $1.54\!\times\!10^{7}$ & $8.89\!\times\!10^{6}$ & $\mathbf{8.06\!\times\!10^{6}}$ & 11 & 0.056260 & 0.106031 & 382 & $2.11\!\times\!10^{4}$\\
8 & 60 & $6.16\!\times\!10^{7}$ & $3.40\!\times\!10^{7}$ & $\mathbf{2.27\!\times\!10^{7}}$ & 21 & 0.046757 & 0.046720 & 324 & $7.01\!\times\!10^{4}$\\
8 & 90 & $1.39\!\times\!10^{8}$ & $7.58\!\times\!10^{7}$ & $\mathbf{4.51\!\times\!10^{7}}$ & 35 & 0.024454 & 0.028091 & 300 & $1.50\!\times\!10^{5}$\\
8 & 120 & $2.46\!\times\!10^{8}$ & $1.34\!\times\!10^{8}$ & $\mathbf{7.52\!\times\!10^{7}}$ & 39 & 0.029164 & 0.022287 & 297 & $2.53\!\times\!10^{5}$\\
\hline\hline
\end{tabular}%
}
\endgroup
\caption{Expected QPE $T$ counts at $\Delta E=0.01tM^2$ for the $M\times M$ square lattice Fermi--Hubbard model. $T_{\mathrm{QPE}}$ is the total $T$ count of the ground state preparation algorithm using qubitized phase estimation. The Pauli LCU baseline $\mathrm P$ follows Refs.~\cite{babbush2018encoding,yoshioka2024hunting}, retaining only the leading cost $T_{\textsc{BE},\mathrm P}=10N_s$. Here $N_s=2M^2$, $\Lambda_{\mathrm P}=M^2(4|t|+|U|/4)$ and $\Delta_{\mathrm P}=0$. Total costs are estimated as $T_{\mathrm{QPE}}\approx N_{\mathrm{QPE}}T_{\textsc{BE}}$. Bold entries indicate the lowest total cost in each row.}
\label{tab:fh-qpe-t-counts}
\end{table*}

\subsection{Time-evolution of the Spin-liquid Hamiltonian}
\label{subsec:numerical_honeycomb}

The LANL application study identifies the simulation of magnetization under applied magnetic fields as the computational kernel for comparing these candidates with MAGLAB measurements~\cite{bartschi2026potential}. The field-dependent response constrains the exchange interactions and magnetic anisotropy. We therefore use time evolution as the application for comparing our honeycomb encodings.

We take the color-independent exchange couplings from the candidate labeled ``Neglecting non-local SOC'' in Fig.~2-7 of the LANL study~\cite{bartschi2026potential}. The tabulated values are bond averaged. The common couplings are given in~\Cref{tab:honeycomb_reference_couplings}.
\begin{table}[ht]
\centering
\begin{tabular}{rrrrrrrr}
\hline\hline
 $J_1$ & $J_2$ & $J_3$ & $K_1$ & $K_2$ & $K_3$ & $\Gamma_1$ & $\Gamma'_1$\\
\hline
 $-1.3$  & $0.25$ & $1.0$ & $-13.3$ & $-0.67$ & $0.1$ & $9.4$ & $-2.3$\\
\hline\hline
\end{tabular}
\caption{Reference exchange couplings for the honeycomb time-evolution comparison in meV. The Kitaev and off-diagonal couplings are independent of bond color. }
\label{tab:honeycomb_reference_couplings}
\end{table}
We consider applied fields along the crystallographic $b$ axis, $\mathbf H=(0,H_b,0)$, with $H_b=10\,\mathrm T$. We use $g_b=2.3$ and convert the field to the spin frame using~\Cref{eq:honeycomb_zeeman}. We use the same rotation-synthesis precision for all encodings. For an illustrative example, we take the rotation synthesis error to be $\epsilon=10^{-10}$. Additionally, the expected cost of one synthesized axial rotation is estimated as
\begin{align*}
T_{\mathrm{rot}}=0.53\log_2(1/\epsilon)+4.86,
\end{align*}
using the repeat-until-success construction of Ref.~\cite{kliuchnikov2023shorter}. Rotation-synthesis errors are assumed negligible in the comparisons below.

For encoding $j$, write $A_j=(H-\Delta_jI)/\Lambda_j$. Setting $\hbar=1$, time evolution for a fixed Hamiltonian is $e^{-iHt}=e^{-i\Delta_jt}e^{-i\Lambda_jtA_j}.$ The shift contributes a known global phase, while the normalization sets the effective evolution time $\Lambda_j|t|$. For target simulation error $\varepsilon_{\mathrm{sim}}$, qubitization with generalized quantum signal processing has the query bound~\cite{low2019qubitization,berry2024doubling},
\begin{align}
Q_{\mathrm{sim},j}
=\mathcal O\!\left(\Lambda_j|t|+\log\frac{1}{\varepsilon_{\mathrm{sim}}}\right).
\label{eq:time_evolution_query_scaling}
\end{align}
We use the same simulation construction and error target for all encodings. For fixed $\varepsilon_{\mathrm{sim}}$ and large $\Lambda_j|t|$, we retain only the leading term $Q_{\mathrm{sim},j}\approx\Lambda_j|t|$, using the improved polynomial-degree bounds of Ref.~\cite{pocrnic2026improved}. Multiplying this query count by the block-encoding cost gives the leading encoding contribution,
\begin{align}
T_{\mathrm{sim},j}
&\approx Q_{\mathrm{sim},j}T_{\textsc{BE},j},\nonumber\\
T_{\mathrm{sim},j}
&\approx |t|\Lambda_jT_{\textsc{BE},j}.
\label{eq:time_evolution_t_cost_comparison}
\end{align}
Thus, the leading comparison is determined by $\Lambda_jT_{\textsc{BE},j}$ when the evolution-time contribution dominates. We assume that this is the case in our comparison. For short evolution times, the precision-dependent contribution must also be retained. 

We compare the bond-wise, anticommuting-star, and symmetry-reduced spectral-star constructions from~\Cref{sec:spin_liquids}. Each has $\Lambda_j=N_l\lambda_{p,j}$, with $N_l=M^2$, so at a fixed lattice size we compare the products $\lambda_{p,j}T_{\textsc{BE},j}$. The shared routing contribution grows with the system size, whereas the extra cost of the local circuit is independent of it. Consequently, a construction with a smaller normalization can become favorable at larger lattice sizes even if its local circuit is more expensive. We also include a baseline using the product-wise Pauli LCU approach from Refs.~\cite{babbush2018encoding,yoshioka2024hunting} constructed in ~\Cref{app:honeycomb_pauli_baseline}, using only the leading cost $T_{\textsc{BE},\mathrm P}=16N_s$. Its normalization is $\Lambda_{\mathrm P}=M^2\lambda_{p,\mathrm P}$, with $\lambda_{p,\mathrm P}=39.4878\,\mathrm{meV}$. 

\Cref{tab:honeycomb_time_evolution_j2_nonzero} reports the global block-encoding costs and normalizations for $M\times M$ unit cells, with $N_s=2M^2$ system qubits. We also report the leading cost score $\mathcal C_j=t\Lambda_jT_{\textsc{BE},j}$,  and highlight its smallest value in each row. The comparison shows substantial savings from combining patch routing with tighter local encodings. Relative to the Pauli LCU baseline, the spectral-star construction reduces the estimated leading time-evolution $T$ cost by approximately $3\times$ at $M=20$, $4.4\times$ at $M=40$, and $5\times$ at $M=80$ and $120$. The crossover also demonstrates the value of having different local encodings available. The baseline has the lowest estimated cost at $M=4$, the anticommuting-star construction at $M=10$, and the spectral-star construction at all reported sizes $M\geq20$. Since local circuit cost is independent of lattice size, larger lattices benefit increasingly from the tighter spectral normalization as we would expect. For these parameters, the bond-wise construction has a larger normalization and a higher cost per query than both star constructions, so it never has the lowest cost. It still improves on the baseline for $M\geq20$.

\begin{table*}[ht]
\centering
\begingroup
\small
\setlength{\tabcolsep}{2pt}
\resizebox{\textwidth}{!}{%
\begin{tabular}{r|rrr||rrr||rrr||rrr}
\hline\hline
 & \multicolumn{3}{c||}
 {Pauli LCU  (\Cref{app:honeycomb_pauli_baseline})}
 & \multicolumn{3}{c||}{Bond-wise (\Cref{subsec:honeycomb_bondwise})}
 & \multicolumn{3}{c||}{Anticommuting star (\Cref{subsec:honeycomb_star})}
 & \multicolumn{3}{c}{Spectral star (\Cref{subsec:honeycomb_star})}\\
$M$
 & $T_{\textsc{BE}}$ & $\Lambda$ & $\mathcal C$
 & $T_{\textsc{BE}}$ & $\Lambda$ & $\mathcal C$
 & $T_{\textsc{BE}}$ & $\Lambda$ & $\mathcal C$
 & $T_{\textsc{BE}}$ & $\Lambda$ & $\mathcal C$\\
\hline
4
 & $5.12\times10^{2}$ & $6.32\times10^{2}$ & $\mathbf{2.43\times10^{7}}$
 & $6.40\times10^{3}$ & $3.51\times10^{2}$ & $1.69\times10^{8}$
 & $2.76\times10^{3}$ & $2.98\times10^{2}$ & $6.18\times10^{7}$
 & $4.67\times10^{3}$ & $2.40\times10^{2}$ & $8.42\times10^{7}$\\
10
 & $3.20\times10^{3}$ & $3.95\times10^{3}$ & $9.50\times10^{8}$
 & $7.90\times10^{3}$ & $2.19\times10^{3}$ & $1.30\times10^{9}$
 & $4.26\times10^{3}$ & $1.86\times10^{3}$ & $\mathbf{5.97\times10^{8}}$
 & $6.17\times10^{3}$ & $1.50\times10^{3}$ & $6.96\times10^{8}$\\
20
 & $1.28\times10^{4}$ & $1.58\times10^{4}$ & $1.52\times10^{10}$
 & $1.28\times10^{4}$ & $8.77\times10^{3}$ & $8.42\times10^{9}$
 & $9.14\times10^{3}$ & $7.45\times10^{3}$ & $5.12\times10^{9}$
 & $1.11\times10^{4}$ & $6.00\times10^{3}$ & $\mathbf{4.99\times10^{9}}$\\
40
 & $5.12\times10^{4}$ & $6.32\times10^{4}$ & $2.43\times10^{11}$
 & $3.21\times10^{4}$ & $3.51\times10^{4}$ & $8.45\times10^{10}$
 & $2.84\times10^{4}$ & $2.98\times10^{4}$ & $6.37\times10^{10}$
 & $3.03\times10^{4}$ & $2.40\times10^{4}$ & $\mathbf{5.47\times10^{10}}$\\
80
 & $2.05\times10^{5}$ & $2.53\times10^{5}$ & $3.89\times10^{12}$
 & $1.09\times10^{5}$ & $1.40\times10^{5}$ & $1.15\times10^{12}$
 & $1.05\times10^{5}$ & $1.19\times10^{5}$ & $9.44\times10^{11}$
 & $1.07\times10^{5}$ & $9.60\times10^{4}$ & $\mathbf{7.74\times10^{11}}$\\
120
 & $4.61\times10^{5}$ & $5.69\times10^{5}$ & $1.97\times10^{13}$
 & $2.37\times10^{5}$ & $3.16\times10^{5}$ & $5.62\times10^{12}$
 & $2.33\times10^{5}$ & $2.68\times10^{5}$ & $4.71\times10^{12}$
 & $2.35\times10^{5}$ & $2.16\times10^{5}$ & $\mathbf{3.82\times10^{12}}$\\
\hline\hline
\end{tabular}%
}
\endgroup
\caption{Time-evolution resource estimates for an $M\times M$ honeycomb lattice of unit cells, with $N_s=2M^2$ spins for the reference couplings reported in~\Cref{tab:honeycomb_reference_couplings}. $T_{\textsc{BE}}$ is the $T$-gate cost per global block-encoding call, $\Lambda$ is the global normalization in meV, and $\mathcal C=t\Lambda T_{\textsc{BE}}$ estimates the leading total time-evolution $T$ cost at $|K_1|t=1000$. The Pauli LCU baseline uses product-wise construction~\cite{babbush2018encoding,yoshioka2024hunting}, retaining only the leading cost $T_{\textsc{BE},\mathrm P}=16N_s$. The spectral-star construction gives the lowest estimated cost at all reported sizes $M\geq20$, reaching an approximately $5\times$ reduction relative to the baseline at $M=120$. Bold entries indicate the lowest $\mathcal C$ in each row.}
\label{tab:honeycomb_time_evolution_j2_nonzero}
\end{table*}

\section{Conclusions and Outlook}

Efficient quantum simulation of many-body systems is central to the prospect of using quantum computers to understand strongly correlated matter beyond the reliable reach of classical computation. We develop a routing-based framework for constructing a hierarchy of block encodings for frustrated lattice Hamiltonians by grouping interactions into local patches. Small interacting patches capture part of the frustration and provide tighter spectral bounds, but realizing these bounds can require expensive local circuits. Our approach makes these circuits practical on large lattices by exploiting coherent routing and translation invariance or repeated structures in the Hamiltonian. Coherent routing brings a selected interaction patch to a fixed location, allowing its local encoding to be applied once per query, with a leading routing cost of $8N_s$ $T$ gates for $N_s$ system qubits. Since the local encoding cost is independent of the lattice size, this makes more sophisticated local encodings affordable and allows frustration within a patch to be incorporated into tighter bounds on the full Hamiltonian. Tighter spectral bounds reduce the overall normalization and can increase the phase separation between low-energy states under the qubitization relation $x=\cos\theta$, thereby reducing the queries needed for energy estimation or eigenstate filtering. In addition to reducing the number of queries, routing also lowers the leading block-encoding cost compared with the product-wise Pauli LCU baseline, leading to compounding improvements. These $T$-count improvements are evident in the resource estimates for the three classes of Hamiltonians we studied (Heisenberg, Fermi--Hubbard, and spin liquids), which we estimated  both analytically and numerically. We provide explicit circuit constructions and rigorous $T$-count bounds for the local encodings, showing how the improvements in shift and normalization can be realized and accounting for their implementation costs. 

To achieve more performant block encodings, we exploited several types of symmetry and structure in the target Hamiltonian. For example, a four-spin plaquette in the Heisenberg construction uses angular-momentum structures within three-spin triangles to attain an optimal shift and normalization, illustrating the value of choosing interaction groups that admit simple spectral encodings. Similarly, the honeycomb constructions use anticommuting Pauli groups and symmetry-reduced spectral decompositions to encode the same routed register, while QET allows larger Heisenberg and Fermi--Hubbard patches to approach tighter spectral bounds using smaller encodings as building blocks. These results motivate a systematic study of how quickly increasing patch size captures the remaining frustration, and what improvement in overall complexity justifies the additional local block-encoding complexity. Even at fixed patch size, redistributing interaction weights and exploiting further symmetries may improve the tradeoff. For example, in the $3\times3$ Heisenberg patch, approximate encodings with certified errors are particularly worth exploring. Classical calculations on manageable cluster sizes could guide these choices, identifying local descriptions that save the most quantum resources on the full lattice.

An important next step is to extend the resource analysis from $T$ count to $T$ depth as a proxy for execution time. The routing networks allow substantial parallelism, whereas larger-patch QET encodings require sequential calls to their underlying block encodings. An encoding with fewer $T$ gates may therefore take longer to execute, depending on the available work qubits, magic-state supply, and measurement latency. Optimizing the encoding together with state preparation, the remainder of the fault-tolerant algorithm, and any logical design choices will determine which encoding strategies yield the greatest practical benefit. Nevertheless, our framework provides a systematic way to construct and compare encoding strategies, forming a basis for optimizing these choices for a given application and fault-tolerant architecture.

Across the parameter ranges studied, our resource estimates at $M=120$ show approximately $15\times$, $3\times$, and $5\times$ reductions in the estimated overall $T$ cost for the Heisenberg, Fermi--Hubbard, and honeycomb models, respectively, when compared with state-of-the-art Pauli LCU constructions that exploit the product structure of interaction terms and use unary iteration for their efficient implementation~\cite{babbush2018encoding,yoshioka2024hunting}. These gains combine a lower cost per block-encoding query with tighter spectral bounds that reduce the number of queries required. The magnitude and consistency of these gains demonstrate the effectiveness of our approach combining coherent routing with tighter local encodings. These savings can support larger lattices and finer energy resolution when distinguishing competing phases, as well as longer time evolution when computing dynamical response. The improvements developed here therefore create room for larger systems, higher accuracy, and more demanding observables within a fixed quantum budget, strengthening the prospect that fault-tolerant quantum computers can address increasingly demanding questions in frustrated magnetism and correlated fermion physics.

\section*{Note added in print}

After preparation of this manuscript, we became aware that Ref.~\cite{summer2026efficient} independently developed a routing-based approach for Hamiltonian simulation.

\section*{Acknowledgments}
We thank Panagiotis Barkoutsos, Mario Berta, Yu-Ping Lin, Mark Nowakowski, and Luning Zhao for  insightful discussions and support. 

\section*{AI Use Statement}

The authors acknowledge the use of large language models from Anthropic and OpenAI in the development of this work. AI models were used in the development of code for numerical calculations, ideation of technical results, and for initial drafting and editing of the manuscript. The authors verified all AI-generated output and take full responsibility for the content presented in this work. All text in the manuscript was revised and edited by the authors. 

\bibliographystyle{apsrev4-2}
\bibliography{citations}

\clearpage
\onecolumngrid

\phantomsection
\addcontentsline{toc}{section}{\protect\numberline{}Appendices}

\startappendixcontents
\appendix
\appendixtableofcontents
\clearpage

\section{Notation and definitions}
\label{app:notation}

\subsection{Lattice and patch quantities}
\begin{description}
\item[$M,N_l$] An $M\times M$ array contains $N_l=M^2$ unit cells. Also, $N_l$ counts the translations of one reference patch in our constructions.
\item[$N_s$] Number of system qubits, excluding signal and work qubits. For square-lattice spin systems, $N_s=N_l$, whereas for square-lattice Fermi--Hubbard and honeycomb lattice spin systems, $N_s=2N_l$.
\item[$q$] Number of system qubits per unit cell in the general routing construction, so $N_s=qN_l$.
\item[$p=(p_x,p_y)$] Coherent address of a translated patch, with
$0\leq p_x,p_y<M$. The coordinates are periodic modulo $M$.
\item[$\mathcal P_p,H_p$] Spatial support of patch $p$ and its weighted Hamiltonian.
\end{description}

\subsection{Spectra, shifts, and normalizations}
\begin{description}
\item[$e_{\min},e_{\max}$] Minimum and maximum eigenvalues of a local operator.
\item[$E_{\min},E_{\max}$] Minimum and maximum eigenvalues of the full Hamiltonian; $E_0=E_{\min}$.
\item[$e_0$] Ground-state energy per site: $E_0/N_s$ for Heisenberg and $E_0/N_l$ for Fermi--Hubbard.
\item[$\ell_p,u_p$] Lower and upper spectral bounds for $H_p$:
$\ell_p\leq e_{\min}(H_p)\leq e_{\max}(H_p)\leq u_p$.
\item[$\delta_p,\lambda_p$] Local shift and normalization: $\delta_p=(u_p+\ell_p)/2$, $\lambda_p=(u_p-\ell_p)/2$.
\item[$\Delta,\Lambda$] Global shift and normalization in $(H-\Delta I)/\Lambda$.
\item[$\Lambda_{\mathrm{eff}}$] Effective normalization for QPE
\end{description}

\subsection{Circuit and resource conventions}
\begin{description}

\item[Signal qubits] Qubits whose specified input and output states define the encoded block~\cite{low2019qubitization}.
\item[Work qubits (ancillas)] Temporary workspace restored before a block-encoding call ends and excluded from the signal projector.
\item[$R$] Routing unitary.
\item[$T_R$] Combined $T$ cost of forward and inverse routing, including the address arithmetic and local order correction specified for the construction.
\item[$\textsc{Prep},\textsc{Sel},B_{\mathrm{ref}}$] Address preparation, routed selection, and reference-patch block encoding.
\item[$T_{\mathrm{PREP}},T_{\mathrm{BE}}^p,T_{\mathrm{BE}}$]
Costs of one address preparation, one local encoding, and one global
encoding, respectively; $T_{\mathrm{BE}}=2T_{\mathrm{PREP}}+T_R+T_{\mathrm{BE}}^p$.
\item[$T_{\mathrm{rot}}(\epsilon)$] $T$ count for synthesizing one axial rotation to error $\epsilon$.
\item[$N_{\mathrm{QPE}},T_{\mathrm{QPE}}$]
Estimated block-encoding queries and $T$ cost for one QPE run.
Initial-state preparation, repetitions, and success amplification are
additional costs unless explicitly included.
\item[$\mathcal C$] Leading time-evolution cost score
$\mathcal C=|t_{\mathrm{evol}}|\Lambda T_{\mathrm{BE}}$
\end{description}

\subsection{Quantum eigenvalue transformation and errors}
\begin{description}
\item[$\beta_p,\rho_p$] QET source normalization and amplification ratio,
$A_p^{(0)}=(H_p-\delta_pI)/\beta_p$ and $\rho_p=\beta_p/\lambda_p>1$.
\item[$s,\widetilde\lambda_p$] QET headroom and resulting implemented
normalization, $\widetilde\lambda_p=\lambda_p/(1-s)$.
Normalization improves when $0<s<1-1/\rho_p$.
\item[$P_d,d,d_\star$] Real odd QET polynomial, its allowed odd degree,
and the minimum feasible degree under the stated interval constraints.
The tabulated convention $d=0$ means bypassing QET and making one source call.
\item[$a$] Number of source signal qubits. 
\item[$\epsilon_{\mathrm{QET}}$] Uniform QET polynomial-approximation error.
\item[$\Delta E,r_{\mathrm{QET}}$] Target energy error and fraction of its budget allocated to QET.
\item[$s_{\mathrm{prep}}$] Precision in bits for uniform address preparation. This is distinct from QET headroom $s$.
\end{description}

\section{Routing cost analysis}
\label{app:routing_cost_accounting}
\subsection{The square lattice $2\times 2$ routing}
\label{app:square_lattice_routing_cost}
Each of the one-dimensional networks routes a qubit from position $r$ to $0$ among $L=M/2$ positions. It is known that they can be constructed using $L-1$ CSWAPs. For the square lattice, the even and odd column networks each route a column containing $M$ qubits, using $M/2-1$ CSWAPs each. The row routing acts only within the two destination columns, giving four further networks of size $M/2-1$. Combining them and multiplying by two for forward and backward routing gives a total of $2\times(2M+4)(M/2-1)=2\,M^2-8$ CSWAP gates. With a clean ancilla, measurement and Clifford corrections a CSWAP costs $4\,T$ gates~\cite{gidney2018halving}. Thus, the swap networks cost $8\,M^2-32$ T-gates.

The even column and row networks each need one computed address ($ (t+s) \bmod M/2)$. For $M/2$ that is a power of two, each address can be computed in a separate register using $\log_2(M/2)-1=\log_2(M)-2$ temporary logical ANDs at four $T$ gates each~\cite{gidney2018halving}. We retain the computed addresses and carry ancillas until reverse routing is complete, keeping the original addresses unchanged. The computed addresses and carries can be erased using CNOTs and measurement-based uncomputation. Thus, computing and uncomputing each address costs $4(\log_2 M-2)$ $T$ gates. For non-power-of-two $L=M/2$, we can first compute $z=t+s$ in a separate $\lceil\log_2 L\rceil$-qubit register using $\lceil\log_2 L\rceil-1$ temporary logical ANDs. There is no overflow because $z\leq L<2^{\lceil\log_2 L\rceil}$. Computing the equality flag $d=[z=L]$ requires at most another $\lceil\log_2 L\rceil-1$ temporary logical ANDs, and the desired modular address is $z\oplus dL$, obtained using CNOTs. Retaining the intermediate work until reverse routing is complete allows measurement-based uncomputation~\cite{gidney2018halving}. Thus, computing and uncomputing each modular address costs at most $8(\lceil\log_2 L\rceil-1)=8(\lceil\log_2 M\rceil-2)$ $T$ gates. Denote the system size or the number of spins as $N_s=M^2$. For a reflection-invariant patch operator, the complete routing cost for every even $M\geq4$ is bounded by
\begin{align}
T_R^{\square}
&\leq8(N_s-4)+16(\lceil\log_2M\rceil-2)\nonumber\\
&=8N_s+16\left\lceil\frac{\log_2N_s}{2}\right\rceil-64.
\label{eq:square_routing_t_cost}
\end{align}
For every even value of $M\geq4$, computing and uncomputing each address costs at most $8(\lceil\log_2 M\rceil-2)$ $T$ gates.
When the patch operator is not reflection invariant, the eight CSWAPs that restore the local order and undo this correction add $32$ $T$ gates. Including these corrections, the complete routing cost for every even $M\geq4$ is bounded by
\begin{align}
T_R^{\square,\mathrm{ord}}
&\leq8(N_s-4)+32+16(\lceil\log_2M\rceil-2)\nonumber\\
&=8N_s+16\left\lceil\frac{\log_2N_s}{2}\right\rceil-32.
\label{eq:square_routing_t_cost_ordered}
\end{align}

\subsection{General lattices, long range patches, and fermionic systems}
\label{app:routing_longrange_fermion}

Consider an $M\times M$ array of unit cells, each represented by $q$ qubits, so that $N_s=qM^2$. This encapsulates fermionic systems, where the number of qubits per unit cell is just the number of spin orbitals per unit cell (we would need to have similar fermionic mode ordering separated by $x,y\bmod U$ to make $\Pi$ calculation easy and then apply $U^2$ controlled-$f_{\mathrm{swap}}$ gates whose cost is small). If each site has a spin-up mode and a spin-down mode such as in the Fermi--Hubbard model, then $q$ is twice the number of sites. For the square lattice Heisenberg model, $q=1$. For the honeycomb lattice spin system, $q=2$. For the square lattice Fermi--Hubbard model, $q=2$.

Assume $U\geq2$, $U\mid M$, and $L=M/U\geq2$. The forward column routing uses $qUM(L-1)$ controlled qubit swaps, and the subsequent row routing uses $qU^2(L-1)$. Thus, forward and reverse routing together cost
\begin{align}
T_{\mathrm{CSWAP}}^{(U,q)}
&=8qU(M+U)(L-1)=8(N_s-qU^2).
\label{eq:uxu_routing_tree_cost}
\end{align}

To count order restoration, write each remainder in $\lceil\log_2 U\rceil$ binary bits. For each bit $s_j$, conditionally rotate the destination columns or rows left by $2^j$ positions modulo $U$. A rotation on $U$ objects by $2^j$ positions is a product of $U-\gcd(U,2^j)$ transpositions. Defining
\begin{align}
S_U=\sum_{j=0}^{\lceil\log_2U\rceil-1}\bigl[U-\gcd(U,2^j)\bigr],
\end{align}
the correction and its inverse, in both coordinate directions, cost $16qUS_U$ $T$ gates. Each transposition exchanges two columns or rows of $U$ cells and therefore uses $qU$ controlled qubit swaps. This construction restores the full local order for every address.

For each coordinate, only the $U-1$ groups $g=0,\ldots,U-2$ can require an incremented address. Compute $b_g=[s>g]$ and then $(t+b_g)\bmod L$ once per group, sharing these registers among all qubits routed with the same address. A comparison with a classical $\lceil\log_2U\rceil$-bit constant uses at most $\lceil\log_2U\rceil-1$ temporary logical ANDs; its intermediate work can be retained for measurement-based uncomputation. The two coordinates therefore contribute at most $8(U-1)(\lceil\log_2U\rceil-1)$ $T$ gates for comparisons. Each conditional modular increment, including its final erasure, costs at most $8(\lceil\log_2 L\rceil-1)$ $T$ gates for every integer $L\geq2$, by the construction in~\Cref{app:square_lattice_routing_cost}.

When $U$ is a power of two, $t$ and $s$ are obtained from the binary address by separating its bits, at no $T$ cost. For other $U$, coherent division by $U$ must also be counted. One explicit bound follows by processing the address bits from most to least significant while retaining a one-hot remainder register. At each step, for $U-1$ remainder values, compute the logical AND of the remainder flag and the next address bit. The remaining branch follows by XOR because the remainder flags are one-hot. The next remainder flags and quotient bit are then obtained using CNOTs. The first input bit requires only Clifford gates. Retaining all intermediate work gives the following bound for the division cost $\mathcal D_U(N_s)$ in both coordinates, including its final erasure:

\begin{align}
\mathcal D_U(N_s)
&\leq8(U-1)\left(\left\lceil\frac{\log_2(N_s/q)}{2}\right\rceil-1\right).
\label{eq:uxu_address_conversion_bound}
\end{align}
If the patch address is instead prepared directly in the mixed-radix registers $(t,s)$, with $p=Ut+s$, this division is unnecessary and $\mathcal D_U$ can be set to zero. Such state preparation is counted in $T_{\textsc{Prep}}$, separately from $T_R$.

Including local order restoration and all of the above address arithmetic, for every integer $L=\sqrt{N_s/q}/U\geq2$ we obtain
\begin{align}
T_R^{(U,q)}
&\leq8N_s+16qUS_U-8qU^2+8(U-1)\left[2\left\lceil\frac{\log_2(N_s/q)}{2}-\log_2U\right\rceil+\lceil\log_2U\rceil-3\right]
+\mathcal D_U(N_s).
\label{eq:uxu_routing_t_cost}
\end{align}
These are constructive upper bounds, rather than claims of optimal arithmetic or local permutation synthesis. For fixed $U$ and $q$, their leading term is $8N_s$, with the local order correction independent of $N_s$ and the arithmetic cost logarithmic in $N_s$. For $U=2$, with $\mathcal D_2=0$, the bounds reduce to the ordered square-lattice formula when $q=1$ and to the honeycomb formula when $q=2$.

\paragraph{$3\times 3$ routing for the square lattice Heisenberg model}
For the practically relevant $U=3$ case, we can improve the order correction. In each destination row, apply the transpositions $(0,1)$ controlled on $s\ne0$, $(1,2)$ controlled on $s=1$, and $(0,2)$ controlled on $s=2$, in that order. They implement the required left cyclic shift for each $s\in\{0,1,2\}$. If $s=s_0+2s_1$, these controls are $s_0\oplus s_1$, $s_0$, and $s_1$, respectively, so they require only Clifford gates. The same controls give $b_0$ and $b_1$ without any $T$ gates. The corrections in both coordinates and their inverses use $36q$ controlled qubit swaps, costing $144q$ $T$ gates. Using this circuit, for every integer $L=\sqrt{N_s/q}/3\geq2$ the complete cost is bounded by
\begin{align}
T_R^{(3,1)}
&\leq8N_s+72q+32\left\lceil\frac{\log_2(N_s/q)}{2}-\log_23\right\rceil
-32+\mathcal D_3(N_s),
\label{eq:3x3_routing_t_cost}
\end{align}
where $q=1$ for the square lattice Heisenberg model.

\paragraph{$2\times 2$ unit cell routing for the Honeycomb lattice spin system and the square lattice Fermi--Hubbard model}
Since the honeycomb lattice has $q=2$, we have essentially two copies of the square lattice routing, while the quantum modular arithmetics required stays the same. notice that for the hexagonal lattice with the same number of cells, it has double the number of spins and therefore $N_s$ is doubled compared to the square lattice) The same holds for the square lattice Fermi--Hubbard model since $q=2$. Additional Clifford gates are required, however. Therefore, for every even $M\geq4$, the complete routing cost for both systems is bounded by (notice that with the same number of cells, they has double the number of qubits compared to the square lattice spin system, and therefore $N_s$ is doubled compared to the square lattice spin system) 
\begin{align}
T_R^{\,\hexago}=T_R^{\mathrm{FH}}
&\leq8N_s+16\left\lceil\frac{\log_2N_s-1}{2}\right\rceil-32.
\label{eq:honeycomb_fh_routing_t_cost}
\end{align}

\subsection{Controlled fermionic-swap gadget}
\label{app:fermionic_gadget_cost}
The circuit of~\Cref{fig:controlled_Gamma} uses one temporary logical AND and its measurement-based inverse, together with Clifford gates. The AND costs four $T$ gates and its inverse is free in $T$ count~\cite{gidney2018halving}; hence a controlled fermionic swap costs $4T$, matching the controlled qubit swap. For $N_s=2M^2$, the forward and inverse trees cost $8N_s-64$. Adding order restoration and address arithmetic gives~\Cref{eq:honeycomb_fh_routing_t_cost}.
Within each routing tier the swap intervals are disjoint. Their parities can be computed in parallel with CNOT depth logarithmic in the interval length and no extra ancilla, or cached between tiers at the expense of work ancillas. The two routed destination columns may be processed sequentially.

\subsection{Reflection-invariant hierarchical patch routing}
\label{app:three_by_three_routing}
We route the nine spins of $\mathcal P_p$ with the construction of~\Cref{subsec:long_range_routing} at $U=3$, which requires $M$ to be divisible by three. Every translated $3\times3$ patch contains exactly one site from each of the nine residue classes $(x\bmod3,y\bmod3)$, and each class is routed by its own network. A class occupies $N/9$ sites, so its swap tree uses $N/9-1$ CSWAPs, as in~\Cref{subsec:square_lattice_routing}. The forward and inverse swap networks therefore use $N-9$ CSWAPs each, compared with $N-4$ for the plaquette router. The routed patch arrives with its columns and rows cyclically shifted by $s_x=p_x\bmod3$ and $s_y=p_y\bmod3$. Since $H^{3\times3}_p$ is invariant under row and column reflections, it is sufficient to restore the order up to a reflection. We exchange columns $1$ and $2$ controlled on $s_x=1$ and columns $0$ and $1$ controlled on $s_x=2$, and treat the rows in the same way. These singly controlled exchanges use $12$ CSWAPs, or $24$ with their inverses. We include these exchanges and the address arithmetic in the routing cost. Four conditional modular increments cost at most $32(\lceil\log_2(M/3)\rceil-1)$, and division of both coordinates by three costs at most $16(\lceil\log_2M\rceil-1)$, including their inverses. Thus,
\begin{align}
T_R^{3\times3}
&\leq8(N-9)+96+32(\lceil\log_2(M/3)\rceil-1)+16(\lceil\log_2M\rceil-1)\nonumber\\
&=8N+16\lceil\log_2M\rceil+32\lceil\log_2(M/3)\rceil-24.
\label{eq:three_by_three_routing_t_cost}
\end{align}

\section{Resource accounting}
\label{app:cost_accounting}

\subsection{Address preparation}
\label{app:cost_conventions}

For translated patches, the ideal address state is uniform. We implement its preparation using the flagged equal-superposition construction of Ref.~\cite{sanders2020compilation} with precision parameter $s=7$ (distinct from the QET headroom $s$ used below). For a non-power-of-two register with $M$ possible values and $\lceil\log_2 M\rceil$ address qubits, their cost is $4\lceil\log_2 M\rceil+2s-13=4\lceil\log_2 M\rceil+1$ Toffoli gates or $16\lceil\log_2 M\rceil+4$ $T$ gates. A power-of-two register instead requires only Hadamards. We prepare the two coordinate registers $p_x$ and $p_y$ independently, so for an $M\times M$ array of unit cells with $M^2$ patches we use the common upper bound
\begin{align}
T_{\textsc{prep}}
&\leq32\lceil\log_2 M\rceil+8
\label{eq:methods_uniform_prep_t_cost}
\end{align}
whether $M$ is a power of two or not. Thus, preparation and its inverse contribute at most $64\lceil\log_2 M\rceil+16$ $T$ gates.

We ignore the cost associated with the reusable phase-gradient resource state which is amortized over the entire QPE sequence. Each coordinate preparation retains a rotated ancilla, which is counted as a signal qubit. The preparation circuit also includes an inequality-success flag; it returns clean when preparation is reversed around the address-preserving selection. With $s_{\mathrm{prep}}=7$, the failure probability is very small~\cite{sanders2020compilation}, so we neglect this failure probability and its effect on normalization, similar to Ref.~\cite{low2025fast}.

\subsection{QET details}
\label{app:qet_cost_accounting}

We give the circuit for the real odd polynomial $P_d$ chosen in~\Cref{subsec:qet}, with $d=d_\star\geq1$ odd. The source $U_0^p$ is a Hermitian unitary, $(U_0^p)^\dagger=U_0^p$ and $(U_0^p)^2=I$, whose block on $a\geq1$ signal qubits is $A_p^{(0)}$ of~\Cref{eq:local_qet_initial_encoding}. The local source constructions used here have this reflection property. Write $\Pi_a=\ket{0^a}\!\bra{0^a}\otimes I$ and $R_a=2\Pi_a-I$. Temporary work ancillas are erased and are excluded from $\Pi_a$.

\subsubsection{Circuit construction}

Choose a symmetric QSP phase vector $\Phi=(\phi_0,\ldots,\phi_d)$, with $\phi_j=\phi_{d-j}$ and fixed Clifford phase-convention offsets absorbed. A real odd bounded polynomial admits this symmetric phase factorization at the same degree~\cite{wang2022energy}, so imposing symmetry does not change the optimization defining $d_\star$ in~\Cref{eq:local_qet_degree_optimization}. Define
\begin{align}
V_\Phi
&=e^{i\phi_0R_a}U_0^p e^{i\phi_1R_a}\cdots U_0^p e^{i\phi_dR_a},\qquad
V_{-\Phi}=V_\Phi^\dagger.
\label{eq:local_qet_symmetric_sequence}
\end{align}
The second identity follows by reversing the symmetric phase sequence and using $(U_0^p)^\dagger=U_0^p$. The phases are synthesized so that the signal block of $(V_\Phi+V_\Phi^\dagger)/2$ equals $P_d(A_p^{(0)})$~\cite{low2019qubitization,gilyen2019quantum}.

Introduce a processing signal qubit $b$ and signed phases $S_j=e^{i\phi_jR_aZ_b}$. Alternating these phases with the source gives
\begin{align}
\mathcal M_\Phi
&=S_0U_0^pS_1\cdots U_0^pS_d\nonumber\\
&=\ket0\!\bra0_b\otimes V_\Phi
+\ket1\!\bra1_b\otimes V_\Phi^\dagger.
\label{eq:local_qet_signed_sequence}
\end{align}
We use the Hermitian real-part completion
\begin{align}
U_{\mathrm{QET}}^{(d)}
&=H_bX_b\mathcal M_\Phi H_b\nonumber\\
&=H_bX_b\bigl(\ket0\!\bra0_b\otimes V_\Phi
+\ket1\!\bra1_b\otimes V_\Phi^\dagger\bigr)H_b.
\label{eq:local_qet_hermitian_completion}
\end{align}
The operator $X_b\mathcal M_\Phi$ is a Hermitian unitary, so $U_{\mathrm{QET}}^{(d)}$ is a reflection suitable for the qubitization walk. Its block on $b=0$ and the original source signal state is $P_d(A_p^{(0)})$. Thus QET uses $d$ source calls and $d+1$ signed phases, and its output signal count is $a_{\mathrm{out}}=a+1$. The completion adds only Clifford gates.

For external control, write $d=2m+1$ and define $L=S_mU_0^pS_{m-1}\cdots U_0^pS_0H_b$, with $L=S_0H_b$ when $m=0$. Since $X_bS_jX_b=S_j^\dagger$, symmetry gives
\begin{align}
U_{\mathrm{QET}}^{(d)}=L^\dagger(X_bU_0^p)L.
\label{eq:local_qet_conjugation}
\end{align}
An external control on $U_{\mathrm{QET}}^{(d)}$ therefore requires control only on the central $U_0^p$ and $X_b$; $L$ and $L^\dagger$ remain unconditional and cancel when the control is off. The controlled $X_b$ is a Clifford CNOT. Only one of the $d$ local block-encoding source calls needs to be controlled, and none of the phase rotations on $b$ needs the external control. The resulting extra $T$ gates come from controlling that one source call and give an additive overhead independent of $d$. We neglect this overhead in our resource estimates and report only the uncontrolled QET cost below.

\subsubsection{Cost and approximation error}

Each signed phase $e^{i\phi R_aZ_b}$ tests only the original $a$ source signal qubits. Compute their all-zero predicate with $a-1$ temporary logical ANDs, apply a CNOT from that flag to $b$, perform one axial rotation on $b$, undo the CNOT and erase the predicate by measurement. The processing signal $b$ is the rotation target and is excluded from this predicate. The phase therefore costs at most $4(a-1)+T_{\mathrm{rot}}$. If one source call costs $T_0^p$, the total cost for odd $d_\star\geq1$ is
\begin{align}
T_{\textsc{BE}}^p
&\leq d_\star T_0^p+(d_\star+1)\bigl[4(a-1)+T_{\mathrm{rot}}\bigr].
\label{eq:local_qet_t_cost}
\end{align}
This expression is reproduced in~\Cref{subsec:qet}. When QET is bypassed, the source is called once, $a_{\mathrm{out}}=a$ and the local cost is $T_0^p$. The later notation $d=0$ denotes this bypass and is not substituted into the finite-degree formula.

The polynomial error in~\Cref{eq:local_qet_error} becomes an error in physical energy after restoring the normalization. For a common local error bound $\epsilon_{\mathrm{QET}}$, the encoded Hamiltonian $\widetilde H$ obeys
\begin{align}
\|\widetilde H-H\|
\leq\sum_p\tilde\lambda_p\epsilon_{\mathrm{QET}}
=\Lambda\epsilon_{\mathrm{QET}}.
\label{eq:local_qet_global_error}
\end{align}
Thus, assigning a fraction $0<r_{\mathrm{QET}}<1$ of an energy-error budget $\Delta E$ to QET permits $\epsilon_{\mathrm{QET}}\leq r_{\mathrm{QET}}\Delta E/\Lambda$.

\section{Strategies for efficient local encodings}
\subsection{Coupled-spin bases and Clebsch--Gordan transforms}
\label{app:clebsch_gordan}

A local patch operator that is a linear combination of squared spins of nested or disjoint subsets of the patch is diagonal in a coupled-spin basis. Its spectral encoding (\Cref{subsec:spectral_encoding}) then needs only the basis change between the computational basis and this coupled-spin basis. Here we collect the identities that give the spectra of such operators and the circuits for this basis change. We use the spin operators $\mathbf S_a=\boldsymbol\sigma_a/2$, denote by $\ket0$ and $\ket1$ the eigenstates of $S^z_a$ with eigenvalues $+\tfrac12$ and $-\tfrac12$, and write $R_y(\phi)=\exp(-i\phi Y/2)$ and $R_z(\phi)=\exp(-i\phi Z/2)$.

\subsubsection{Coupled-spin operators}
\label{app:cg_operators}

For a subset $\mathcal A$ of the spins in a patch, let $\mathbf S_{\mathcal A}=\sum_{a\in\mathcal A}\mathbf S_a$ denote its total spin. For disjoint subsets $\mathcal A$ and $\mathcal B$,
\begin{align}
\mathbf S_{\mathcal A}\cdot\mathbf S_{\mathcal B}=\frac12\left(\mathbf S_{\mathcal A\cup\mathcal B}^2-\mathbf S_{\mathcal A}^2-\mathbf S_{\mathcal B}^2\right).
\label{eq:cg_coupling_identity}
\end{align}
The squared spin $\mathbf S_{\mathcal A}^2$ commutes with every component of $\mathbf S_{\mathcal A}$ and with every operator that acts outside $\mathcal A$. Hence, the squared spins of nested or disjoint subsets commute with each other and with the total $z$ component $S^z$. A coupling scheme merges the spins two subsets at a time until it reaches the full patch. Its squared spins, with eigenvalues $s_{\mathcal A}(s_{\mathcal A}+1)$, together with $S^z$, with eigenvalue $m$, label an orthonormal basis of the patch, the coupled-spin basis. Any linear combination of these operators is diagonal in this basis, with eigenvalues given by the same linear combination of their eigenvalues.

For two spins, $\mathbf S_a^2=\mathbf S_b^2=\tfrac34I$ and~\Cref{eq:cg_coupling_identity} gives $\mathbf S_a\cdot\mathbf S_b=(\mathbf S_{ab}^2-\tfrac32I)/2$, where $\mathbf S_{ab}^2$ is $0$ on the singlet and $2$ on the triplet. Equivalently, $\boldsymbol\sigma_a\cdot\boldsymbol\sigma_b=2\mathbf S_{ab}^2-3I=2P_{ab}-I$, where $P_{ab}$ is the SWAP operator.

For three spins $1,2,3$, we couple spins $1$ and $3$ first and then add spin $2$, i.e., $\mathbf L=\mathbf S_1+\mathbf S_3$ and $\mathbf F=\mathbf L+\mathbf S_2$. Then
\begin{align}
\mathbf S_1\cdot\mathbf S_3=\frac{\mathbf L^2-\tfrac32I}{2},\qquad
\mathbf S_2\cdot(\mathbf S_1+\mathbf S_3)=\frac{\mathbf F^2-\mathbf L^2-\tfrac34I}{2}.
\label{eq:cg_three_spin_identities}
\end{align}
With $\ell\in\{0,1\}$ and $f\in\{\tfrac12,\tfrac32\}$, the operators $\mathbf L^2$ and $\mathbf F^2$ have the three joint eigenspaces in~\Cref{tab:cg_three_spin_sectors}, and $F^z=S^z$ resolves the states within each of them. Any operator $c_1\,\mathbf S_1\cdot\mathbf S_3+c_2\,\mathbf S_2\cdot(\mathbf S_1+\mathbf S_3)+c_3F^z$ is therefore diagonal in the basis $\ket{\ell,f,m}$, with eigenvalues read off from~\Cref{tab:cg_three_spin_sectors} plus $c_3m$.

\begin{table}[ht]
\centering
\begin{tabular}{c|c|c|cc}
\hline\hline
$(\ell,f)$ & dimension & $m$ & $\mathbf S_1\cdot\mathbf S_3$ & $\mathbf S_2\cdot(\mathbf S_1+\mathbf S_3)$\\
\hline
$(0,\tfrac12)$ & $2$ & $\pm\tfrac12$ & $-\tfrac34$ & $0$\\
$(1,\tfrac12)$ & $2$ & $\pm\tfrac12$ & $\tfrac14$ & $-1$\\
$(1,\tfrac32)$ & $4$ & $\pm\tfrac12,\pm\tfrac32$ & $\tfrac14$ & $\tfrac12$\\
\hline\hline
\end{tabular}
\caption{Joint eigenspaces of $\mathbf L^2$ and $\mathbf F^2$ for three spins and the values of the two pair operators in~\Cref{eq:cg_three_spin_identities} on them.}
\label{tab:cg_three_spin_sectors}
\end{table}

For four spins coupled as two pairs, $\mathbf L_1=\mathbf S_1+\mathbf S_3$, $\mathbf L_2=\mathbf S_2+\mathbf S_4$ and $\mathbf J_{\mathrm{tot}}=\mathbf L_1+\mathbf L_2$,~\Cref{eq:cg_coupling_identity} expresses the four cross terms in $\mathbf L_1\cdot\mathbf L_2$ through $\mathbf J_{\mathrm{tot}}^2$, $\mathbf L_1^2$ and $\mathbf L_2^2$.

\subsubsection{Clebsch--Gordan transforms}
\label{app:cg_transforms}

A Clebsch--Gordan transform $V_{\mathrm{CG}}$ is a unitary that maps each computational basis state to a state of the coupled-spin basis. For an operator that is diagonal in this basis, the basis change of~\Cref{eq:local_multiplexed_dilation} is $V_p=V_{\mathrm{CG}}^\dagger$, and each transformed projector $Q_j$ is a sum of computational basis projectors. Efficient circuits for this transform exist for any number of spins~\cite{bacon2006efficient}. For the small patches in this work, we use the explicit circuits below. Only the eigenspaces enter the dilation, so any orthonormal basis and any phases can be chosen within a degenerate eigenspace to simplify the circuit.

The circuits are built from the number-preserving Givens rotation $G_{a,b}(\theta)$ on qubits $a$ and $b$, which acts on the one-excitation subspace as
\begin{align}
G_{a,b}(\theta)\ket{1_a0_b}&=\cos\theta\ket{1_a0_b}+\sin\theta\ket{0_a1_b},\nonumber\\
G_{a,b}(\theta)\ket{0_a1_b}&=-\sin\theta\ket{1_a0_b}+\cos\theta\ket{0_a1_b},
\label{eq:cg_givens}
\end{align}
and leaves both $\ket{0_a0_b}$ and $\ket{1_a1_b}$ unchanged.

\paragraph{Two-spin transform}
When an operator is diagonal in the Bell basis $\ket{\Phi^\pm}=(\ket{00}\pm\ket{11})/\sqrt2$, $\ket{\Psi^\pm}=(\ket{01}\pm\ket{10})/\sqrt2$, as is the case for any function of $\mathbf S_{12}^2$ and for any exchange interaction $J_xX_1X_2+J_yY_1Y_2+J_zZ_1Z_2$, the Clifford circuit $V_p$ that applies $\mathrm{CNOT}_{1\to2}$ and then a Hadamard gate on qubit $1$ maps $\ket{\Phi^+},\ket{\Psi^+},\ket{\Phi^-},\ket{\Psi^-}$ to $\ket{00},\ket{01},\ket{10},\ket{11}$. The second bit then flags the $\ket{\Psi^\pm}$ sector and the first bit gives the sign, so the singlet $\ket{\Psi^-}$ is the single string $11$. If the basis must also diagonalize $S^z$, the transform $V_{\mathrm{CG}}=G_{1,2}(\pi/4)$ maps $\ket{10}$ and $\ket{01}$ to the triplet state $\ket{\Psi^+}$ and the singlet $\ket{\Psi^-}$, respectively, and fixes the triplet states $\ket{00}$ and $\ket{11}$, so $V_p=G_{1,2}(-\pi/4)$.

\paragraph{Three-spin transform}
For the coupling scheme of~\Cref{eq:cg_three_spin_identities}, the circuit
\begin{align}
V_{\mathrm{CG}}=\mathrm{CZ}_{1,2}\,\mathrm{CZ}_{1,3}\,G_{1,3}\!\left(\frac{\pi}{4}\right)G_{1,2}\!\left(\arcsin\frac{1}{\sqrt{3}}\right)
\label{eq:cg_three_spin_circuit}
\end{align}
maps the computational basis states $\ket{b_1b_2b_3}$ to the coupled states in~\Cref{tab:cg_three_spin_map}. The joint eigenspaces of $\mathbf L^2$ and $\mathbf F^2$ therefore correspond to the sets of bit strings $\{000,011,100,111\}$ for $(1,\tfrac32)$, $\{010,101\}$ for $(1,\tfrac12)$, and $\{001,110\}$ for $(0,\tfrac12)$. From the last two columns of~\Cref{tab:cg_three_spin_map}, we see that two parities identify each set uniquely. The parity $u=b_2\oplus b_3$ vanishes exactly on the $f=\tfrac32$ set, and on the two $f=\tfrac12$ sets the parity $v=b_1\oplus b_2$ is $1$ for $\ell=1$ and $0$ for $\ell=0$. Two CNOTs controlled by qubit $2$ compute $u$ and $v$ in place on qubits $3$ and $1$. Since $V_{\mathrm{CG}}$ preserves the Hamming weight, the coupled state $V_{\mathrm{CG}}\ket{b_1b_2b_3}$ has $m=\tfrac32-(b_1+b_2+b_3)$.
\begin{table}[ht]
\centering
\begin{tabular}{c|c|c|cc}
\hline\hline
$\ket{b_1b_2b_3}$ & $V_{\mathrm{CG}}\ket{b_1b_2b_3}$ & $(\ell,f,m)$ & $u$ & $v$\\
\hline
$\ket{000}$ & $\ket{000}$ & $(1,\tfrac32,+\tfrac32)$ & $0$ & $0$\\
$\ket{100}$ & $(\ket{100}+\ket{010}+\ket{001})/\sqrt3$ & $(1,\tfrac32,+\tfrac12)$ & $0$ & $1$\\
$\ket{011}$ & $(\ket{011}+\ket{101}+\ket{110})/\sqrt3$ & $(1,\tfrac32,-\tfrac12)$ & $0$ & $1$\\
$\ket{111}$ & $\ket{111}$ & $(1,\tfrac32,-\tfrac32)$ & $0$ & $0$\\
$\ket{010}$ & $(2\ket{010}-\ket{100}-\ket{001})/\sqrt6$ & $(1,\tfrac12,+\tfrac12)$ & $1$ & $1$\\
$\ket{101}$ & $(\ket{011}+\ket{110}-2\ket{101})/\sqrt6$ & $(1,\tfrac12,-\tfrac12)$ & $1$ & $1$\\
$\ket{001}$ & $(\ket{001}-\ket{100})/\sqrt2$ & $(0,\tfrac12,+\tfrac12)$ & $1$ & $0$\\
$\ket{110}$ & $(\ket{011}-\ket{110})/\sqrt2$ & $(0,\tfrac12,-\tfrac12)$ & $1$ & $0$\\
\hline\hline
\end{tabular}
\caption{Action of the three-spin Clebsch--Gordan circuit in~\Cref{eq:cg_three_spin_circuit} on the computational basis, with the quantum numbers of the image states and the parities $u=b_2\oplus b_3$ and $v=b_1\oplus b_2$ of the input bits.}

\label{tab:cg_three_spin_map}
\end{table}

\subsubsection{Gate cost}
\label{app:cg_gate_cost}

The two Pauli operators in the generator of the Givens rotation commute, so
\begin{align}
G_{a,b}(\theta)=\exp\left(-\frac{i\theta}{2}X_aY_b\right)\exp\left(\frac{i\theta}{2}Y_aX_b\right).
\label{eq:cg_givens_pauli}
\end{align}
Each factor is a Clifford-conjugated single-qubit rotation $R_z(\pm\theta)$, so $G_{a,b}(\theta)$ needs two axial rotations. For $\theta=\pi/4$, these are exact $T$ or $T^\dagger$ gates. Hence, the three-spin transform in~\Cref{eq:cg_three_spin_circuit} costs $2+2T_{\mathrm{rot}}(\epsilon)$, and the pair $V_{\mathrm{CG}},V_{\mathrm{CG}}^\dagger$ costs $4+4T_{\mathrm{rot}}(\epsilon)$. The controlled-$Z$ gates, the two-spin Bell transform and the parity computations are Clifford. A sector identified by a single parity, such as $f=\tfrac32$, can control an operation directly. A sector identified by both parities is flagged with one temporary logical AND, which costs four $T$ gates and is uncomputed by measurement~\cite{gidney2018halving}. For larger patches, a Clebsch--Gordan transform can be built from successive coupling steps of the same kind~\cite{bacon2006efficient}. It diagonalizes a patch operator only when the operator is a function of the squared spins of one coupling scheme and of $S^z$. A spin-isotropic patch operator of a more general form, such as a Heisenberg model on a larger patch, only splits into blocks of fixed total spin in this basis, and these blocks still need a generic diagonalization. This remaining basis change and the number of distinct eigenvalues grow quickly with the patch size, which motivates the QET construction of~\Cref{subsec:qet}.

\subsection{Eigenbasis construction for real symmetric exchange interactions}
\label{app:real_symmetric_exchange}

A two-spin interaction can contain cross terms such as $S_i^xS_j^y$, which make its eigenbasis less apparent. When the exchange matrix is real and symmetric, a common rotation of the two spins removes these terms. For spin-$\tfrac12$, the resulting interaction is diagonal in the Bell basis. This gives an eigenbasis construction for any interaction of the form
\begin{align}
H_{ij}=\sum_{a,b\in\{x,y,z\}}M_{ab}S_i^aS_j^b,
\qquad \mathbf M=\mathbf M^T\in\mathbb R^{3\times3}.
\label{eq:general_symmetric_exchange}
\end{align}

We first diagonalize $\mathbf M$ classically. Its orthonormal eigenvectors can be chosen as the columns of a matrix $O$ with $\det O=1$. We can always change the sign of one eigenvector to fix the determinant if needed. Then,
\begin{align}
O^T\mathbf M O&=\operatorname{diag}(\tilde J_x,\tilde J_y,\tilde J_z),\qquad M_{ab}=\sum_c\tilde J_cO_{ac}O_{bc}.
\label{eq:general_exchange_matrix_diagonalization}
\end{align}
Since $O$ is a proper rotation, using the $SU(2)\xrightarrow{} SO(3)$ correspondence, it determines a single-spin unitary $u$, up to an overall sign, through
\begin{align}
uS^cu^\dagger=\sum_aO_{ac}S^a.
\label{eq:general_exchange_spin_rotation}
\end{align}
Applying the same rotation to both spins, with $U=u_i\otimes u_j$, gives
\begin{align}
U^\dagger H_{ij}U
=\tilde J_xS_i^xS_j^x+\tilde J_yS_i^yS_j^y+\tilde J_zS_i^zS_j^z.
\label{eq:general_exchange_xyz}
\end{align}
The three terms commute and share the Bell eigenbasis. Let $C_{\mathrm{Bell}}$ be the Clifford circuit whose columns are $(\ket{\Psi^-},\ket{\Phi^-},\ket{\Phi^+},\ket{\Psi^+})$. Thus,
\begin{align}
V=(UC_{\mathrm{Bell}})^\dagger
\label{eq:general_exchange_eigenbasis}
\end{align}
maps the eigenstates of $H_{ij}$ to computational basis states. The bond energies are the corresponding Bell eigenvalues of the XYZ interaction. They determine the shift, normalization, and ancilla rotations for the spectral encoding in~\Cref{subsec:spectral_encoding}. A general $u$ can be implemented using three Euler rotations, for example $u=R_z(\alpha)R_y(\beta)R_z(\chi)$~\cite{barenco1995elementary}. Additional structure in $\mathbf M$ can fix some of its eigenvectors and reduce the number of variable rotations. When the encoding is selected by an LCU flag, $V$ and $V^\dagger$ can remain unconditional and only the ancilla operations need the selection control.

\subsection{Circuits for anticommuting Pauli groups}
\label{app:anticommuting_pauli_groups}

A linear combination of pairwise anticommuting Hermitian Pauli strings can be normalized to a Hermitian unitary. We can construct a circuit for this unitary by conjugating a reference Pauli string with a sequence of rotations~\cite{zhao2020measurement} as shown next. Let
\begin{align}
G=\sum_{j=0}^{m-1}c_jP_j,\qquad
\beta=\sqrt{\sum_jc_j^2}>0,\qquad
S=G/\beta.
\label{eq:anticommuting_group_unitary}
\end{align}
Since $G^2=\beta^2I$, we have $S^2=I$. We choose $P_0$ as the reference string, absorbing the sign of $c_0$ into $P_0$ so that $c_0\geq0$, and seek a circuit $U$ such that $S=UP_0U^\dagger$.

For each remaining string, define
\begin{align}
Q_j=iP_0P_j,\qquad
R_j=\exp\!\left(\frac{i\alpha_j}{2}Q_j\right),
\qquad j=1,\ldots,m-1.
\label{eq:anticommuting_generators}
\end{align}
Here $Q_j$ is a Hermitian Pauli string up to a sign. It anticommutes with $P_0$ and $P_j$, and commutes with every other string in the group. Consequently,
\begin{align}
R_jP_0R_j^\dagger
&=\cos\alpha_j\,P_0+\sin\alpha_j\,P_j,\nonumber\\
R_jP_jR_j^\dagger
&=\cos\alpha_j\,P_j-\sin\alpha_j\,P_0,\nonumber\\
R_jP_kR_j^\dagger
&=P_k,\qquad k\notin\{0,j\}.
\label{eq:anticommuting_rotation_action}
\end{align}
Thus, each rotation mixes only the coefficients of $P_0$ and $P_j$. We choose the angles so that successive conjugations by $R_j^\dagger$ eliminate the coefficients of $P_j$. Writing $\hat c_j=c_j/\beta$ and $r_0=\hat c_0$, the required angles are
\begin{align}
r_j=\sqrt{r_{j-1}^2+\hat c_j^2},\qquad
\alpha_j=\operatorname{atan2}(\hat c_j,r_{j-1}).
\label{eq:anticommuting_rotation_angles}
\end{align}
If both arguments vanish, we set $\alpha_j=0$. After the last step, the remaining coefficient of $P_0$ is one. Reversing the elimination gives $U=R_1R_2\cdots R_{m-1}$ and hence $S=UP_0U^\dagger$.

Each $R_j$ uses a Clifford basis change, a CNOT ladder, one $R_z$ rotation, and the inverse Clifford operations~\cite{Whitfield2011Simulation}. Thus, applying $U$ and its inverse requires at most $2(m-1)$ synthesized rotations. In a controlled application of $S$, both basis changes can remain uncontrolled, since they cancel on the branch where $S$ is not selected. Only $P_0$ needs the selection control, and a singly controlled Pauli string is Clifford. Given a computed selection flag, the group therefore contributes at most $2(m-1)T_{\mathrm{rot}}(\epsilon)$ to the $T$-cost.

\section{Heisenberg encodings}
\label{app:heisenberg_cost_accounting}

In this section we give the gate-level circuits of the encodings in~\Cref{sec:heisenberg} and their $T$ gate counts. Each temporary logical AND costs four $T$ gates and is erased by measurement at no $T$ cost~\cite{gidney2018halving}, and each CSWAP costs four $T$ gates, as in the routing analysis.

\subsection{Bond-wise encoding}
\label{app:heisenberg_bondwise_cost_accounting}

 The local preparation selects an NN bond with total probability $J_1/(J_1+J_2)$ and an NNN bond with total probability $J_2/(J_1+J_2)$. We use three qubits for the bond-selection register. The first selects NN or NNN interactions, while the other two are prepared with Hadamards to select the specific bond,
\begin{align}
\textsc{Prep}^p\ket{000}
=\left(\sqrt{\frac{J_1}{J_1+J_2}}\ket0
+\sqrt{\frac{J_2}{J_1+J_2}}\ket1\right)\ket+\ket+.
\label{eq:heisenberg_bond_local_preparation}
\end{align}
Applying the bond-controlled SWAPs and reversing $\textsc{Prep}^p$ gives the local block-encoding circuit.

The local preparation in~\Cref{eq:heisenberg_bond_local_preparation} uses an $R_y$ rotation on the first bond-selection qubit and Hadamards on the other two, so $T_{\textsc{Prep}}^p=T_{\mathrm{rot}}(\epsilon)$. On the NN branch, the last two qubits select one of the four edges. On the NNN branch, the second qubit selects one of the two diagonals and the third qubit is ignored (each diagonal appears twice and gives the correct probability). Each NN selection combines its three controls into a flag with two temporary logical ANDs, applies the CSWAP, and erases the flag by measurement, which costs $12$ $T$ gates. Each NNN selection uses only the first two selection qubits, so it needs one temporary AND and one CSWAP, or $8$ $T$ gates. With four NN and two NNN selections, the local encoding cost is
\begin{align}
T_{\textsc{BE}}^p&=2T_{\textsc{Prep}}^p+T_{\textsc{Sel}}^p
=64+2T_{\mathrm{rot}}(\epsilon).
\label{eq:heisenberg_bond_local_t_cost}
\end{align}
The three selection qubits form the local signal register. Adding~\Cref{eq:heisenberg_bond_local_t_cost} to the routing upper bound in~\Cref{eq:square_routing_t_cost} and the address preparation cost in~\Cref{eq:methods_uniform_prep_t_cost}, the total cost for even $M\geq4$ is bounded by
\begin{equation}
T_{\textsc{BE},b}^\square
\leq8N+80\lceil\log_2M\rceil
+2T_{\mathrm{rot}}(\epsilon)+16,
\label{eq:heisenberg_bond_total_t_cost}
\end{equation}
where $N=M^2$.

\subsection{Bond-wise encoding with exchange anisotropy}
\label{app:heisenberg_anisotropic}

The bond-wise construction of~\Cref{sec:heisenberg_bond} extends naturally to an anisotropic exchange. We replace every $\boldsymbol\sigma_i\cdot\boldsymbol\sigma_j$ in~\Cref{eq:heisenberg_pauli_convention} by the XXZ interaction, so that a bond with coupling $J>0$ becomes
\begin{align}
H^b_{ij}(J,\kappa)=J\left(X_iX_j+Y_iY_j+\kappa Z_iZ_j\right)=J(2P_{ij}-I)+(\kappa-1)JZ_iZ_j.
\label{eq:heisenberg_xxz_bond}
\end{align}
The anisotropy $\kappa\in\mathbb R$ is the same for the NN and NNN bonds, and $\kappa=1$ recovers $H^b_{ij}(J)$. Both $P_{ij}$ and $Z_iZ_j$ are diagonal in the Bell basis $\ket{\Phi^\pm}=(\ket{00}\pm\ket{11})/\sqrt2$, $\ket{\Psi^\pm}=(\ket{01}\pm\ket{10})/\sqrt2$, where $Z_iZ_j$ is $+1$ on $\ket{\Phi^\pm}$ and $-1$ on $\ket{\Psi^\pm}$. Hence, the bond has the eigenvalues
\begin{align}
\kappa J\ \text{on}\ \ket{\Phi^\pm},\qquad
(2-\kappa)J\ \text{on}\ \ket{\Psi^+},\qquad
-(2+\kappa)J\ \text{on}\ \ket{\Psi^-}.
\label{eq:heisenberg_xxz_spectrum}
\end{align}
For $\kappa=1$, we recover the triplet and singlet eigenvalues $J$ and $-3J$. The lowest eigenvalue is $-(2+\kappa)J$ for $\kappa\geq-1$ and $\kappa J$ for $\kappa<-1$, and the highest is $(2-\kappa)J$ for $\kappa\leq1$ and $\kappa J$ for $\kappa>1$. The optimal shift and normalization are therefore
\begin{align}
\delta_b(J,\kappa)=-J\operatorname{sign}(\kappa)\min(1,|\kappa|),\qquad
\lambda_b(J,\kappa)=J\left[1+\max(1,|\kappa|)\right].
\label{eq:heisenberg_xxz_normalization}
\end{align}
A Pauli LCU of the three terms has normalization $(2+|\kappa|)J$, which exceeds $\lambda_b(J,\kappa)$ by $J\min(1,|\kappa|)$. For $|\kappa|\leq1$, the shifted and normalized bond operator is
\begin{align}
A^b_{ij}(\kappa)=\ket{\Psi^+}\!\bra{\Psi^+}-\ket{\Psi^-}\!\bra{\Psi^-}+\kappa\left(\ket{\Phi^+}\!\bra{\Phi^+}+\ket{\Phi^-}\!\bra{\Phi^-}\right).
\label{eq:heisenberg_xxz_A}
\end{align}
For $|\kappa|>1$, the normalized eigenvalue on the $\ket{\Phi^\pm}$ sector becomes $\operatorname{sign}(\kappa)$. The intermediate eigenvalue $\operatorname{sign}(\kappa)(3-|\kappa|)/(1+|\kappa|)$ then lies on $\ket{\Psi^+}$ for $\kappa>1$ and on $\ket{\Psi^-}$ for $\kappa<-1$, and the remaining Bell state stays at $-\operatorname{sign}(\kappa)$.

Unless $\kappa=\pm1$, $A^b_{ij}(\kappa)$ is not a unitary, and we use the spectral encoding of~\Cref{eq:local_multiplexed_dilation}. Its basis change $V_p$ is the Clifford circuit that applies $\mathrm{CNOT}_{i\to j}$ and then a Hadamard gate on qubit $i$. It maps $\ket{\Phi^+},\ket{\Psi^+},\ket{\Phi^-},\ket{\Psi^-}$ to $\ket{b_ib_j}=\ket{00},\ket{01},\ket{10},\ket{11}$ (\Cref{app:cg_transforms}), where $b_a$ is the computational-basis bit of qubit $a$. Hence, qubit $j$ flags the $\ket{\Psi^\pm}$ sector and qubit $i$ gives the sign. In this encoding, the eigenvalue $+1$ gives $R_y(0)=I$ and needs no gate, and the eigenvalue $-1$ gives $R_y(2\pi)=-I$, a Clifford sign flip on the system. Only the intermediate eigenvalue needs a synthesized rotation of the signal qubit $w$, controlled on the Bell sector that carries it.

With these bit labels, the rotation for $|\kappa|\leq1$ is $R_y(2\arccos\kappa)$ controlled on qubit $j$ being in $\ket0$, and the sign flip on $\ket{\Psi^-}$ is $\mathrm{CZ}_{i,j}$. For $\kappa>1$, the rotation is controlled on $b_ib_j=01$, and the sign flip on $\ket{\Psi^-}$ is again $\mathrm{CZ}_{i,j}$. For $\kappa<-1$, the rotation is controlled on $b_ib_j=11$, and the sign flip on $\ket{\Phi^\pm}$ is $X_jZ_jX_j$. In both cases, one temporary logical AND combines the two controls of the rotation.

Since $A^b_{ij}(\kappa)$ does not change when $J$ is rescaled, the six bonds of the patch share the same normalized operator. We therefore route the selected bond to the reference pair $(0,0),(1,0)$ with three controlled exchanges and apply the encoding once, using the routing construction of~\Cref{eq:methods_select} within the patch. The local preparation is still given by~\Cref{eq:heisenberg_bond_local_preparation}, since $\lambda_b(J,\kappa)$ is proportional to $J$. The XXZ bonds are symmetric under $i\leftrightarrow j$, and all bonds of the same type (NN or NNN) carry the same $\kappa$, so the patch Hamiltonian stays invariant under row and column reflections and the global routing needs no ordering correction. Summing over the patch gives
\begin{align}
\delta_{p,b}(\kappa)=-2(J_1+J_2)\operatorname{sign}(\kappa)\min(1,|\kappa|),\qquad
\lambda_{p,b}(\kappa)=2(J_1+J_2)\left[1+\max(1,|\kappa|)\right],
\label{eq:heisenberg_xxz_patch_normalization}
\end{align}
and the global constants are $\Delta_b(\kappa)=N\delta_{p,b}(\kappa)$ and $\Lambda_b(\kappa)=N\lambda_{p,b}(\kappa)$. At $\kappa=1$, they reduce to $\Delta_b$ and $\Lambda_b$.

The bond routing applies three exchanges in the following order. Controlled on the third selection qubit, one CSWAP exchanges sites $(1,0)$ and $(0,1)$. Controlled on the second, two CSWAPs exchange the two rows. Controlled on the first, one CSWAP exchanges sites $(1,0)$ and $(1,1)$. This sequence brings each of the four edges and each of the two diagonals to the reference pair. On the NNN branch, the exchange controlled on the third qubit maps each diagonal to itself, so this qubit remains irrelevant. The three selection qubits and $w$ form the local signal register.

If the NN and NNN bonds carry different anisotropies $\kappa_1$ and $\kappa_2$, the preparation angle changes unless $\max(1,|\kappa_1|)=\max(1,|\kappa_2|)$. The rotation of $w$ then depends on the first selection qubit, and one rotation of $w$ that is uniformly controlled on this qubit and on the qubits $i$ and $j$ covers both bond types~\cite{mottonen2005transformation}. The sign flip also depends on the first selection qubit when exactly one of $\kappa_1$ and $\kappa_2$ is below $-1$, since it is $\mathrm{CZ}_{i,j}$ for $\kappa\geq-1$ and $X_jZ_jX_j$ for $\kappa<-1$, and the same uniformly controlled rotation includes these sign flips. A fully anisotropic XYZ exchange is also diagonal in the Bell basis, and we treat it together with off-diagonal exchange for the honeycomb models in~\Cref{sec:spin_liquids}.

The local preparation is that of~\Cref{eq:heisenberg_bond_local_preparation}, so $T_{\textsc{Prep}}^p=T_{\mathrm{rot}}(\epsilon)$. The bond routing and its inverse use eight CSWAPs, or $32$ $T$ gates. The basis change and the sign flip are Clifford, and the controlled rotation of $w$ uses two CNOTs and two rotations. For $|\kappa|>1$, its two controls need one temporary logical AND. Hence,
\begin{align}
T_{\textsc{BE}}^p(\kappa)=2T_{\textsc{Prep}}^p+T_{\textsc{Sel}}^p=
\begin{cases}
32+4T_{\mathrm{rot}}(\epsilon),&|\kappa|\leq1,\\
36+4T_{\mathrm{rot}}(\epsilon),&|\kappa|>1.
\end{cases}
\label{eq:heisenberg_xxz_local_t_cost}
\end{align}
At $\kappa=\pm1$, the rotation is $R_y(0)=I$ or $R_y(2\pi)=-I$, and at $\kappa=0$ it is $R_y(\pi)=-iY$, whose controlled version is Clifford. At these points, $T_{\textsc{BE}}^p=32+2T_{\mathrm{rot}}(\epsilon)$. For different anisotropies, the uniformly controlled rotation of $w$ with three controls uses eight CNOTs and eight rotations~\cite{mottonen2005transformation}, so $T_{\textsc{BE}}^p(\kappa_1,\kappa_2)\leq32+10T_{\mathrm{rot}}(\epsilon)$. Adding~\Cref{eq:heisenberg_xxz_local_t_cost} to the routing upper bound in~\Cref{eq:square_routing_t_cost} and the address preparation cost in~\Cref{eq:methods_uniform_prep_t_cost}, the total cost for even $M\geq4$ is bounded by
\begin{equation}
T_{\textsc{BE},b}^\square(\kappa)
\leq8N+80\lceil\log_2M\rceil+4T_{\mathrm{rot}}(\epsilon)+
\begin{cases}
-16,&|\kappa|\leq1,\\
-12,&|\kappa|>1,
\end{cases}
\label{eq:heisenberg_xxz_total_t_cost}
\end{equation}
where $N=M^2$. At $\kappa=1$, this circuit applies a single SWAP to the reference pair, so it also encodes the isotropic bond, with the local cost $32+2T_{\mathrm{rot}}(\epsilon)$ instead of~\Cref{eq:heisenberg_bond_local_t_cost}. For the lattices of~\Cref{sec:overall-cost}, this saving is below $0.5\%$ of $T_{\textsc{BE},b}^\square$, so the estimates there keep the circuit of~\Cref{app:heisenberg_bondwise_cost_accounting}.

\subsection{Triangular encoding}
\label{app:heisenberg_triangular_cost_accounting}
In the circuit of~\Cref{sec:triangle_interaction}, the transformed projectors $Q_{(\ell,f)}=V_p\Pi_{(\ell,f)}V_p^\dagger$ project onto the computational basis states with the bit strings of~\Cref{tab:cg_three_spin_map}, in the bit order $b_ib_jb_k$,
\begin{align}
Q_{(0,\frac12)}:\ \{001,110\},\qquad
Q_{(1,\frac12)}:\ \{010,101\},\qquad
Q_{(1,\frac32)}:\ \{000,011,100,111\}.
\label{eq:triangle_sector_sets}
\end{align}
As shown in~\Cref{app:cg_transforms}, the two parities $u=b_j\oplus b_k$ and $v=b_i\oplus b_j$ identify these sets. The top sector has $u=0$, while $(1,\tfrac12)$ and $(0,\tfrac12)$ have $u=1$ with $v=1$ and $v=0$, respectively. At $J_2=J_1/2$, the reflection $2Q_{(1,\frac32)}-I$ is therefore the Pauli operator $Z_jZ_k$, since $u$ flags the two sectors with $f=\tfrac12$. We compute $u$ and $v$ in place with two CNOTs. The middle-sector flag is the logical AND of $u$ with $\bar v$ for $r\leq\frac12$ and with $v$ for $r>\frac12$, and it controls $R_y(2\arccos\xi)$ on $w$. The low-sector sign flip acts on the same two bits. It is $\mathrm{CZ}_{u,v}$ for $r\leq\frac12$ and $X_v\mathrm{CZ}_{u,v}X_v$ for $r>\frac12$. Finally, we uncompute the parities and apply $V_p^\dagger=V_{\mathrm{CG}}$. Together with the $Z_w$ that~\Cref{eq:local_multiplexed_dilation} applies first, this gives the encoding of $A^\triangle_{i,j,k}$.

The local preparation applies a Hadamard to each of the two triangle-selection qubits. These qubits control the column and row reflections, which we call the triangle-selection reflections. Together with the dilation qubit $w$, they form the local signal register, so the encoding has three local signal qubits. 

The two Hadamards give $T_{\textsc{Prep}}^p=0$. The triangle-selection reflections use two CSWAPs each, so they and their inverses cost $32$ $T$ gates. The basis changes $V_p$ and $V_p^\dagger$ cost $4+4T_{\mathrm{rot}}(\epsilon)$ (\Cref{app:cg_gate_cost}). The middle-sector flag uses one temporary AND, and the singly controlled $R_y$ uses two rotations. The parities and the low-sector sign flip are Clifford. Hence,
\begin{align}
T_{\textsc{BE}}^p=2T_{\textsc{Prep}}^p+T_{\textsc{Sel}}^p=40+6T_{\mathrm{rot}}(\epsilon).
\label{eq:heisenberg_triangle_local_t_cost}
\end{align}Adding~\Cref{eq:heisenberg_triangle_local_t_cost} to the routing upper bound in~\Cref{eq:square_routing_t_cost} and the address preparation cost in~\Cref{eq:methods_uniform_prep_t_cost}, the total cost for even $M\geq4$ is bounded by
\begin{equation}
T_{\textsc{BE},\triangle}^\square
\leq8N+80\lceil\log_2M\rceil
+6T_{\mathrm{rot}}(\epsilon)-8,
\label{eq:heisenberg_triangle_total_t_cost}
\end{equation}
where $N=M^2$.

\subsection{Triangle encoding with an on-site field}
\label{app:heisenberg_onsite}

The triangle construction of~\Cref{sec:heisenberg_triangle} extends to an on-site field
\begin{align}
H_h=h\sum_i Z_i
\label{eq:onsite_field}
\end{align}
added to~\Cref{eq:heisenberg_pauli_convention}, where we take $h\geq0$ without loss of generality since a global spin flip maps $h$ to $-h$. The field is a one-body term, so its support already lies inside the routed patch. Every site belongs to twelve triangles, three in each of its four plaquettes, so each triangle carries $h/12$ of the field on each of its sites. Since $Z_i+Z_j+Z_k=2F^z$, where $F^z$ is the $z$ component of $\mathbf F$, the triangle term becomes $H^\triangle_{i,j,k}+(h/6)F^z$.

The operator $F^z$ commutes with $\mathbf L^2$ and $\mathbf F^2$ (\Cref{app:cg_operators}) and is diagonal in the coupled-spin basis $\ket{\ell,f,m}$ of~\Cref{tab:cg_three_spin_map}. The same basis change $V_p=V_{\mathrm{CG}}^\dagger$ therefore diagonalizes the local operator, and the field splits each sector,
\begin{align}
e(\ell,f,m)=e_{(\ell,f)}+\frac{h}{6}m,
\label{eq:triangle_field_eigenvalues}
\end{align}
where $e_{(\ell,f)}$ is the field-free value of the sector $(\ell,f)$ in~\Cref{eqn:triangular_patch_eigenvalues}. For generic $h>0$ and $J_2\neq J_1/2$, the eight eigenvalues are distinct. The extreme eigenvalues are
\begin{align}
e_{\mathrm{max}}(h)=e_3+\frac{h}{4},
\qquad
e_{\mathrm{min}}(h)=\min\left(e_{\mathrm{min}}-\frac{h}{12},\ e_3-\frac{h}{4}\right),
\label{eq:triangle_field_extremes}
\end{align}
where $e_{\mathrm{min}}$ is the field-free minimum. The two candidates for the minimum cross at $h=12\lambda_\triangle$, which gives
\begin{align}
h\leq 12\lambda_\triangle:&\quad
\delta_\triangle(h)=\delta_\triangle+\frac{h}{12},
&\lambda_\triangle(h)&=\lambda_\triangle+\frac{h}{6},\nonumber\\
h> 12\lambda_\triangle:&\quad
\delta_\triangle(h)=e_3,
&\lambda_\triangle(h)&=\frac{h}{4}.
\label{eq:triangle_field_normalization}
\end{align}
The uniform field keeps the patch invariant under row and column reflections, so the triangle selection is unchanged. The global constants are $\Delta_\triangle(h)=4N\delta_\triangle(h)$ and $\Lambda_\triangle(h)=4N\lambda_\triangle(h)$.

Since every coupled state $V_{\mathrm{CG}}\ket b$ is an eigenvector, we can take each $Q_j$ in~\Cref{eq:local_multiplexed_dilation} to be a single projector $\ket b\!\bra b$, also when two eigenvalues coincide. The controlled rotations then become
\begin{align}
\sum_{b\in\{0,1\}^3}R_y(2\arccos x_b)\otimes\ket b\!\bra b,\qquad x_b=\frac{e(b)-\delta_\triangle(h)}{\lambda_\triangle(h)},
\label{eq:triangle_field_be}
\end{align}
where $e(b)$ is the eigenvalue of $V_{\mathrm{CG}}\ket b$, whose quantum numbers are listed in~\Cref{tab:cg_three_spin_map}. We implement this as a uniformly controlled rotation of $w$ with the three qubits of the reference triangle as controls. It has eight angles, including those of the two extreme eigenvalues, and uses eight CNOTs and eight rotations~\cite{mottonen2005transformation}.

Together with the basis changes and the triangle selection, the local encoding cost is
\begin{align}
T_{\textsc{BE}}^p(h)=36+12T_{\mathrm{rot}}(\epsilon).
\label{eq:triangle_field_local_t_cost}
\end{align}
Adding this to the routing upper bound in~\Cref{eq:square_routing_t_cost} and the address preparation cost in~\Cref{eq:methods_uniform_prep_t_cost}, the total cost for even $M\geq4$ is bounded by
\begin{equation}
T_{\textsc{BE},\triangle}^\square(h)
\leq8N+80\lceil\log_2M\rceil
+12T_{\mathrm{rot}}(\epsilon)-12.
\label{eq:triangle_field_total_t_cost}
\end{equation}

\subsection{Optimal shift and normalization of the plaquette operator}
\label{appendix:optimal_plaquette_normalization}

We show that the triangle decomposition of~\Cref{sec:heisenberg_triangle} attains the optimal shift and normalization of the plaquette operator $H^\square_{\mathrm{ref}}$. Write $\mathbf S_{x,y}=\boldsymbol\sigma_{x,y}/2$ for the spin at site $(x,y)$, and couple the two spins on each diagonal of the plaquette,
\begin{align*}
\mathbf L_1=\mathbf S_{0,0}+\mathbf S_{1,1},\qquad
\mathbf L_2=\mathbf S_{0,1}+\mathbf S_{1,0},\qquad
\mathbf J_{\mathrm{tot}}=\mathbf L_1+\mathbf L_2.
\end{align*}
The four NN bonds are the four cross terms of $\mathbf L_1\cdot\mathbf L_2$, and the two NNN bonds couple the two spins within $\mathbf L_1$ and within $\mathbf L_2$. Hence
\begin{align*}
H^\square_{\mathrm{ref}}
&=2J_1\,\mathbf L_1\cdot\mathbf L_2+4J_2\left(\mathbf S_{0,0}\cdot\mathbf S_{1,1}+\mathbf S_{0,1}\cdot\mathbf S_{1,0}\right)\\
&=J_1\mathbf J_{\mathrm{tot}}^2+(2J_2-J_1)\left(\mathbf L_1^2+\mathbf L_2^2\right)-6J_2I,
\end{align*}
where the second equality uses~\Cref{eq:cg_coupling_identity}. The operators $\mathbf L_1^2$, $\mathbf L_2^2$ and $\mathbf J_{\mathrm{tot}}^2$ commute. Their eigenvalues are $\ell_1(\ell_1+1)$, $\ell_2(\ell_2+1)$ and $j(j+1)$, with $\ell_1,\ell_2\in\{0,1\}$ and $|\ell_1-\ell_2|\leq j\leq\ell_1+\ell_2$. The eigenvalues in these sectors of $H^\square_{\mathrm{ref}}$ are therefore
\begin{align*}
\begin{array}{c|ccccc}
\hline\hline
(\ell_1,\ell_2,j)&(0,0,0)&(1,0,1),\,(0,1,1)&(1,1,0)&(1,1,1)&(1,1,2)\\
\hline
E&-6J_2&-2J_2&2J_2-4J_1&2J_2-2J_1&2J_1+2J_2\\
\hline\hline
\end{array}
\end{align*}
For $J_1,J_2>0$, the largest eigenvalue is $2(J_1+J_2)$, on the fully polarized sector $j=2$. The smallest eigenvalue is $2J_2-4J_1$ for $2J_2\leq J_1$ and $-6J_2$ for $2J_2>J_1$. The optimal shift $\delta_2$ and normalization $\lambda_2$ of $H^\square_{\mathrm{ref}}$ are the midpoint and the half-width of its spectrum,
\begin{align*}
2J_2\leq J_1:&\quad\delta_2=2J_2-J_1,&\lambda_2&=3J_1,\\
2J_2>J_1:&\quad\delta_2=J_1-2J_2,&\lambda_2&=J_1+4J_2.
\end{align*}
Comparison with~\Cref{eq:A_triangle_normalization} gives $\delta_2=4\delta_\triangle$ and $\lambda_2=4\lambda_\triangle$, so the triangle encoding attains these values. Equivalently, each extreme eigenvalue of $H^\square_{\mathrm{ref}}$ is four times the corresponding extreme eigenvalue of a triangle term in~\Cref{eqn:triangular_patch_eigenvalues}. Every ground state of $H^\square_{\mathrm{ref}}$ therefore minimizes all four triangle terms at once, so the plaquette operator is frustration free with respect to its triangle decomposition.

\subsection{Hierarchical block encoding}
\label{app:heisenberg_hierarchical}

This appendix gives the gate-level circuit of the hierarchical encoding of~\Cref{sec:heisenberg_three_by_three} and its $T$ count. The circuit acts on the nine system qubits after the $3\times3$ routing. For every $r$, we take $U_2$ to be the triangular circuit of~\Cref{app:heisenberg_triangular_cost_accounting} with its dilation qubit $w$. At $J_2=J_1/2$, where $\xi=-1$, its middle-sector rotation is $R_y(2\pi)=-I$, so the construction and the bounds below also hold there.

\subsubsection{Plaquette averaging}

\begin{figure}[!ht]
\centering
\includegraphics[width=0.7\textwidth]{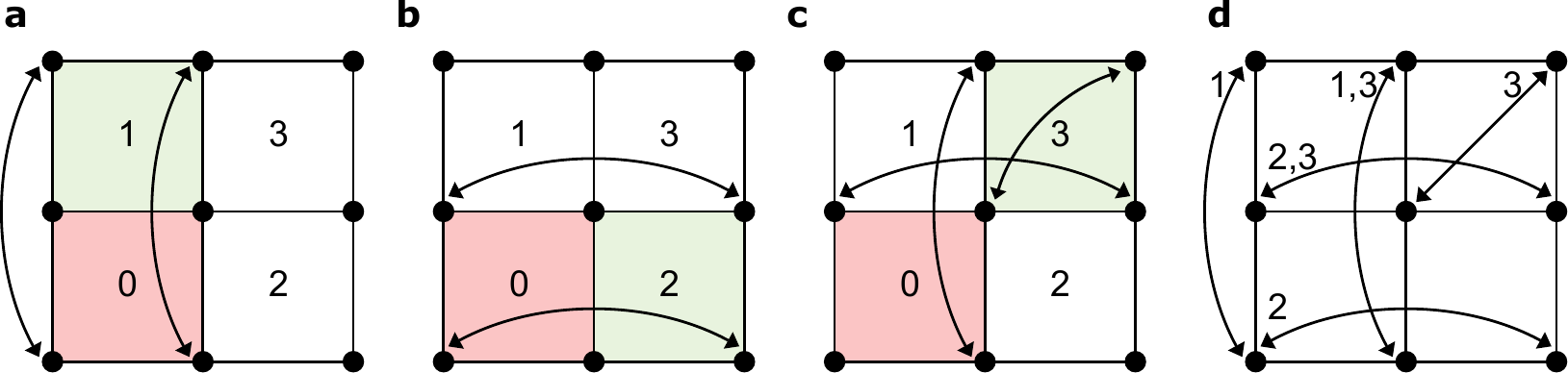}
\caption{\textbf{Swap operations needed to build the $3\times 3$ patch block encoding from the $2\times 2$ plaquette block encoding.} \textbf{a,b,c} Swaps required to map the target green plaquette to the red plaquette where the $2\times 2$ encoding circuit is applied. \textbf{d} For each swap, the index of the target green tile for which it is applied is labeled.
}
\label{fig:3x3_plaquette_swap}
\end{figure}

The preparation $\textsc{Prep}_q$ applies two Hadamards to the subplaquette register, so that $\textsc{Prep}_q\ket{00}=\frac12\sum_{q\in\{0,1\}^2}\ket{q}$. Let $R_{\mathrm{sub}}$ denote the controlled subplaquette routing of~\Cref{fig:3x3_plaquette_swap}, which moves subplaquette $q$ onto the four input qubits of $U_2$. The averaged encoding is
\begin{align*}
U_{\mathrm{av}}=\textsc{Prep}_q^\dagger R_{\mathrm{sub}}^\dagger U_2R_{\mathrm{sub}}\textsc{Prep}_q.
\end{align*}
This is the routed construction of~\Cref{eq:methods_select}, applied inside the patch. Hence $U_{\mathrm{av}}$ is a reflection, and its signal block is $A_{\mathrm{av}}$ of~\Cref{eq:hierarchical_average_signal_operator}.

Each of the seven CSWAPs of $R_{\mathrm{sub}}$ is controlled by one flag, which marks the subplaquettes for which the swap acts (\Cref{fig:3x3_plaquette_swap}d). A temporary logical AND of $q_x$ and $q_y$ gives the $(1,1)$ flag, and CNOTs give the remaining flags. We keep the $(1,1)$ flag through the forward and inverse subplaquette routing and erase it once by measurement.

\subsubsection{Recentering}

The circuit $B^{(0)}_3$ implements the two-term LCU of~\Cref{eq:hierarchical_recentered_operator} with a recentering qubit $\mathrm{rc}$. One rotation prepares
\begin{align*}
\textsc{Prep}_{\mathrm{rc}}\ket{0}
=\sqrt{\frac{\lambda_2}{\beta_3}}\ket{0}
+\sqrt{\frac{\delta_3-\delta_2}{\beta_3}}\ket{1},
\end{align*}
and
\begin{align*}
B^{(0)}_3=\textsc{Prep}_{\mathrm{rc}}^\dagger\left(\ket0\!\bra0_{\mathrm{rc}}\otimes U_{\mathrm{av}}-\ket1\!\bra1_{\mathrm{rc}}\otimes Z_w\right)\textsc{Prep}_{\mathrm{rc}}.
\end{align*}
The operator $-Z_w$ has the signal block $-I$ on $w=0$, so the signal block of $B^{(0)}_3$ is $A^{(0)}_3$. Both branches are reflections, so $B^{(0)}_3$ is a reflection.

A $Z$ gate on $\mathrm{rc}$ gives the branch sign, so the circuit for the $\mathrm{rc}=1$ branch must apply $Z_w$. We obtain it from $U_{\mathrm{av}}$ by disabling the middle-sector rotation and the low-sector sign flip of $U_2$ on this branch. The remaining gates of $U_{\mathrm{av}}$ then cancel with their inverses, except the factor $Z_w$ of~\Cref{eq:local_multiplexed_dilation}. These gates are the preparations, the subplaquette routing, the triangle-selection reflections, the basis changes and the parity CNOTs, and they stay uncontrolled. To disable the two operations, we compute $g=\overline{\mathrm{rc}}\wedge u$ with one temporary logical AND, where $u=b_j\oplus b_k$ is the triangle parity of~\Cref{app:heisenberg_triangular_cost_accounting}. We then use $g$ in place of $u$ in the middle-sector flag and in the low-sector sign flip.

\subsubsection{Local QET}

The source signal register of $B^{(0)}_3$ consists of $\mathrm{rc}$, the two subplaquette qubits, the two triangle-selection qubits and $w$, so $a=6$. We apply the QET circuit of~\Cref{app:qet_cost_accounting} to this source, with the degree $d_\star=d_\star(\rho_3,s,\epsilon_{\mathrm{QET}})$ of~\Cref{eq:local_qet_degree_optimization}. Each QET phase $S_j=e^{i\phi_jR_aZ_b}$ (\Cref{eq:local_qet_signed_sequence}) applies $e^{i\phi_jZ_b}$ to the processing signal $b$ when all six source signal qubits are in $\ket0$, and $e^{-i\phi_jZ_b}$ otherwise. With $b$, the output signal register has $a_{\mathrm{out}}=7$ qubits.

\subsubsection{Cost analysis}

The source contains the triangular encoding $U_2$, which costs $40+6T_{\mathrm{rot}}(\epsilon)$ (\Cref{eq:heisenberg_triangle_local_t_cost}). The subplaquette routing of~\Cref{fig:3x3_plaquette_swap} uses seven CSWAPs in each direction and one temporary AND for the $(1,1)$ flag, so it costs $4+2\times7\times4=60$ $T$ gates. The AND for the branch-control flag $g$ adds four $T$ gates, and $\textsc{Prep}_{\mathrm{rc}}$ and its inverse use one rotation each. Thus,
\begin{align}
T_0^p=104+8T_{\mathrm{rot}}(\epsilon).
\label{eq:hierarchical_initial_t_cost}
\end{align}
Each QET phase uses five temporary ANDs, which flag the all-zero state of the $a=6$ source signal qubits, and one rotation of $b$, so the $d_\star+1$ phases cost $T_{\mathrm{ph}}^p=(d_\star+1)(20+T_{\mathrm{rot}}(\epsilon))$. By~\Cref{eq:local_qet_t_cost}, the local QET cost is at most $124d_\star+20+(9d_\star+1)T_{\mathrm{rot}}(\epsilon)$.

The two coordinate preparations and their inverses cost $2T_{\textsc{Prep}}\leq64\lceil\log_2M\rceil+16$ (\Cref{eq:methods_uniform_prep_t_cost}). Combining~\Cref{eq:local_qet_t_cost,eq:methods_t_cost,eq:three_by_three_routing_t_cost}, the global block-encoding cost for $d_\star\geq1$ is bounded by
\begin{align}
T_{\textsc{BE},3\times3}^\square
&\leq2T_{\textsc{Prep}}+T_R^{3\times3}+d_\star T_0^p+T_{\mathrm{ph}}^p\nonumber\\
&\leq8N+80\lceil\log_2M\rceil+32\lceil\log_2(M/3)\rceil+124d_\star+12+(9d_\star+1)T_{\mathrm{rot}}(\epsilon).
\label{eq:hierarchical_t_cost}
\end{align}
Without QET, the source is used directly with six output signal qubits, and the global cost is at most $8N+80\lceil\log_2M\rceil+32\lceil\log_2(M/3)\rceil+96+8T_{\mathrm{rot}}(\epsilon)$.

\section{Fermi--Hubbard encodings}
\label{app:fh_cost_accounting}

\subsection{Bond-wise block encoding}

\subsubsection{Diagonalization of the Hamiltonian}
We first need to construct the diagonalization circuit $V$ of $H_b$. As mentioned in \Cref{sec:nn-fh-bondwise}, the spin-up and spin-down number conservation allows to diagonalize nine number sectors. The result is shown in Eq.~\ref{eqn:FH_bond_sectors}.
\begin{align}\label{eqn:FH_bond_sectors}
\begin{array}{c|c|c}
(N_\uparrow,N_\downarrow)&\text{occupation basis}&H_b\text{ in that sector}\\ \hline
(0,0)&|0000\rangle&+a\\
(2,2)&|1111\rangle&+a\\
(0,2)&|0101\rangle&-a\\
(2,0)&|1010\rangle&-a\\
(0,1)&|0100\rangle,|0001\rangle&-tX\\
(1,0)&|1000\rangle,|0010\rangle&-tX\\
(1,2)&|1101\rangle,|0111\rangle&+tX\\
(2,1)&|1110\rangle,|1011\rangle&+tX\\
(1,1)&|1100\rangle,|0110\rangle,|1001\rangle,|0011\rangle&
\begin{pmatrix}
 a & t & -t & 0\\
 t &-a & 0  & t\\
-t & 0 &-a  &-t\\
 0 & t &-t  & a
\end{pmatrix}
\end{array}
\end{align}
Therefore, we need four rotations to diagonalize the $(0,1),(1,0),(1,2),(2,1)$ sectors individually. Since $H_b$ in those sectors is $\pm tX$, this can be accomplished by $\pi/4$ rotations between the occupation basis states in those sectors (which maps the two basis states to the two egeinstates of $X$, as a Hadamard does).

We still need to further diagonalize the $(1,1)$ sector. Define
\begin{align}
|D_\pm\rangle=\frac{|1100\rangle\pm|0011\rangle}{\sqrt2},\quad
|S\rangle=\frac{|1001\rangle-|0110\rangle}{\sqrt2},\quad
|T_0\rangle=\frac{|1001\rangle+|0110\rangle}{\sqrt2},
\end{align}
where $D$ stands for the doublon-holon configuration, $S$ stands for singlet, and $T_0$ stands for the zero-magnetization component of the triplet. Then, we have
\begin{align}
H_b^{(1,1)}
\cong
\left.
\begin{pmatrix}
-a&-2t\\
-2t&a
\end{pmatrix}
\right|_{\{|S\rangle,|D_+\rangle\}}
\oplus(-a)_{|T_0\rangle}
\oplus(a)_{|D_-\rangle}.
\end{align}
Since the new basis consists of equal superpositions of $|1100\rangle,|0011\rangle$ or equal superposition of $|1001\rangle,|0110\rangle$, transforming $H_b^{(1,1)}$ into this form rotations. Consider a $\pi/4$ rotation on $|1100\rangle,|0011\rangle$ and a $-\pi/4$ rotation on $|1001\rangle,|0110\rangle$. These rotations map $|1100\rangle,|0011\rangle,|1001\rangle,|0110\rangle$ to $|D_+\rangle,-|D_-\rangle,|S\rangle,|T_0\rangle$, respectively. Their inverse transforms the Hamiltonian into the form above.

Finally, to diagonalize $H_b$ in the $\{|S\rangle,|D_+\rangle\}$ basis, which is now transformed into the $\{|1001\rangle,|1100\rangle\}$ basis, we need to apply a $\phi$ rotation between $|1001\rangle,|1100\rangle$, where
\begin{align}
\cos(2\phi)=\frac{a}{\sqrt{a^2+4t^2}},\qquad
\sin(2\phi)=\frac{2t}{\sqrt{a^2+4t^2}}.
\end{align}

We now summarize the result. Define a \(R_{p,q}(\theta)\) by
\begin{align}
R_{p,q}(\theta)|p\rangle
&=\cos\theta|p\rangle+\sin\theta|q\rangle,\\
R_{p,q}(\theta)|q\rangle
&=-\sin\theta|p\rangle+\cos\theta|q\rangle,
\end{align}
and let it act as identity on the other fourteen basis states. Its implementation is discussed in Appendix~\ref{sec:R_p_q_implementation}. Define
\begin{align}
V=R_{1000,0010}(\pi/4)
R_{0100,0001}(\pi/4) R_{1110,1011}(\pi/4)
R_{1101,0111}(\pi/4) R_{1001,0110}(-\pi/4)
R_{1100,0011}(\pi/4) R_{1001,1100}(\phi).
\end{align}
Then, in the regular lexicographical order,
\begin{align}
V^\dagger H_b V
=\operatorname{diag}(&a,t,t,a,-t,-a,-a,-t,-t,-\lambda_b,-a,-t,\lambda_b,t,t,a).
\end{align}

We encode the positive normalized magnitudes $|a|/\lambda_b$ and $|t|/\lambda_b$ with rotations on the signal qubit $w$, selected by their eigenspace flags. Negative eigenvalues are obtained by sign-dependent $X_w$ conjugations. The eigenvalues $\pm\lambda_b$ are covered by the common $Z_w$ and the same sign conjugations. This is because $\pm \lambda_b$ are the endpoint eigenvalues (maximum absolute value).

\subsubsection{Implementation of $R_{p,q}(\theta)$}\label{sec:R_p_q_implementation}

For \(R_{p,q}(\theta)\), let $d=p\oplus q$ and select any pivot bit \(k\) with \(d_k=1\).
Construct a reversible basis permutation \(P\):
\begin{enumerate}
    \item Apply \(X_i\) for every bit where \(p_i=1\). This maps \(p\) to \(0000\) and \(q\) to \(d\).
    \item For every \(i\neq k\) with \(d_i=1\), apply $\operatorname{CNOT}_{k\rightarrow i}$. This maps \(d\) to the single-bit string \(e_k\).
\end{enumerate}
Consequently,
\[
P|p\rangle=|0000\rangle,\qquad
P|q\rangle=|e_k\rangle.
\]Now apply \(R_y(2\theta)\) to qubit \(k\), controlled on all other three qubits being zero:
\begin{align*}
R_{p,q}(\theta)=P^\dagger\left[C^3_0R_{y,k}(2\theta)\right]P.
\end{align*}
Every spectator state maps to a string with at least one nonpivot bit equal to one, so it fails the control predicate and is restored exactly by \(P^\dagger\).
For example, the interacting rotation \(R_{1001,1100}(\phi)\) uses
\[
p\oplus q=0101.
\]Choose \(k=1\). Then:
\begin{itemize}
    \item Apply \(X_0X_3\).
    \item Apply \(\operatorname{CNOT}_{1\rightarrow3}\).
    \item Apply \(R_y(2\phi)\) on qubit \(1\), controlled on \(q_0=q_2=q_3=0\).
    \item Undo the CNOT and \(X_0X_3\).
\end{itemize}
The three zero-controls can be converted to positive controls with \(X\) gates. With two clean reusable work qubits, compute their AND, apply a singly controlled \(R_y(2\theta)\), and uncompute. A controlled rotation can be decomposed into two CNOTs and \(R_y(\pm\theta)\).

In total, two logical ANDs are required to reduce the control on three qubits onto one ancilla, costing $8$ $T$ gates. If the rotation angle is arbitrary, the total cost would be $2T_{\mathrm{rot}}+8$. If the rotation angle is $\pi/4$, the total cost would be $2+8=10$, where the $2$ comes from the two rotations $R_y(\pm\pi/4)$.

\subsubsection{Selecting a bond within the routed plaquette}

The above method describes how to encode one bond-wise interaction. Since the routing network routes a plaquette, we would like to block encode the full plaquette. Instead of repeating the bond-wise circuits four times, we can route the bonds on the plaquette to a fixed bond.

\begin{figure}[h]
\centering
\includegraphics[width=0.5\textwidth]{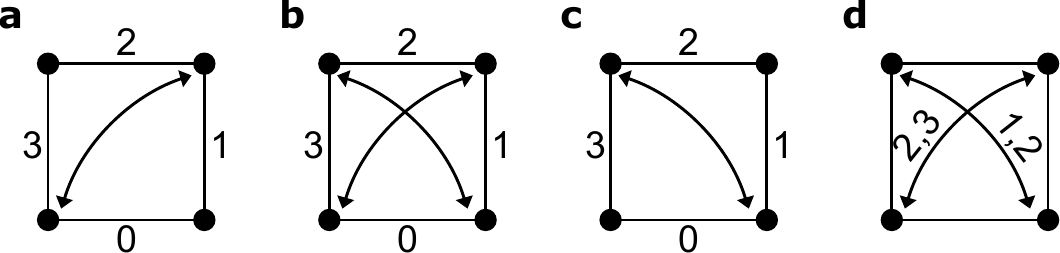}
\caption{\textbf{Swap operations needed to build the bond-wise block encoding.} \textbf{a,b,c} Swaps required to map the target bond to bond zero where the bond-wise block-encoding circuit is applied. \textbf{d} For each swap, the index of the bond for which it is applied is labeled.
}
\label{fig:FH_square_swap}
\end{figure}

We prepare two local signal qubits in $\ket+\ket+$ to select the four NN bonds uniformly. After fermionic routing brings the $2\times2$ plaquette to the reference register, the controlled bond-routing circuit in~\Cref{fig:FH_square_swap} brings the selected bond to the fixed input pair in each spin sector. We apply $V^\dagger$, the selected dilation of $H_b/\lambda_b$, and $V$, then reverse the bond routing and the two Hadamards. This executes the bond interaction only once.

The source signal register consists of the two bond-selection qubits and $w$. For a controlled source, combine the external control with the two nontrivial-magnitude dilation flags. The sign-dependent $X_w$ conjugations remain unconditional and cancel on the inactive branch, while the common $Z_w$ becomes a CZ. Preparation, bond routing and the eigenbasis changes remain unconditional.

\subsubsection{Cost analysis}
In the preparation oracle, the cost of preparing the address equal superposition is bounded by $32\lceil\log_2M\rceil+8$ $T$ gates as given by \Cref{eq:methods_uniform_prep_t_cost}. Additionally, we need to specify which bond in the local patch to choose, and since there are four NN bonds, we need two qubits in uniform superpositions in $\textsc{Prep}$, which can be accomplished by two Hadamards.

For $\textsc{Sel}$, fermionic routing has a cost given by \Cref{eq:honeycomb_fh_routing_t_cost}. To route and unroute the bond as shown in Fig.~\ref{fig:FH_square_swap}\textbf{d}, there are two CSWAPs controlled on the bond selection index (a multiplicity of two comes from having two spins), and each can be controlled by a single signal qubit, so we only require eight CSWAP gates (including unswapping), costing $32$ $T$ gates.

For the implementation of the single bond-wise interaction, we invoke the diagonalization operator $V$ twice and the diagonal block encoding once. The implementation of $V$ consists of rotations between different sets of two computational basis states, six of which have angle $\pi/4$ and one has an arbitrary angle. As discussed in Appendix~\ref{sec:R_p_q_implementation},  the cost is $2T_{\mathrm{rot}}+8$ for arbitrary angle and $10$ for $\pi/4$. The total cost for each $V$ is therefore $2T_{\mathrm{rot}}+68$.

For the diagonal block encoding, each rotation is controlled on some collection of possible four qubit computational basis states, and each of the sixteen states only controls one rotation. Specifying control on each basis state requires three logical ANDs, so there are a total of $16\times 3=48$ logical ANDs, and $48$ measurement-based inverse-logical ANDs must also be used after the respective controlled rotations are implemented. Since there are four arbitrary angle rotations here, the total cost of the implementation of $H_b$ is $4T_{\mathrm{rot}}+192$.

The local bond-average block encoding therefore costs $T_0^p=32+2(2T_{\mathrm{rot}}+68)+(4T_{\mathrm{rot}}+192)=360+8T_{\mathrm{rot}}$. Its local signal register contains the two bond-selection qubits and the dilation qubit $w$, so $a=3$. For a controlled source, the two nontrivial-magnitude dilation flags acquire one extra temporary AND each. The sign-dependent $X_w$ conjugations remain unconditional and cancel on the inactive branch; endpoint eigenvalues are covered by the common $Z_w$, which becomes a CZ. Preparation, bond routing and $V,V^\dagger$ remain unconditional. This gives $C_0^p\leq368+8T_{\mathrm{rot}}$.

Including local order restoration and address arithmetic from~\Cref{eq:honeycomb_fh_routing_t_cost}, the selection cost is bounded by
\begin{align}
T_{\textsc{Sel},b}\leq16M^2+16\lceil\log_2M\rceil+8T_{\mathrm{rot}}+328.
\label{eq:fh-bond-sel-cost}
\end{align}
The total block-encoding cost is therefore
\begin{align}
T_{\textsc{Be},b}
&\leq16M^2+80\lceil\log_2M\rceil+8T_{\mathrm{rot}}+344.
\label{eq:fh-bond-be-cost}
\end{align}

\subsection{$2\times 2$ block encoding}

\subsubsection{Verification of the norm}\label{sec:fh-2x2-verification}
To verify Eq.~\eqref{eq:nn-fh-plaquette-closed-form} analytically, we use conservation of $N_\uparrow$ and $N_\downarrow$ to decompose the eight-mode Fock space into 25 sectors of dimension $\binom{4}{N_\uparrow}\binom{4}{N_\downarrow}$. 
We construct the Hamiltonian in the occupation basis, including the fermionic signs, and factor the characteristic polynomial of each sector using exact symbolic arithmetic in SymPy. 
The $(N_\uparrow,N_\downarrow)=(2,2)$ sector contains a three-dimensional invariant block with characteristic polynomial $f(E)=E^3-(4\tau^2+v^2)E-2\tau^2v$. 
An algebraic comparison with the remaining characteristic-polynomial factors shows that no eigenvalue has magnitude greater than the largest root of $f(E)$. 
Together with the spectral symmetry established above, this identifies the extreme eigenvalues of the full plaquette Hamiltonian as $\pm\lambda_\square$. 
Solving the cubic by substituting $E=2\sqrt{(4\tau^2+v^2)/3}\cos\theta$ gives Eq.~\eqref{eq:nn-fh-plaquette-closed-form}, with the principal branch of $\arccos$ selecting the largest root. 
The expression also reproduces the limits $\lambda_\square(t,0)=2|t|$ and $\lambda_\square(0,U)=|U|/4$, and extends continuously to $\lambda_\square(0,0)=0$.

For an independent numerical check, we construct the full $256\times256$ Hamiltonian using Jordan--Wigner creation and annihilation matrices and compute all its eigenvalues with a double-precision Hermitian eigensolver. 
For nonzero couplings, we rescale the matrix by $\max(\tau,v)$ before diagonalization. 
We test 231 parameter pairs, including 121 logarithmically spaced ratios $v/\tau$ between $10^{-12}$ and $10^{12}$, 100 randomly sampled coupling pairs with both signs allowed, and additional representative points and zero-coupling cases. 
At every nonzero test point, the relative discrepancy between the numerical spectral norm and Eq.~\eqref{eq:nn-fh-plaquette-closed-form} is below $7\times10^{-15}$. 
We further diagonalize the 36-dimensional $(2,2)$ sector at seven parameter ratios using 90-digit arithmetic; the largest eigenvalue agrees with the closed-form expression Eq.~\eqref{eq:nn-fh-plaquette-closed-form} to a relative discrepancy below the double precision error. 
These calculations provide an independent numerical check of the analytical result across weak and strong coupling.

\subsubsection{Cost analysis}

The $2\times2$ block encoding discussed in the main text uses QET with the bond-average unitary counted above as its source, with $T_0^p=360+8T_{\mathrm{rot}}$ and $a=3$. Its two bond-selection signals and $w$ enter each QET phase predicate; the processing signal $b$ gives $a_{\mathrm{out}}=4$. By~\Cref{eq:local_qet_t_cost}, the cost for odd $d_\star\geq1$ is
\begin{align}
T_{\textsc{BE},\square}^{p}
&\leq d_\star(360+8T_{\mathrm{rot}})+(d_\star+1)(8+T_{\mathrm{rot}})\nonumber\\
&=368d_\star+8+(9d_\star+1)T_{\mathrm{rot}}.
\label{eq:fh_plaquette_local_qet_cost}
\end{align}
The global encoding consequently has cost
\begin{align}
T_{\textsc{BE},\square}
&\leq16M^2+80\lceil\log_2 M\rceil+368d_\star-8+(9d_\star+1)T_{\mathrm{rot}}.
\label{eq:fh_plaquette_global_qet_cost}
\end{align}

\section{Honeycomb encodings}
\label{app:honeycomb_cost_accounting}

\subsection{Bond-wise encoding}
\label{app:subsec:honeycomb_bondwise}

Each NN matrix is invariant under exchanging the two axes perpendicular to $\gamma$. Hence, $(e_{\alpha_\gamma}-e_{\beta_\gamma})/\sqrt2$ is an eigenvector with eigenvalue $J_1-\Gamma_{1,\gamma}$. In the remaining basis $\{(e_{\alpha_\gamma}+e_{\beta_\gamma})/\sqrt2,e_\gamma\}$, the matrix is
\begin{align}
\begin{pmatrix}
J_1+\Gamma_{1,\gamma} & \sqrt2\Gamma'_{1,\gamma}\\
\sqrt2\Gamma'_{1,\gamma} & J_1+K_{1,\gamma}
\end{pmatrix}.
\label{eq:honeycomb_exchange_two_dimensional_block}
\end{align}
Only one variable angle is needed to diagonalize this real $2\times2$ block. The single-spin rotation in~\Cref{app:real_symmetric_exchange} can therefore be written as
\begin{align}
u=C_\gamma R_z(-\pi/4)R_x(\theta_\gamma),
\label{eq:honeycomb_structured_spin_rotation}
\end{align}
where the Clifford $C_\gamma$ maps $(x,y,z)$ cyclically to $(\alpha_\gamma,\beta_\gamma,\gamma)$ and $\theta_\gamma$ diagonalizes the remaining block. Each spin uses one variable rotation and one exact $T^\dagger$ gate, together with Clifford gates. Thus, $U=u_i\otimes u_j$ and its inverse use four synthesized rotations and four exact $T$ gates in total. The Bell transforms are Clifford. For second- and third-neighbor bonds, the exchange matrices are already diagonal, so we use $U=I$.

We use LCU to combine the 15 bond encodings within the patch. We use five signal qubits for the bond-selection register. The first two select the interaction range $n=1,2,3$, the next two select the bond color $\gamma=x,y,z$, and the last qubit is prepared with a Hadamard to select the specific bond. For NN and second-neighbor interactions, there are two bonds of each color. For third-neighbor interactions, there is only one bond of each color, so the last qubit is ignored. The total selection probabilities for NN, second-neighbor, and third-neighbor interactions are proportional to $\sum_\gamma\lambda_b(\gamma,1)$, $2\sum_\gamma\lambda_b(\gamma,2)$, and $\sum_\gamma\lambda_b(\gamma,3)$, respectively. Then, the local preparation is
\begin{align}
\textsc{Prep}^p\ket{00000}
=\frac{1}{\sqrt{\lambda_{p,b}}}\sum_{\gamma=x,y,z}
\left(\sqrt{\lambda_b(\gamma,1)}\ket1
+\sqrt{2\lambda_b(\gamma,2)}\ket2
+\sqrt{\lambda_b(\gamma,3)}\ket3\right)\ket\gamma\ket+.
\label{eq:honeycomb_bond_patch_preparation}
\end{align}
Selecting the bond encodings and reversing this preparation gives $B_{\mathrm{ref}}$, with shift and normalization as in~\Cref{eq:honeycomb_bond_patch_normalization}. The last qubit selects between two bonds for each NN or second-neighbor color. On a third-neighbor branch it is ignored, so the two values represent the same bond with half its probability on each label.

Preparing the interaction-range register requires one rotation and one controlled rotation, costing at most three synthesized rotations. For the color register, uniformly controlled rotations use at most four rotations for its first qubit and eight for its second~\cite{mottonen2005transformation}. The pair qubit uses a Hadamard. Hence, local preparation and its inverse contribute at most $2(3+4+8)T_{\mathrm{rot}}(\epsilon)=30T_{\mathrm{rot}}(\epsilon)$. Unary iteration over the fifteen distinct bonds costs at most $4(15-1)=56$ $T$ gates~\cite{babbush2018encoding}; the pair qubit is ignored on the third-neighbor branches.

For each selected bond, the eigenbasis changes remain unconditional and cancel on inactive branches. The spectrum of $A_{ij}$ has endpoints $+1$ and $-1$, leaving at most two eigenvalues with nontrivial ancilla rotations. A Clifford permutation of the two eigenvalue-label bits places these two states on one branch of a label bit $a$. Given the bond-selection flag $f$, compute $q=f\land a$ and $r=q\land b$ using two temporary ANDs, where $b$ is the other label bit. The flags $r$ and $q\oplus r$ select the two rotations. The complementary flag $f\oplus q$ and the appropriate value of $b$ select the negative endpoint using a Clifford controlled phase. The common ancilla $Z$ also needs only a Clifford control. Thus, each bond contributes at most eight $T$ gates for the two ANDs and four synthesized rotations for the two controlled $R_y$ gates. This construction also gives an upper bound when the spectrum has fewer distinct eigenvalues.

The six NN basis changes and their inverses contribute $24T_{\mathrm{rot}}(\epsilon)+24$, while the remaining nine bonds use only Clifford Bell transforms. Adding the preparation, unary iteration, and eigenvalue selection gives
\begin{align}
T_{\textsc{BE}}^p
&\leq\left[30+24+15(4)\right]T_{\mathrm{rot}}(\epsilon)
+56+24+15(8)\nonumber\\
&=200+114T_{\mathrm{rot}}(\epsilon).
\label{eq:honeycomb_bond_local_t_cost}
\end{align}
Combining this with the routing and \textsc{Prep} costs give,
\begin{align}
T_{\textsc{BE},b}^{\hexago}
&\leq2\left(32\lceil\log_2M\rceil+8\right)+\left(8N_s+16\lceil\log_2M\rceil-32\right)
+200+114T_{\mathrm{rot}}(\epsilon)\nonumber\\
&=8N_s+80\left\lceil\frac{\log_2N_s-1}{2}\right\rceil
+114T_{\mathrm{rot}}(\epsilon)+184.
\label{eq:honeycomb_bond_total_t_cost}
\end{align}

\subsection{Zeeman-field extension}
\label{app:subsec:honeycomb_bondwise_zeeman}

For the Zeeman extension, we reuse the exchange-matrix analysis for $w_g\mathbf M$, whose eigenvalues are $\tilde J_c$. We suppress the labels $(\gamma,n)$ for a fixed bond. The rotation $O$ is unchanged by this positive rescaling and also rotates the field. Applying $U=u_i\otimes u_j$ gives
\begin{align}
U^\dagger K^b_{ij}U
=\sum_{c\in\{x,y,z\}}\tilde J_cS_i^cS_j^c
+\sum_{c\in\{x,y,z\}}f_c(S_i^c+S_j^c),
\qquad
\mathbf f=\eta_g O^T\mathbf h.
\label{eq:honeycomb_rotated_field_bond}
\end{align}
The Bell transform separates this operator into the unchanged singlet energy $E_-$ and the triplet matrix $\mathbf T$ in~\Cref{eq:honeycomb_field_block}. The three triplet energies have closed-form expressions~\cite{kopp2008efficient} of the form, 
\begin{align}
E_k=q+2r_T\cos\left(\phi-\frac{2\pi k}{3}\right),\qquad k=0,1,2.
\label{eq:honeycomb_zeeman_eigenvalues}
\end{align}
where,
\begin{align}
q=\frac{\operatorname{tr}\mathbf T}{3},\qquad
\mathbf C=\mathbf T-qI_3,\qquad
r_T=\sqrt{\frac{\operatorname{tr}(\mathbf C^2)}{6}}, \qquad
\phi=\frac13\arccos\left(\frac{\det\mathbf C}{2r_T^3}\right)
\end{align}
and $r_T>0$. If $r_T=0$, all three energies equal $q$.

We reuse the local preparation oracle with the normalizations of $K^b_{ij}$ as the LCU weights, and sum their shifts and normalizations over the patch. The NN normalizations $\lambda_Z(\gamma,1)$ already include the bond weight $1/2$. Hence, in~\Cref{eq:honeycomb_bond_patch_preparation}, we replace $\sqrt{\lambda_b(\gamma,1)}$ by $\sqrt{2\lambda_Z(\gamma,1)}$ and normalize by the new total patch weight. The second- and third-neighbor coefficients are unchanged.

After choosing eigenvector phases, we decompose $Q$ into complex Givens rotations~\cite{barenco1995elementary}. Each is implemented with two $R_z$ rotations and a controlled $R_y$. The transformations $Q$ and $Q^\dagger$ remain unconditional and cancel on inactive branches. They are needed only for NN bonds. There are at most three complex Givens rotations for the three-dimensional triplet block. Each controlled $R_y$ uses two synthesized rotations and two CNOTs, so each embedded Givens rotation costs at most four synthesized rotations. Thus, $Q$ and $Q^\dagger$ add at most $2(3)(4)T_{\mathrm{rot}}(\epsilon)=24T_{\mathrm{rot}}(\epsilon)$ per NN bond. The same eigenvalue-selection controls apply, since the normalized spectrum still has endpoints $+1$ and $-1$.

The six NN bonds therefore add at most $6(24)T_{\mathrm{rot}}(\epsilon)$ to the field-free local cost. The preparation, bond selection, and eigenvalue-selection bounds are unchanged, giving
\begin{align}
T_{\textsc{BE},Z}^{p}
&\leq200+\left[114+6(24)\right]T_{\mathrm{rot}}(\epsilon)\nonumber\\
&=200+258T_{\mathrm{rot}}(\epsilon).
\label{eq:honeycomb_zeeman_local_t_cost}
\end{align}
Adding the same spatial preparation and routing cost bounds gives
\begin{align}
T_{\textsc{BE},Z}^{\hexago}
&\leq2\left(32\lceil\log_2M\rceil+8\right)+\left(8N_s+16\lceil\log_2M\rceil-32\right)
+200+258T_{\mathrm{rot}}(\epsilon)\nonumber\\
&=8N_s+80\left\lceil\frac{\log_2N_s-1}{2}\right\rceil
+258T_{\mathrm{rot}}(\epsilon)+184.
\label{eq:honeycomb_zeeman_global_t_cost}
\end{align}

\subsection{Star encoding using anticommuting groups}
\label{app:subsec:honeycomb_anti}

For the anticommuting star encoding, the ten group unitaries are implemented using~\Cref{app:anticommuting_pauli_groups}. The conjugating rotations remain unconditional, and only the reference Pauli needs the group-selection control.

We prepare the LCU weights using four qubits and leveraging the repeated group normalizations. The first two select NN, second-neighbor, or field interactions, with total weights $W_{\mathrm{NN}}=\beta_1+2\beta_2$, $W_2=\beta_4+2\beta_5$, and $W_h=4\beta_7$, respectively. The remaining two qubits select the groups within each family, giving,
\begin{align}
\textsc{Prep}^\star\ket{0000}
=\frac{1}{\sqrt{\beta^\star}}\Big[\ket{00}\left(\sqrt{\beta_1}\ket0+\sqrt{2\beta_2}\ket1\right)
+\ket{01}\left(\sqrt{\beta_4}\ket0+\sqrt{2\beta_5}\ket1\right)
+\ket{10}\sqrt{2\beta_7}\left(\ket0+\ket1\right)
\Big]\ket+.
\label{eq:honeycomb_star_group_preparation}
\end{align}
On the NN branch, the third qubit selects $\mathcal G_1$ or the pair $\{\mathcal G_2,\mathcal G_3\}$. The last qubit selects between the pair and is ignored for $\mathcal G_1$. The second-neighbor branch works the same way for $\mathcal G_4,\mathcal G_5,\mathcal G_6$. On the field branch, the last two qubits select the four site groups uniformly. 

Prepare the first two qubits with a rotation and a controlled rotation. They control an $R_y$ rotation on the third qubit with angles $\phi_{\mathrm{NN}}=2\arccos\sqrt{\beta_1/W_{\mathrm{NN}}}$ for $\ket{00}$, $\phi_2=2\arccos\sqrt{\beta_4/W_2}$ for $\ket{01}$, and $\pi/2$ for $\ket{10}$. Choose the unused-state angle as $\phi_{11}=\phi_{\mathrm{NN}}+\phi_2-\pi/2$ in the uniformly controlled rotation construction~\cite{mottonen2005transformation}. Apply a Hadamard to the last qubit, and use unary iteration to generate the group-selection flags.
The first two qubits require at most three synthesized rotations. The chosen unused-state angle makes one elementary angle in the third-qubit uniformly controlled rotation zero, leaving at most three synthesized rotations for that step. Thus, the preparation costs at most $6T_{\mathrm{rot}}(\epsilon)$, and its inverse has the same cost. Zero-weight choices of interaction range are omitted.

The three NN groups contain at most nine strings each, while the three second-neighbor groups and four field groups contain at most three strings each. The selected group circuits therefore contribute at most $2[3(9-1)+3(3-1)+4(3-1)]T_{\mathrm{rot}}(\epsilon)=76T_{\mathrm{rot}}(\epsilon)$. Unary iteration over the ten distinct groups costs at most $4(10-1)=36$ $T$ gates~\cite{babbush2018encoding}. For $\mathcal G_1$ and $\mathcal G_4$, the last selection qubit is ignored, so their duplicated labels do not require separate group circuits. Combining these costs gives
\begin{align}
T_{\textsc{BE}}^\star
&\leq36+\left[2(6)+76\right]T_{\mathrm{rot}}(\epsilon)\nonumber\\
&=36+88T_{\mathrm{rot}}(\epsilon).
\label{eq:honeycomb_star_anti_t_cost}
\end{align}

\subsection{Symmetry reduction of the four-spin star}
\label{appendix:star_additional_symmetries}

We construct the eigenbasis transformation for the exchange Hamiltonian $H^\star_{\mathcal S,0}$ under the color-independent coupling assumptions of the main text. The cyclic symmetry first separates the Hamiltonian into three sectors. Reflection and time reversal then reduce the diagonalization to two real $2\times2$ matrices and one real $6\times6$ matrix. We use these smaller matrices to construct the circuit and account for its $T$-cost.

First, define the leaf permutation as,
\begin{align}
P_\ell\ket a_c\ket{b_x,b_y,b_z}_\ell
=\ket a_c\ket{b_z,b_x,b_y}_\ell,\qquad
P_\ell=\operatorname{SWAP}_{\ell_x,\ell_y}\operatorname{SWAP}_{\ell_y,\ell_z}.
\label{eq:honeycomb_star_leaf_cycle}
\end{align}
The rightmost gate acts first. Combine this permutation with a $2\pi/3$ rotation of every spin about $(1,1,1)/\sqrt3$,
\begin{align}
u_3=\frac12[I-i(X+Y+Z)],\qquad C_3=P_\ell u_3^{\otimes4}.
\label{eq:appendix_star_cyclic_symmetry}
\end{align}
The rotation cycles the spin axes as $X\to Y\to Z\to X$, and the leaf permutation moves each rotated interaction to the bond of the next color. The coupling matrices obey the same cycle, so $[C_3,H^\star_{\mathcal S,0}]=0$. Since $u_3^3=-I$ and there are four spins, $C_3^3=I$. Its eigenvalues are therefore $1,\omega,\omega^2$, with $\omega=e^{2\pi i/3}$. The corresponding projectors are,
\begin{align}
\Pi_m=\frac13\sum_{r=0}^2\omega^{-mr}C_3^r,\qquad m=0,1,2.
\label{eq:honeycomb_star_c3_projectors}
\end{align}
Commutation with $H^\star_{\mathcal S,0}$ ensures that each sector is preserved. The leaf cycle contributes $\operatorname{tr}(u_3^3)=-2$ to $\operatorname{tr}C_3$, while the center contributes $\operatorname{tr}u_3=1$. Similarly, $\operatorname{tr}C_3^2=-2$. Hence,
\begin{align}
\operatorname{rank}\Pi_m
=\frac13\left[16-2\omega^{-m}-2\omega^{-2m}\right],\qquad
(\operatorname{rank}\Pi_0,\operatorname{rank}\Pi_1,\operatorname{rank}\Pi_2)=(4,6,6).
\label{eq:honeycomb_star_c3_ranks}
\end{align}

We diagonalize the spin rotation and leaf permutation separately. First define,
\begin{align}
W=R_z\!\left(\frac\pi4\right)R_y\!\left(\arccos\frac1{\sqrt3}\right),\qquad
W^\dagger u_3W=e^{-i\pi Z/3}.
\label{eq:honeycomb_star_c3_axis_change}
\end{align}
Applying $(W^\dagger)^{\otimes4}$ makes the spin-rotation factor diagonal and commutes with $P_\ell$, since all leaves receive the same rotation. Number the leaf wires $1,2,3$ in the order $\ell_x,\ell_y,\ell_z$, and use the Clebsch--Gordan circuit $V_\ell$ from~\Cref{app:clebsch_gordan} for the three-spin case. Here $V_\ell$ maps computational labels to the coupled-spin states. The operator $V_\ell^\dagger P_\ell V_\ell$ fixes the labels $000,011,100,111$. On the ordered pairs $(001,010)$ and $(101,110)$, its matrices are $e^{+2\pi iY/3}$ and $e^{-2\pi iY/3}$, respectively. We diagonalize these pairs using,
\begin{align}
K_\ell=\exp\!\left[-\frac{i\pi}{8}(X_2X_3+Y_2Y_3)\right],\nonumber\\
K_\ell V_\ell^\dagger P_\ell V_\ell K_\ell^\dagger
=\operatorname{diag}(1,\omega,\omega^2,1,1,\omega^2,\omega,1).
\label{eq:honeycomb_star_leaf_cycle_diagonal}
\end{align}
Both leaf transforms preserve Hamming weight, so they leave the diagonal collective spin rotation unchanged. The full cyclic-symmetry transform is therefore,
\begin{align}
F_{C_3}=(I_c\otimes K_\ell V_\ell^\dagger)(W^\dagger)^{\otimes4}.
\label{eq:honeycomb_star_c3_transform}
\end{align}
In the four-bit order $(c,\ell_x,\ell_y,\ell_z)$, the sector labels are $(1,3,12,13)$ for $m=0$, $(0,2,6,7,9,11)$ for $m=1$, and $(4,5,8,10,14,15)$ for $m=2$. Integers denote their four-bit computational strings. Direct sums below group these labels for presentation and do not require an explicit permutation.

The exchange Hamiltonian has a further symmetry that exchanges the $x$ and $y$ bonds. Define,
\begin{align}
u_s=-\frac{i}{\sqrt2}(X-Y),\qquad
\mathcal R=\operatorname{SWAP}_{\ell_x,\ell_y}u_s^{\otimes4}.
\label{eq:honeycomb_star_reflection}
\end{align}
The spin rotation sends $X\to-Y$, $Y\to-X$, and $Z\to-Z$. Each exchange term contains two spin operators, so the minus signs cancel. Swapping the two leaves then restores the bond positions, including those in the leaf triangle. Thus,
\begin{align}
[\mathcal R,H^\star_{\mathcal S,0}]=0,\qquad
\mathcal R^2=I,\qquad
\mathcal R C_3\mathcal R^\dagger=C_3^\dagger.
\label{eq:honeycomb_star_dihedral_relations}
\end{align}
Together, $C_3$ and $\mathcal R$ generate the dihedral group $D_3$. Reflection preserves the $m=0$ sector and exchanges $m=1$ with $m=2$. Within $m=0$, the reflection-parity projectors are $\Pi_{0,\pm}=\Pi_0(I\pm\mathcal R)/2$. Since $\operatorname{tr}(\Pi_0\mathcal R)=0$, each parity sector has dimension two. If $\{\ket{b_j}\}$ is a basis for $m=1$, choose $\{\mathcal R\ket{b_j}\}$ for $m=2$. Commutation with $\mathcal R$ makes the exchange matrices identical in these two bases. Thus, a symmetry basis gives,
\begin{align}
F_{D_3}H^\star_{\mathcal S,0}F_{D_3}^\dagger
=A_+\oplus A_-\oplus B\oplus B,
\label{eq:appendix_star_dihedral_blocks}
\end{align}
where $A_\pm$ are $2\times2$ matrices and $B$ is a $6\times6$ matrix.

Time-reversal symmetry allows the reduced blocks to be chosen real. Define
$\Theta=Y^{\otimes4}\mathcal K$, where $\mathcal K$ denotes complex
conjugation. Within the reflection-parity sectors, we choose a basis
invariant under $\Theta$. Within one six-dimensional sector, we instead
use $\mathcal R\Theta$, which preserves that sector. Choosing the other
sector's basis by reflection then gives two identical real matrices $B$. Starting from $F_{C_3}$, this basis change uses diagonal phase gates and
two two-level Hadamards that separate the reflection parities. We denote
the resulting transformation by $F_{D_3}$.

We diagonalize $A_\pm$ analytically using real plane rotations $O_\pm$. For $B$, we compute a real orthogonal eigenvector matrix $O_B$ classically. Choosing its column signs to give determinant one allows a decomposition into at most fifteen real plane rotations. To apply this circuit once for both copies, we permute the symmetry-basis labels so that one qubit distinguishes the two copies of $B$, while the other three use the same six labels within each copy. The remaining two labels in each sector hold $A_+$ and $A_-$, respectively. This allows us to apply the eigenbasis transformation $O_B$ once on the three-qubit register.  Hence, define $\widehat O_B=O_B\oplus I_2$ on the three-bit register and $U_A=\ket0\bra0\otimes(I_6\oplus O_+)+\ket1\bra1\otimes(I_6\oplus O_-).$

The shared $\widehat O_B$ leaves the two parity blocks unchanged, while $U_A$ acts only on those blocks. The complete eigenbasis transformation is therefore,
\begin{align}
O_\pm^T A_\pm O_\pm&=D_\pm,\qquad O_B^TBO_B=D_B,\nonumber\\
V^\star_0&=U_A^T(I_2\otimes\widehat O_B^T)\mathcal P F_{D_3},\nonumber\\
V^\star_0H^\star_{\mathcal S,0}V^{\star\dagger}_0
&=(D_B\oplus D_+)\oplus(D_B\oplus D_-).
\label{eq:honeycomb_star_exchange_eigenbasis}
\end{align}
The transformation $V^\star_0$ maps eigenstates to computational labels. It uses at most seventeen coupling-dependent plane rotations per direction (fifteen for the shared $O_B$ and one for each of $O_+$ and $O_-$). The sector bit is one of the four system qubits, so this doesn't require additional ancillas.

The permutation $\mathcal P$ can be implemented using three Toffoli gates, one CSWAP, and Clifford gates. In the measurement-assisted model used throughout, each Toffoli and CSWAP costs four $T$ gates. Thus, $\mathcal P$ and its inverse together cost thirty-two $T$ gates.

Let $T_{\mathrm{rot}}(\epsilon)$ denote the $T$-cost of synthesizing one single-qubit axial rotation to error $\epsilon$. Clifford gates contribute no $T$ gates, and a temporary logical AND costs four $T$ gates to compute and none to erase using measurement~\cite{gidney2018halving}. We first count the fixed symmetry transforms and their inverses. The pair $F_{C_3},F_{C_3}^\dagger$ uses at most twelve synthesized rotations and sixteen exact $T$ gates. Eight synthesized rotations come from the four $R_y$ factors of $W$ in both directions, and four come from the Givens gate at angle $\arcsin(1/\sqrt3)$ in $V_\ell$ and its inverse. The $R_z(\pi/4)$ factors, the Givens gate at angle $\pi/4$, and $K_\ell$ contribute the sixteen exact $T$ gates. The diagonal phase adjustment uses at most fifteen rotations about nonidentity $Z$ strings per direction, giving thirty synthesized rotations in total. Each of the two two-level Hadamards, together with its associated phase correction, costs at most thirteen $T$ gates: eight for the selection predicate, two for the controlled Hadamard, and three for the controlled phase. Including both directions gives fifty-two $T$ gates.
\begin{align}
T_{F_{D_3}}+T_{F_{D_3}^\dagger}
\leq(12+30)T_{\mathrm{rot}}(\epsilon)+16+4(13)
=42T_{\mathrm{rot}}(\epsilon)+68.
\label{eq:honeycomb_star_symmetry_basis_cost}
\end{align}
The shared $O_B$ uses fifteen real two-level rotations on three qubits per direction. After a Clifford relabeling within that register, each is an $R_y$ on one target with a two-bit condition. One temporary AND computes this condition at four $T$ gates, and the resulting singly controlled $R_y$ uses two synthesized rotations. Each of the two parity rotations acts on four qubits and retains a three-bit condition, costing eight $T$ gates and two synthesized rotations. Including both directions and the permutation gives,
\begin{align}
T_{V^\star_0}+T_{V^{\star\dagger}_0}
&\leq42T_{\mathrm{rot}}(\epsilon)+100 +30\left[2T_{\mathrm{rot}}(\epsilon)+4\right]
+4\left[2T_{\mathrm{rot}}(\epsilon)+8\right]\nonumber\\
&=110T_{\mathrm{rot}}(\epsilon)+252.
\label{eq:honeycomb_star_exchange_basis_cost}
\end{align}

\subsection{Star encoding using symmetry-reduced spectral method}
\label{app:subsec:honeycomb_symm_spec}

For the symmetry-reduced star encoding, we use three qubits to prepare the local LCU weights,
\begin{align}
\textsc{Prep}^\star\ket{000}
=\left(\sqrt{\frac{\lambda^\star_0}{\lambda^\star}}\ket0
+\sqrt{\frac{\lambda^\star_h}{\lambda^\star}}\ket1\right)\ket+\ket+.
\label{eq:honeycomb_symmetry_star_preparation}
\end{align}
The first qubit selects exchange or field. The last two select the field site and are ignored on the exchange branch. Preparing the first qubit requires one synthesized rotation, while the last two use Hadamards. Thus, the preparation and its inverse contribute $2T_{\mathrm{rot}}(\epsilon)$ to the complete star cost.

The exchange dilation is implemented by a uniformly controlled rotation on $w$, with the four system qubits as controls. On the exchange branch, control the multiplexor rotations and CNOTs with the exchange-selection flag, and replace the common $Z_w$ by a CZ. The eigenbasis changes remain unconditional and cancel on the inactive branch. The field branches use the group circuits already constructed, with their selection flags generated by shared temporary ANDs.

The basis change and its inverse cost at most $110T_{\mathrm{rot}}(\epsilon)+252$ by~\Cref{eq:honeycomb_star_exchange_basis_cost}. The four-control multiplexor adds at most sixteen synthesized rotations and sixteen CNOTs~\cite{mottonen2005transformation}. Controlling the multiplexor replaces each rotation by a controlled rotation, requiring thirty-two synthesized rotations in total, and each CNOT by a Toffoli, adding sixty-four $T$ gates. The controlled $Z_w$ is Clifford. Hence,
\begin{align}
T^\star_{\textsc{BE},0}
&\leq(110+16)T_{\mathrm{rot}}(\epsilon)+252\nonumber\\
&=126T_{\mathrm{rot}}(\epsilon)+252,
\label{eq:honeycomb_exchange_star_encoding_cost}\\
T_{\mathrm{c}B^\star_0}
&\leq(110+32)T_{\mathrm{rot}}(\epsilon)+252+64\nonumber\\
&=142T_{\mathrm{rot}}(\epsilon)+316.
\label{eq:honeycomb_star_controlled_exchange_cost}
\end{align}
The four field-group circuits contribute sixteen synthesized rotations. Their selection flags use three shared temporary ANDs: one combines the field flag with the first site bit, and two more split the resulting two prefixes using the second site bit. Complementary flags are obtained using CNOTs. These computations cost twelve $T$ gates~\cite{gidney2018halving}. Adding the field circuits and the preparation and its inverse gives
\begin{align}
T_{\textsc{BE}}^\star
&\leq(142+16+2)T_{\mathrm{rot}}(\epsilon)+316+12\nonumber\\
&=160T_{\mathrm{rot}}(\epsilon)+328.
\label{eq:honeycomb_symmetry_star_encoding_cost}
\end{align}

\subsection{Complete patch encoding}
\label{app:subsec:honeycomb_combined}

To combine the two stars and three third-neighbor bonds through LCU, we prepare the local LCU weights on three qubits as,
\begin{align}
\textsc{Prep}^p_j\ket{000}
=\frac{1}{\sqrt{\lambda_{p,j}}}\Big[
&\sqrt{2\lambda^\star_j}\ket{00}
+\sqrt{\lambda_{3,x}}\ket{01} + \sqrt{\lambda_{3,y}}\ket{10}
+\sqrt{\lambda_{3,z}}\ket{11}\Big]\ket+.
\label{eq:honeycomb_two_star_patch_preparation}
\end{align}
The first two qubits select a star encoding or one of the three bonds. On the star branch, the last qubit selects between the two stars and is ignored on the bond branches. We compute the flag for the second star and use four CSWAPs to bring its center and color-matched leaves to the first star's register. After applying the shared star circuit, we reverse these swaps. Thus, we implement the star circuit once. Both star constructions use four inner encoding ancillas, which can also be used by the bond encodings, giving seven local encoding ancillas in total, in addition to temporary work qubits.

Preparing the first two qubits costs at most three synthesized rotations, while the last uses a Hadamard. Unary iteration over the four outer branches costs at most twelve $T$ gates. Computing the second-star flag costs one temporary AND, and the eight forward and reverse CSWAPs cost thirty-two $T$ gates. For the anticommuting star, we start its inner unary iteration with the outer star flag, retaining the bound in~\Cref{eq:honeycomb_star_anti_t_cost}. For the spectral star, let $s$ be the outer star flag and $t$ its inner field-selection bit. One temporary AND computes $s\land t$ for the field branch, and $s\oplus(s\land t)$ gives the exchange flag. Thus, enabling the complete spectral star adds four $T$ gates to~\Cref{eq:honeycomb_symmetry_star_encoding_cost}. Combining these costs with the three bond encodings and the outer preparation and its inverse gives,
\begin{align}
T_{\textsc{BE},j}^p
&=2T_{\textsc{Prep},j}^p+T_{\textsc{Sel},j}^p\leq\begin{cases}
100T_{\mathrm{rot}}(\epsilon)+108,&j=\mathrm{anti},\\
172T_{\mathrm{rot}}(\epsilon)+404,&j=\mathrm{spec}.
\end{cases}
\label{eq:honeycomb_two_star_local_cost_derivation}
\end{align}
Adding the address preparation and routing costs give
\begin{align}
T_{\textsc{BE},j}
&\leq2\left(32\lceil\log_2M\rceil+8\right)+\left(8N_s+16\lceil\log_2M\rceil-32\right)
+T_{\textsc{BE},j}^p.
\end{align}
Substituting the two local bounds and using $N_s=2M^2$ gives~\Cref{eq:honeycomb_two_star_anti_total_t_cost,eq:honeycomb_two_star_spec_total_t_cost}.

\subsection{Baseline using product-wise Pauli LCU approach~\cite{yoshioka2024hunting,babbush2018encoding}}
\label{app:honeycomb_pauli_baseline}

To the best of our knowledge, there is no existing baseline for the spin-liquid Hamiltonian using the current state-of-art approaches that use Pauli LCU along with the product-wise structure of the interaction terms and implemented efficiently using unary iteration~\cite{yoshioka2024hunting,babbush2018encoding}. To construct a simple baseline, we expand the honeycomb Hamiltonian directly in
Pauli strings. Combining coefficients of identical strings before
taking absolute values, define
\begin{align}
b_{1,\gamma}
&=\frac{|J_1+K_{1,\gamma}|+2|J_1|
+2|\Gamma_{1,\gamma}|+4|\Gamma'_{1,\gamma}|}{4},\\
b_{r,\gamma}
&=\frac{|J_r+K_{r,\gamma}|+2|J_r|}{4},
\qquad r=2,3.
\end{align}
The resulting shift and Pauli LCU normalization are
\begin{align}
\Delta_{\mathrm P}&=\frac{AN_s}{4},\\
\Lambda_{\mathrm P}
&=\frac{N_s}{2}
\left[
\sum_{\gamma=x,y,z}
\left(b_{1,\gamma}+2b_{2,\gamma}+b_{3,\gamma}\right)
+\|\mathbf h\|_1
\right].
\label{eq:honeycomb_pauli_normalization}
\end{align}
The constant $\Delta_{\mathrm P}I$ is omitted from the circuit.

Since the Hamiltonian is 2-local, we can use the same product-wise trick as in Ref.~\cite{yoshioka2024hunting}. We have two address registers for the two interacting bodies. For \textsc{Sel}, any arbitrary indexed one-body Pauli can be implemented by two unary iterations, one implementing $X$ controlled on $u_1$ and another implementing $Z$ controlled on $v_1$. Identity and $Y$ can be implemented when $u_1=v_1=0$ and $u_1=v_1=1$, respectively, after taking care of the phase. This is done with the other address as well with controls $u_2,v_2$, so there are four controlled unary iterations in total. Each indexed operation costs at most $4(N_s-1)$ $T$ gates. For \textsc{Prep}, suitable two address and $u_1,v_1,u_2,v_2$ superpositions can be implemented so that the block encoding is correct. Ignoring everything other than unary iteration in \textsc{Sel}, the cost of the block encoding is approximately
\begin{align}
T_{\textsc{BE},\mathrm P}\approx16N_s.\label{eq:honeycomb_pauli_baseline_cost}
\end{align}
We use this leading order cost as the baseline for the spin liquid Hamiltonian.  

\end{document}